\documentclass[prx, twocolumn, superscriptaddress]{revtex4-2}
\usepackage{bm, amsmath, amsfonts, amssymb, mathtools, braket, bbm}
\usepackage{multirow}
\usepackage{graphicx}
\usepackage{float, color, xcolor}
\usepackage[caption = false]{subfig}

\usepackage[
pagebackref=false,
colorlinks=true,
linkcolor=blue,
urlcolor=blue,
filecolor=black,
citecolor=red,
pdfstartview=FitV,
pdftitle={},
pdfauthor={},
pdfsubject={},
pdfkeywords={},
pdfpagemode=None,
bookmarksopen=true
]{hyperref}

\begin{document}

\newcommand{\ii}{\text{i}}

\newcommand{\magenta}[1]{{\textcolor{magenta}{#1}}}
\newcommand{\orange}[1]{{\textcolor{orange}{#1}}}
\newcommand{\teal}[1]{{\textcolor{teal}{#1}}}
\newcommand{\violet}[1]{{\textcolor{violet}{#1}}}

\title{Symmetry of Open Quantum Systems: Classification of Dissipative Quantum Chaos}

\author{Kohei Kawabata}
\thanks{The authors are listed in alphabetical order.}
\affiliation{Department of Physics, Princeton University, Princeton, New Jersey 08544, USA}
\affiliation{Institute for Solid State Physics, University of Tokyo, Kashiwa, Chiba 277-8581, Japan}

\author{Anish Kulkarni}
\thanks{The authors are listed in alphabetical order.}
\affiliation{Department of Physics, Princeton University, Princeton, New Jersey 08544, USA}

\author{Jiachen Li}
\thanks{The authors are listed in alphabetical order.}
\affiliation{Department of Physics, Princeton University, Princeton, New Jersey 08544, USA}

\author{Tokiro Numasawa}
\thanks{The authors are listed in alphabetical order.}
\affiliation{Institute for Solid State Physics, University of Tokyo, Kashiwa, Chiba 277-8581, Japan}

\author{Shinsei Ryu}
\thanks{The authors are listed in alphabetical order.}
\affiliation{Department of Physics, Princeton University, Princeton, New Jersey 08544, USA}

\date{\today}

\begin{abstract}
We develop a theory of symmetry in open quantum systems.
Using the operator-state mapping, we characterize symmetry of Liouvillian superoperators for the open quantum dynamics by symmetry of operators in the double Hilbert space and apply the 38-fold internal-symmetry classification of non-Hermitian operators.
We find rich symmetry classification due to the interplay between symmetry in the corresponding closed quantum systems and symmetry inherent in the construction of the Liouvillian superoperators.
As an illustrative example of open quantum bosonic systems, we study symmetry classes of dissipative quantum spin models.
For open quantum fermionic systems, we develop the $\mathbb{Z}_4$ classification of fermion parity symmetry and antiunitary symmetry in the double Hilbert space, which contrasts with the $\mathbb{Z}_8$ classification in closed quantum systems.
We also develop the symmetry classification of open quantum fermionic many-body systems---a dissipative generalization of the Sachdev-Ye-Kitaev (SYK) model described by the Lindblad master equation.
We establish the periodic tables of the SYK Lindbladians and elucidate the difference from the SYK Hamiltonians.
Furthermore, from extensive numerical calculations, we study its complex-spectral statistics and demonstrate dissipative quantum chaos enriched by symmetry.
\end{abstract}

\maketitle


\section{Introduction}

Symmetry underlies a variety of phenomena and plays a pivotal role in physics.
Spontaneous breaking of symmetry characterizes phases of matter and enables the universal description of phase transitions and critical phenomena~\cite{Goldenfeld-textbook, Cardy-textbook, Sachdev-textbook}.
In general, Hermitian operators are classified by 10-fold internal symmetry classes based on time reversal, charge conjugation (particle-hole transformation), and chiral transformation~\cite{Wigner-51, *Wigner-58, Dyson-62, AZ-97}.
These 10-fold symmetry classes
also lead to the classification of the Anderson transitions~\cite{Evers-review} and topological insulators and superconductors~\cite{Schnyder-08, *Ryu-10, Kitaev-09, CTSR-review} in closed quantum systems.
Even in the many-body case, symmetry protects and enriches topological phases of matter.
Another prime application of symmetry is the characterization of quantum chaos~\cite{Haake-textbook}, which is fundamentally relevant to the foundations of statistical mechanics~\cite{Rigol-review}.
In a number of model calculations,
the spectrum of a nonintegrable quantum system obeys the random-matrix statistics~\cite{BGS-84} while the spectrum of an integrable quantum system obeys the Poisson statistics~\cite{Berry-Tabor-77}.
Here, the universality classes of the random-matrix statistics are determined solely by the 10-fold symmetry and do not rely on any specific details of the system.
Recently, the Sachdev-Ye-Kitaev (SYK) model has attracted widespread interest as a prototype that exhibits 
quantum chaotic behavior
~\cite{Sachdev-Ye-93, Kitaev-KITP15, Sachdev-15, Polchinski-Rosenhaus-16, Maldacena-Stanford-16, Gu-17, Song-17, Rosenhaus-review, Sachdev-review}.
The symmetry classification of the SYK model was developed in terms of the number $N$ of fermion flavors and the number $q$ of many-body interactions
~\cite{You-17, Fu-16, GarciaGarcia-16, Cotler-17, Li-17, Kanazawa-17, Behrends-19, Sun-20}.
This symmetry classification
is also closely related to the $\mathbb{Z}_8$ classification of fermionic topological phases in one dimension~\cite{Fidkowski-Kitaev-10, *Fidkowski-Kitaev-11, Turner-11}.

Meanwhile, recent years have seen remarkable development in the physics of open quantum systems.
In contrast to closed quantum systems, open quantum systems are coupled to the external environment and are no longer described by Hermitian Hamiltonians~\cite{Nielsen-textbook, Breuer-textbook, Rivas-textbook}.
Even though dissipation can destroy quantum coherence and wash out quantum phenomena, engineered dissipation was shown to be useful
in
quantum computation and state preparation~\cite{Verstraete-09, Diehl-08, *Diehl-11, Barreiro-11}.
Researchers have also found rich phenomena in the physics of non-Hermitian Hamiltonians~\cite{Bender-review, Konotop-review, Christodoulides-review}, which
effectively describe open quantum systems subject to continuous monitoring and postselection of the null measurement outcome~\cite{Carmichael-textbook, Plenio-review, Daley-review}, as well as open classical systems.
For example, non-Hermiticity gives rise to new types of the Anderson transitions~\cite{Hatano-Nelson-96, *Hatano-Nelson-97, Efetov-97, Feinberg-97, Brouwer-97, Tzortzakakis-20, Huang-20, KR-21, Luo-21L, *Luo-21B, *Luo-22R, Liu-Fulga-21}
and topological phases~\cite{Bergholtz-review, Okuma-Sato-review} that have no analogs in closed quantum systems.
Notably, non-Hermiticity also changes the nature of symmetry.
While Hermitian operators are classified according to the 10-fold internal symmetry~\cite{AZ-97}, non-Hermitian operators are classified according to the 38-fold internal symmetry~\cite{Bernard-LeClair-02, KSUS-19, Zhou-Lee-19}.
In a similar manner to closed quantum systems, this 38-fold symmetry determines the universality classes of non-Hermitian random matrices, as well as the Anderson transitions and topological phases of non-Hermitian systems.

Furthermore, the characterization of chaotic behavior in open quantum systems has attracted growing attention~\cite{Grobe-88, *Grobe-89, Xu-19, Hamazaki-19, *Hamazaki-22, Denisov-19, Can-19PRL, *Can-19JPhysA, Hamazaki-20, Akemann-19, Sa-20, Wang-20, JiachenLi-21, Tarnowski-21, Cornelius-22, Prasad-22, GJ-22, Xiao-22, Shivam-22}.
Similarly to quantum chaos in closed systems, the complex spectra of several nonintegrable open quantum systems were numerically shown to obey the spectral statistics of non-Hermitian random matrices~\cite{Ginibre-65, Girko-85, Haake-textbook}, 
while the integrable counterparts were shown to obey the Poisson statistics for complex numbers.
Thus, complex-spectral statistics give a measure to quantify dissipative quantum chaos.
Several recent works have proposed prototypes of open quantum many-body systems that exhibit dissipative quantum chaos, such as random bosonic Lindbladians~\cite{Denisov-19, Wang-20}.
Furthermore, as prototypical open quantum systems of strongly correlated fermions, 
the non-Hermitian SYK Hamiltonians (i.e., SYK Hamiltonians with complex-valued parameters)~\cite{ChunxiaoLiu-21, GarciaGarcia-22PRL, PengfeiZhang-21, *Jian-21, GarciaGarcia-22PRX}
and the SYK Lindbladians (i.e., SYK Hamiltonians coupled to Markovian reservoirs described by the Lindblad master equation)~\cite{Sa-22, Kulkarni-22}
have been proposed.
While the symmetry classification of the non-Hermitian SYK Hamiltonians was developed recently~\cite{GarciaGarcia-22PRX}, the symmetry classification of the SYK Lindbladians has yet to be developed.
Correspondingly, 
symmetry-enriched behaviors of dissipative quantum chaos have been largely unexplored.

In general, the dynamics of open quantum systems is described by Liouvillian superoperators that map a density operator to another density operator~\cite{Nielsen-textbook, Breuer-textbook, Rivas-textbook}. 
Symmetry of such Liouvillian superoperators was formulated~\cite{Buca-12, Albert-14, Minganti-18, Lieu-20, Altland-21, Roberts-21}.
Liouvillian superoperators can respect unitary symmetry and possess conserved charges, which are relevant to the open quantum dynamics and steady-state properties~\cite{Buca-12, Albert-14, Minganti-18}.
In addition to unitary symmetry, antiunitary symmetry of Liouvillian superoperators was recently studied~\cite{Lieu-20, Altland-21}.
In these previous works, however, only a part of the 38-fold symmetry classification~\cite{Bernard-LeClair-02, KSUS-19, Zhou-Lee-19} was shown to appear as symmetry of Liouvillian superoperators.
It is still elusive whether all the 38 symmetry classes of non-Hermitian operators can appear in Liouvillian superoperators.
This issue is relevant to a wide range of open quantum phenomena, including dissipative quantum chaos 
and topological phenomena.

In this work, we develop a theory of symmetry in open quantum systems.
To identify unitary and antiunitary symmetry 
of open quantum systems, we map Liouvillian superoperators for the open quantum dynamics to non-Hermitian operators in the double Hilbert space.
After this operator-state mapping, we apply the 38-fold symmetry classification of non-Hermitian operators to characterize the symmetry of the open quantum system. 
Our approach is similar to the analysis of unitary symmetry in Refs.~\cite{Buca-12, Albert-14, Minganti-18} but different from the analysis of antiunitary symmetry in Refs.~\cite{Lieu-20, Altland-21}, which enables us to find the symmetry classes that were shown not to appear in Refs.~\cite{Lieu-20, Altland-21}.
We find rich symmetry classification of Liouvillian superoperators.
On the one hand, Liouvillians respect symmetry that is reminiscent of symmetry in the corresponding closed quantum systems isolated from the environment.
On the other hand, Liouvillians respect additional symmetry inherent in their construction.
Furthermore, we develop the $\mathbb{Z}_4$ classification of fermion parity symmetry and antiunitary symmetry in the double Hilbert space of open quantum systems, which contrasts with the corresponding $\mathbb{Z}_8$ classification in closed quantum systems~\cite{Fidkowski-Kitaev-10, *Fidkowski-Kitaev-11, Turner-11}.
This $\mathbb{Z}_4$ symmetry classification should also be fundamental for the topological classification of open quantum systems.

As an illustrative example of open quantum fermionic many-body systems, we develop the symmetry classification of the SYK Lindbladians.
We establish the periodic tables (Tables~\ref{tab: SYK Lindbladian p=1} and \ref{tab: SYK Lindbladian p=2}) of the SYK Lindbladians and elucidate how they differ from the periodic tables of the SYK Hamiltonians.
Owing to their many-body nature, the SYK Lindbladians respect symmetry 
that cannot appear in noninteracting (quadratic) Lindbladians~\cite{Lieu-20}.
Furthermore, with extensive numerical calculations, we study the complex-spectral statistics and demonstrate dissipative quantum chaos enriched by symmetry.
Our results provide a general theory of symmetry in open quantum systems and lead to a unified understanding of open quantum physics including dissipative quantum chaos.
While we focus on open quantum systems described by the Lindblad master equation, our theory is also applicable to a variety of open quantum systems, such as non-Markovian Liouvillians and discrete quantum channels.

The rest of this work is organized as follows.
In Secs.~\ref{sec: general - boson} and \ref{sec: general - fermion}, we provide the general symmetry analysis of open quantum systems of bosons and fermions, respectively.
On the basis of the operator-state mapping, we apply the 38-fold symmetry classification of non-Hermitian operators to Liouvillian superoperators for the open quantum dynamics.
We elucidate modular conjugation symmetry that is inherent in the construction of Liouvillian superoperators.
As typical examples of open quantum bosonic systems, we study symmetry classes of dissipative quantum spin models.
Furthermore, we develop the $\mathbb{Z}_4$ classification of fermion parity symmetry and antiunitary symmetry in the double Hilbert space of open quantum systems (Table~\ref{tab: PSA}).
In Sec.~\ref{sec: Lindblad SYK symmetry}, we develop the symmetry classification of the SYK Lindbladians as a prototype of open quantum fermionic many-body systems.
Our symmetry classification is summarized in the periodic tables~\ref{tab: SYK Lindbladian p=1} and \ref{tab: SYK Lindbladian p=2}.
In Sec.~\ref{sec: dissipative quantum chaos}, we numerically study the complex-spectral statistics of the SYK Lindbladians and demonstrate their dissipative quantum chaos for several different symmetry classes.
In Sec.~\ref{sec: conclusion}, we conclude this work with several outlooks.
In Appendix~\ref{asec: 38-fold symmetry}, we summarize the 38-fold symmetry classification of non-Hermitian operators.
In Appendix~\ref{asec: NH RMT TRS}, we discuss the level statistics of non-Hermitian random matrices with time-reversal symmetry in a similar manner to the Wigner surmise.
In Appendix~\ref{asec: symmetry classification - SYK Hamiltonian}, we describe the symmetry classification of the SYK Hamiltonians.

\section{Bosonic open quantum systems}
    \label{sec: general - boson}

We discuss the general strategy to identify unitary and antiunitary symmetry of open quantum systems.
In this section, we focus on open quantum bosonic systems.
We begin with mapping a Liouvillian superoperator for the open quantum dynamics to a non-Hermitian operator in the double Hilbert space, as explained in Sec.~\ref{subsec: operator-state mapping}.
Then, we apply the symmetry classification of non-Hermitian operators~\cite{Bernard-LeClair-02, KSUS-19, Zhou-Lee-19} to characterize the symmetry of open quantum systems.
Open quantum systems respect additional antiunitary symmetry of modular conjugation that arises from its construction, as discussed in Sec.~\ref{subsec: modular conjugation}.
In Sec.~\ref{subsec: unitary symmetry}, we clarify the role of unitary symmetry and corresponding conserved quantities.
Moreover, since Lindbladians are required to be contractive, the real part of the complex spectrum is always negative.
As a consequence of this constraint, a constant shift of the Lindbladians is important, as discussed in Sec.~\ref{subsec: shift}.
In Sec.~\ref{subsec: boson}, as prototypical open quantum bosonic systems, we study dissipative quantum spin models and identify their symmetry classes.

\subsection{Operator-state mapping}
    \label{subsec: operator-state mapping}

In general, an operation of an open quantum system is described by a superoperator 
$\Phi$ that maps a density operator $\rho$ to another density operator $\rho' = \Phi \left( \rho \right)$~\cite{Nielsen-textbook, Breuer-textbook, Rivas-textbook}: 
\begin{equation}
    \rho \mapsto \rho' = \Phi \left( \rho \right).
\end{equation}
The superoperator $\Phi$ contains all information about the open quantum system in a similar manner to Hamiltonians for closed quantum systems.
Thus, symmetry of the open quantum system is described by symmetry of the superoperator $\Phi$.
In particular, if the open quantum dynamics is Markovian and time homogeneous, it is generally described by the Lindblad master equation~\cite{GKS-76, Lindblad-76}
\begin{equation}
    \frac{d\rho}{dt} = \mathcal{L} \left( \rho \right),
        \label{eq: Lindblad equation}
\end{equation}
where the superoperator $\mathcal{L}$ is defined by
\begin{equation}
    \mathcal{L} \left( \rho \right) = -\ii \left[ H, \rho \right] + \sum_{m} \left[ L_m \rho L_m^{\dag} - \frac{1}{2} \left\{ L_m^{\dag} L_m, \rho \right\} \right].
        \label{eq: Lindbladian}
\end{equation}
Here, $H$ is a Hermitian Hamiltonian that describes the coherent dynamics, and $L_m$'s are dissipators that describe the dissipative coupling to the external environment.
While our discussions below are generally applicable to a wide variety of open quantum systems, including non-Markovian Liouvillians and discrete time evolution of the Kraus representation~\cite{Nielsen-textbook, Breuer-textbook, Rivas-textbook}, we here focus on symmetry of open quantum systems described by the Lindbladians $\mathcal{L}$.

In contrast to Hamiltonians for closed quantum systems, Lindbladians $\mathcal{L}$ are superoperators.
To resolve the difficulty in analyzing superoperators,
it is useful to double the Hilbert space and map $\mathcal{L}$ and $\rho$ to an operator and a state, respectively.
We first map the density operator $\rho = \sum_{i, j} \rho_{ij} \ket{i} \bra{j}$ to a pure state $\ket{\rho} = \sum_{i, j} \rho_{ij} \ket{i} \ket{j}$ in the double Hilbert space.
Through this mapping, the identity operator $I$, which can also be considered to be the thermal equilibrium state with infinite temperature, is mapped to the maximally-entangled state $\ket{I} = \sum_{i} \ket{i} \ket{i}$.
Similarly, we map superoperators that act on the density operator $\rho$ to operators that act on the state $\ket{\rho}$ in the double Hilbert space.
This vectorization procedure
$\rho \rightarrow |\rho\rangle$
can 
also be stated as
\begin{align}
|\rho\rangle = \rho^+ \otimes I^- | I\rangle,
\end{align}
where $O^{\pm}$ denotes
an operator acting on the
ket and bra 
spaces, respectively,
defined from an arbitrary operator $O$
acting on the original Hilbert space.
Then, the Lindblad equation in Eqs.~(\ref{eq: Lindblad equation}) and (\ref{eq: Lindbladian}) reads
\begin{align}
    \frac{\partial}{\partial t} \ket{\rho} = \mathcal{L} \ket{\rho}
        \label{eq: imaginary Schrodinger equation}
\end{align}
and
\begin{align}
    &{\cal L} = -\ii \left( {H} \otimes I^{-} - I^{+} \otimes {H}^{*} \right) \nonumber \\
    &~+ \sum_{m} \left[ {L}_{m} \otimes {L}_{m}^{*} - \frac{1}{2} ({L}_{m}^{\dag} {L}_{m} \otimes I^{-}) - \frac{1}{2} (I^{+} \otimes {L}_{m}^{T} {L}_{m}^{*}) \right]
        \label{eq: bosonic Lindbladian}
\end{align}
with the identity operators $I^{+}$ and $I^{-}$ in the ket and bra spaces, respectively.
Through this operator-state mapping, the Lindblad equation effectively reduces to the Schr\"odinger equation with the non-Hermitian Hamiltonian $\ii \mathcal{L}$.
In this representation, while the Hamiltonian part acts only on the individual ket or bra space, the dissipation term couples these two spaces.
We can generally identify symmetry of Lindbladians by considering symmetry of the corresponding non-Hermitian operators in the double Hilbert space.
Such an operator-state mapping is also useful to find exact solutions of integrable open quantum systems~\cite{Medvedyeva-16, Shibata-19-Kitaev, *Shibata-19-AshkinTeller} and characterize topological phases of anomalous unitary operators~\cite{Liu-Ryu-21}.

\subsection{Modular conjugation}
    \label{subsec: modular conjugation}

The Lindbladian $\mathcal{L}$ in the double Hilbert space 
is not an arbitrary non-Hermitian operator
as it descends from a quantum channel,
which is completely positive and trace preserving
\footnote{
The implication of complete positivity
is not addressed in the current formalism. 
To address this point, using 
the Choi-Jamiołkowski isomorphism,
we can map a quantum channel to a density operator 
in the double Hilbert space.}.
We now discuss constraints (symmetries) inherent in the construction of the Lindbladians.  
First,
it can readily be checked that 
$\mathcal{L}^{\dag}|I\rangle = \langle I|\mathcal{L} = 0$.
This implies the trace-preserving condition, 
$\langle I | \mathcal{L} | \rho\rangle = 0$
or
$\langle I | e^{t \mathcal{L}} |\rho\rangle = \langle I| \rho(t)\rangle = \mathrm{tr}\, \rho(t) = 1$.
Second, the Lindbladian should preserve 
Hermiticity of the density operator, 
$\rho(t) 
= \rho(t)^{\dag} 
= [e^{ t \mathcal{L} }
(\rho(0))]^{\dag}$. 
The Hermiticity condition leads to
antiunitary symmetry of 
Lindbladians  
based on modular conjugation.
Modular conjugation is an antiunitary,
involutive operation,
and satisfies 
\begin{align}
\label{mod conj def}
    \mathcal{J} O^+ | I \rangle =
    (O^{+})^{\dag} |I\rangle,
\end{align}
as its defining relation~\cite{Ojima-80, Witten-18-review}~\footnote{
In our examples, the modular operator $\Delta$ is the identity, or equivalently, the modular Hamiltonian is zero (i.e., $\log \Delta = 0$).
Hence, modular conjugation coincides with $\mathcal{S} \coloneqq \mathcal{J}\Delta^{1/2}$.
For $\Delta \neq I$, however, we should use $\mathcal{S}$ in Eq.~(\ref{mod conj def}) instead of $\mathcal{J}$.
}.
Hermiticity of the density operator $\rho$
then implies that $|\rho\rangle$ is invariant 
under $\mathcal{J}$,
\begin{align}
    \mathcal{J} |\rho\rangle =
    |\rho\rangle.
\end{align}
It then follows, for any $t$,
\begin{align}
    | \rho(t)\rangle
    &= \mathcal{J} 
    | \rho(t)\rangle \nonumber \\
    &=
    \mathcal{J} 
    e^{t\mathcal{L}}
    | \rho(0)\rangle \nonumber \\ 
    & =
    \mathcal{J} 
    e^{t\mathcal{L}}
    \mathcal{J}^{-1}
    \cdot
    \mathcal{J}
    | \rho(0)\rangle
    \nonumber \\
    &=
     \mathcal{J} 
    e^{t\mathcal{L}}
    \mathcal{J}^{-1}
    |\rho(0)\rangle,
\end{align}
which implies $\mathcal{J} e^{t\mathcal{L}} \mathcal{J}^{-1} = e^{t\mathcal{L}}$ and hence symmetry
\begin{equation}
    \mathcal{J} \mathcal{L} \mathcal{J}^{-1}= \mathcal{L}.
        \label{eq: modular conjugation}
\end{equation}
This modular conjugation symmetry is respected for arbitrary Lindbladians.
It is also respected for more generic superoperators as long as they preserve Hermiticity of density operators.

For bosonic operators, modular conjugation $\mathcal{J}$ is explicitly constructed by
\begin{align}
    \mathcal{J} \left( A \otimes B \right) \mathcal{J}^{-1} = B \otimes A,
    \quad
    \mathcal{J} z \mathcal{J}^{-1}
    = z^*,
        \label{eq: boson modular conjugation 1}
\end{align}
which exchanges the ket and bra spaces ($z\in \mathbb{C}$).
Then, all bosonic Lindbladians in Eq.~(\ref{eq: bosonic Lindbladian}) are invariant under modular conjugation [i.e., Eq.~(\ref{eq: modular conjugation})] regardless of the details of the Hamiltonian $H$ and the dissipators $L_m$'s.
This symmetry can be effectively considered to be time-reversal symmetry 
(or equivalently, particle-hole symmetry$^{\dag}$ for $\ii \mathcal{L}$)
in the 38-fold symmetry classification of non-Hermitian operators (see Appendix~\ref{asec: 38-fold symmetry} for details)~\cite{KSUS-19}.
As a result of this symmetry, eigenvalues generally form complex-conjugate pairs, and a  
number of real eigenvalues 
can
appear in the complex spectrum of the Lindbladians $\mathcal{L}$ (see Appendix~\ref{asec: NH RMT TRS} for details).

\subsection{Conserved quantities}
    \label{subsec: unitary symmetry}
    
Suppose that a Lindbladian $\mathcal{L}$ is invariant under a unitary operation $\mathcal{U}$:
\begin{equation}
    \mathcal{U}\,\mathcal{L}\,\mathcal{U}^{-1} = \mathcal{L}.
        \label{eq: unitary}
\end{equation}
Then, $\mathcal{L}$ is block-diagonalized into different subspaces characterized by the different conserved charges of $\mathcal{U}$.
Prime examples of such conserved charges include the particle number.
In the presence of such unitary symmetry, we need to consider the subspace with a fixed conserved charge to study the symmetry classification and the spectral statistics. 
In fact, if we study, for example, the statistics of the entire spectrum of the Lindbladian $\mathcal{L}$ even in the presence of the unitary symmetry in Eq.~(\ref{eq: unitary}), we only have the Poisson statistics since the different subspaces with different conserved charges are independent of each other.
We note that such unitary symmetry is not included in the internal-symmetry classification for Hermitian~\cite{AZ-97, Evers-review, CTSR-review} and non-Hermitian~\cite{Bernard-LeClair-02, KSUS-19, Zhou-Lee-19} operators.

It is also notable that Lindbladians can be invariant even under a unitary operation in each ket or bra space, i.e.,
\begin{equation}
    U_{\pm}\,\mathcal{L}\,U_{\pm}^{-1} = \mathcal{L},
        \label{eq: unitary - strong}
\end{equation}
where $U_{\pm}$ is the unitary operator that belongs only to the ket or bra space.
The symmetry condition in Eq.~(\ref{eq: unitary - strong}) is stricter than that in Eq.~(\ref{eq: unitary}), and hence Eqs.~(\ref{eq: unitary}) and (\ref{eq: unitary - strong}) were respectively called weak and strong symmetry in Ref.~\cite{Buca-12}.
For bosonic operators, unitary symmetry that commutes with Lindbladians, as well as its consequence on the open quantum dynamics, was studied also in Ref.~\cite{Albert-14}.

In addition to unitary operators,
antiunitary operators can also act on 
Lindbladians and the double Hilbert space. 
With an antiunitary operator,
Lindbladians can be symmetric
in the form of 
time reversal or time reversal$^{\dag}$
in the terminology of the 38-fold classification of non-Hermitian operators~\cite{KSUS-19}. 
To discuss antiunitary symmetry, 
we need a proper shift of Lindbladians, as we discuss momentarily.
In passing, we note that we can combine 
an antiunitary operator
and modular conjugation
to form a unitary operator.
This type of unitary operator (symmetry)
is called 
the Kubo-Martin-Schwinger (KMS) symmetry~\cite{Kubo-57, Martin-Schwinger-59, Sieberer_2015, Glorioso_2018}.
The KMS symmetry was also utilized to 
diagnose topological properties
and quantum anomalies  
of time-reversal-symmetric Floquet systems~\cite{Liu-Ryu-21}.

\subsection{Shifted Lindbladian}
    \label{subsec: shift}

In addition to the trace-preserving and Hermiticity-preserving 
conditions, 
Lindbladians are required to be contractive by construction.
The real part of the Lindbladian spectrum is constrained to be negative, which 
appears to be incompatible with some internal symmetry.
For example, if a Lindbladian $\mathcal{L}$ respected time-reversal symmetry $\mathcal{T} \left( \ii \mathcal{L} \right) \mathcal{T}^{-1} = \ii \mathcal{L}$ with an antiunitary operator $\mathcal{T}$, the complex spectrum of $\mathcal{L}$ would be required to be symmetric around the imaginary axis.
This would be incompatible with the aforementioned constraint unless the Lindbladian $\mathcal{L}$ does not include dissipation and its spectrum is pure imaginary.
For this reason, quadratic noninteracting Lindbladians were argued to fall into only the 10-fold symmetry class (Altland-Zirnbauer$^{\dag}$ symmetry class in Table~\ref{tab: AZ}) out of the 38-fold symmetry class for generic non-Hermitian operators~\cite{Lieu-20}.
However, this constraint can be lifted by introducing a constant shift $\mathrm{tr}\,\mathcal{L}/\mathrm{tr}\,I \leq 0$ to the Lindbladian spectrum so that it will be traceless.
Here, $I = I^{+} \otimes I^{-}$ is the identity operator in the double Hilbert space.
When the Hamiltonian $H$ and the dissipators $L_m$'s are traceless, we have
\begin{align}
    \mathrm{tr}\,\mathcal{L} = - \left( \mathrm{tr}\,I^{\pm} \right) \sum_{m} \mathrm{tr} \left[ L_{m}^{\dag} L_{m} \right] \leq 0.
\end{align}
This constant shift $\mathrm{tr}\,\mathcal{L}/\mathrm{tr}\,I$ does not affect the eigenvectors of $\mathcal{L}$, and if we focus on the shifted Lindbladian 
\begin{align}
\tilde{\cal L} \coloneqq \mathcal{L} - \frac{\mathrm{tr}\,\mathcal{L}}{\mathrm{tr}\,I}\,I 
\end{align}
instead of $\mathcal{L}$, we can in principle reproduce all the 38-fold symmetry, including time-reversal symmetry $\mathcal{T}\,( \ii \tilde{\cal L} )\,\mathcal{T}^{-1} =  \ii \tilde{\cal L}$.
One may think that this constant shift of the spectrum is similar to a constant shift of Fermi energy in Hermitian Hamiltonians, by which particle-hole transformations can be introduced.
Such a constant shift of the Lindbladian spectrum is important for parity-time symmetry~\cite{Prosen-12, Huber-20, Shibata-20, Nakanishi-Sasamoto-22} and sublattice symmetry~\cite{Kawasaki-22} in open quantum systems.
In the following sections, 
we show symmetry of shifted Lindbladians for the dissipative quantum spin models and the SYK Lindbladians.

\subsection{Examples: dissipative quantum spin models}
    \label{subsec: boson}

As prototypical examples of open quantum bosonic systems, we study symmetry classes of dissipative quantum spin models described by the Lindblad master equation in Eqs.~(\ref{eq: Lindblad equation}) and (\ref{eq: Lindbladian}).

\subsubsection{Dissipative quantum Ising model}

The Hamiltonian is chosen to be the quantum Ising model
\begin{align}  
    H = - \sum_{i, j} J_{ij} Z_{i} Z_{j} - \sum_{i} \left( h_{i}^{x} X_{i} + h_{i}^{z} Z_{i} \right), 
\end{align}
where $X_i$, $Y_i$, and $Z_i$ are Pauli matrices at site $i$, $J_{ij} \in \mathbb{R}$ is the interaction strength between the quantum spins at sites $i$ and $j$, and $h_{i}^{x (z)} \in \mathbb{R}$ is the magnetic field along the $x$ ($z$) direction at site $i$.
Since the discussions below are concerned only with internal symmetry, the sites $i, j$ in the sum $\sum_{i, j}$ can be arbitrary.
The dissipators are chosen to be dephasing
\begin{align}
    L_{i} = \sqrt{\gamma_{i}}\,Z_{i}
        \label{eq: spin-dephasing}
\end{align}
or damping
\begin{align}
    L_{i} = \frac{\sqrt{\gamma_{i}} \left( X_{i} + \ii Y_{i} \right)}{2},
        \label{eq: spin-damping}
\end{align}
where $\gamma_i \geq 0$ is the dissipation strength at site $i$~\cite{Nielsen-textbook, Breuer-textbook, Rivas-textbook}.
Similar dissipative spin models and their complex spectral statistics were recently studied~\cite{Hamazaki-20, Akemann-19, Sa-20}.

Through the operator-state mapping in Eq.~(\ref{eq: bosonic Lindbladian}), the above Lindbladian is represented as a non-Hermitian operator $\mathcal{L}$ in the double Hilbert space, which enables us to study its symmetry.
First, as discussed in Sec.~\ref{subsec: modular conjugation}, $\mathcal{L}$ is invariant under modular conjugation in Eq.~(\ref{eq: boson modular conjugation 1}).
This symmetry can be considered as time-reversal symmetry, or equivalently particle-hole symmetry$^{\dag}$ (see Appendix~\ref{asec: 38-fold symmetry} for details)~\cite{KSUS-19}, which changes the level statistics around and on the real axis~\cite{Ginibre-65, Xiao-22}.

In addition to modular conjugation symmetry, the Lindbladian $\mathcal{L}$ respects time-reversal symmetry$^{\dag}$
\begin{equation}
    \mathcal{T} \mathcal{L}^{\dag} \mathcal{T}^{-1} = \mathcal{L}
        \label{eq: spin-TRS-dag}
\end{equation}
for the dephasing in Eq.~(\ref{eq: spin-dephasing}).
Here, the antiunitary operator $\mathcal{T}$ is chosen to be complex conjugation $\mathcal{K}$.
In contrast to time-reversal symmetry discussed above, this symmetry is relevant to the level statistics of generic complex eigenvalues even away from the real axis~\cite{Hamazaki-20}.
On the other hand, the damping in Eq.~(\ref{eq: spin-damping}) breaks this symmetry, resulting in the local spectral correlations same as the Ginibre ensemble of non-Hermitian random matrices~\cite{Ginibre-65}.
Consequently, the above dissipative spin model belongs to class BDI$^{\dag}$ for the dephasing in Eq.~(\ref{eq: spin-dephasing}) and class AI (or equivalently class D$^{\dag}$) for the damping in Eq.~(\ref{eq: spin-damping}).

\subsubsection{Dephasing XYZ model with magnetic fields}

More symmetry classes appear in different dissipative quantum spin models.
As another example, we study the XYZ model with a magnetic field
\begin{align}
    H = - \sum_{i, j} \left( J_{ij}^{x} X_{i} X_{j} + J_{ij}^{y} Y_{i} Y_{j} + J_{ij}^{z} Z_{i} Z_{j}\right) - \sum_{i} h_i^{x} X_{i},
        \label{eq: XYZ}
\end{align}
where $J_{ij}^{x}$, $J_{ij}^{y}$, $J_{ij}^{z} \in \mathbb{R}$ are the interaction strength between the quantum spins at sites $i$ and $j$. 
The XYZ model respects time-reversal symmetry
\begin{align}
    T H T^{-1} = H
        \label{eq: XYZ-TRS}
\end{align}
with $T = \mathcal{K}$.
In addition, it is also invariant under the global spin flip and hence respects the $\mathbb{Z}_2$ unitary symmetry
\begin{align}
    U H U^{-1} = H,\quad U \coloneqq \prod_{i} X_i.
        \label{eq: XYZ-Z2-flip}
\end{align}

Now, let us add dephasing in Eq.~(\ref{eq: spin-dephasing}) and investigate symmetry of the Lindbladian.
The dissipation term reads
\begin{align}
    \mathcal{D} &= \sum_{i} \left[ L_{i} \otimes L_{i}^{*} - \frac{1}{2}\,( L_{i}^{\dag} L_{i} \otimes I^{-} ) - \frac{1}{2}\,( I^{+} \otimes L_{i}^{T} L_{i}^{*} ) \right] \nonumber \\
    &= \sum_{i} \gamma_i Z_{i}^{+} Z_{i}^{-} - \left( \sum_{i} \gamma_i \right) I,
\end{align}
where $Z_{i}^{+}$ and $Z_{i}^{-}$ are the Pauli matrices in the ket and bra spaces, respectively.
Symmetry of this model was investigated also in Ref.~\cite{Prosen-12A}.
As discussed above, the Lindbladian $\mathcal{L}$ is invariant under modular conjugation because of its Hermiticity-preserving nature.
In addition, symmetry in the original XYZ model can survive even in the presence of dissipation.
In fact, time-reversal symmetry of the XYZ model in Eq.~(\ref{eq: XYZ-TRS}) survives, and the Lindbladian $\mathcal{L}$ respects time-reversal symmetry$^{\dag}$ in Eq.~(\ref{eq: spin-TRS-dag}). 
Furthermore, similarly to the $\mathbb{Z}_2$ unitary symmetry in Eq.~(\ref{eq: XYZ-Z2-flip}), the Lindbladian $\mathcal{L}$ is invariant under the global spin flip in the double Hilbert space and respects
\begin{align}
    \mathcal{U} \mathcal{L} \mathcal{U}^{-1} = \mathcal{L},\quad \mathcal{U} \coloneqq \prod_{i} X_{i}^{+} X_{i}^{-}.
        \label{eq: spin-total-Z2}
\end{align}
By contrast, the dephasing in Eq.~(\ref{eq: spin-dephasing}) breaks the spin-flip symmetry in the individual ket or bra space.
Nevertheless, under the combination of the spin flip and time reversal, the Lindbladian respects 
\begin{align}
    (U^{\pm} \mathcal{K})\,(\mathcal{L} + \sum_{i}\gamma_i)\,(U^{\pm} \mathcal{K})^{-1} = - (\mathcal{L} + \sum_{i}\gamma_i)
        \label{eq: spin-PHS-dag}
\end{align}
with the spin flip $U^{\pm} \coloneqq \prod_{i} X_{i}^{\pm}$ in the individual ket or bra space.
This symmetry is considered as particle-hole symmetry$^{\dag}$ (or equivalently, time-reversal symmetry) in the 38-fold symmetry classification of non-Hermitian operators.
The constant shift $\mathrm{tr}\,\mathcal{L}/\mathrm{tr}\,I = - \sum_{i}\gamma_i$ is needed to capture this symmetry, as discussed in Sec.~\ref{subsec: shift}.

In summary, the XYZ model with the dephasing respects modular conjugation symmetry in Eq.~(\ref{eq: modular conjugation}), time-reversal symmetry$^{\dag}$ in Eq.~(\ref{eq: spin-TRS-dag}), and particle-hole symmetry$^{\dag}$ in Eq.~(\ref{eq: spin-PHS-dag}).
Notably, this Lindbladian is invariant under the global spin flip in the double Hilbert space [i.e., Eq.~(\ref{eq: spin-total-Z2})], and we should consider symmetry in the subspace of this $\mathbb{Z}_2$ unitary symmetry.
Since all the above antiunitary symmetries commute with this $\mathbb{Z}_2$ unitary symmetry, they remain to be symmetries in this subspace.
Consequently, the dephasing XYZ model belongs to class BDI$^{\dag} + \mathcal{S}_{++}$ in the 38-fold symmetry classification of non-Hermitian operators (or equivalently, class BDI + $\mathcal{S}_{++}$;
see Appendix~\ref{asec: 38-fold symmetry} for the definitions of the symmetry classes).

\begin{figure}[t]
\centering
\includegraphics[width=86mm]{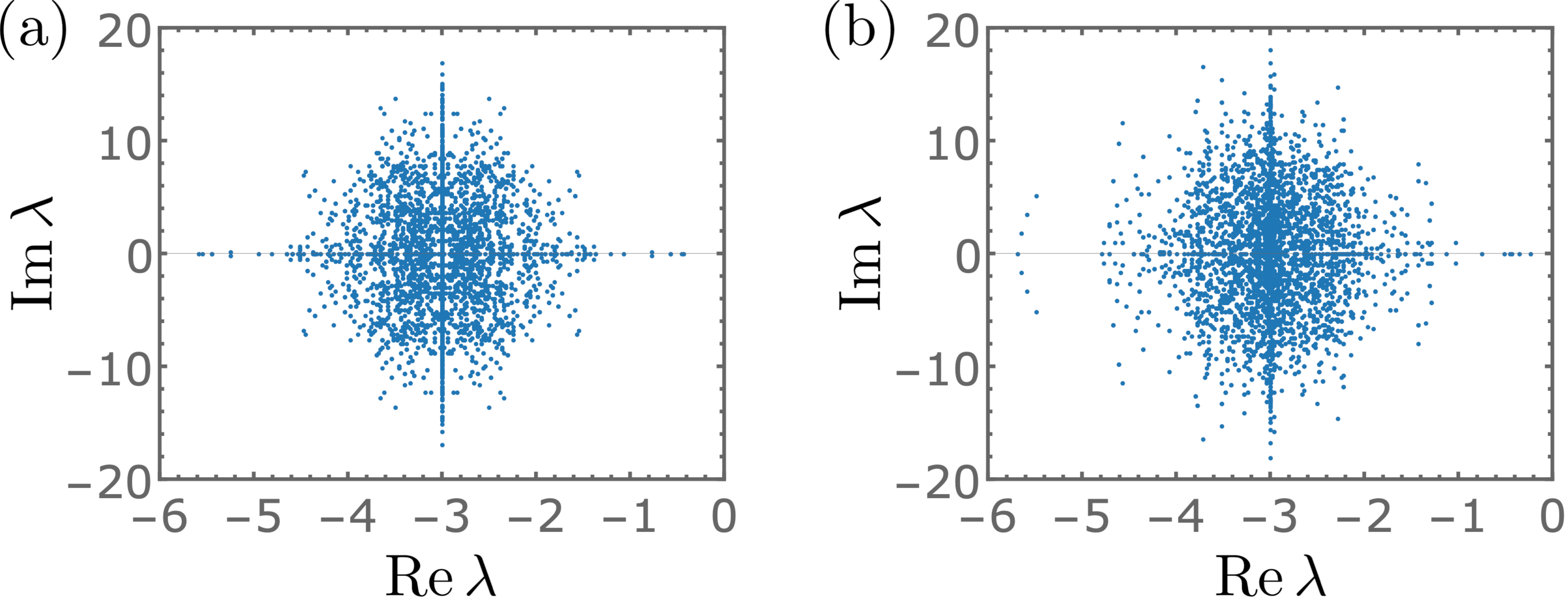} 
\caption{Dissipative XYZ model in one dimension with periodic boundaries and nearest-neighbor coupling ($L=6$, $J_{ij}^{x} = 1.0$, $J_{ij}^{y} = 0.7$, $J_{ij}^{z} = 0.9$, $\gamma_{i} = 0.5$). 
(a)~Complex spectrum in the presence of a magnetic field $h^{x}_{i} = 0.2$. (b)~Complex spectrum in the presence of magnetic fields $h^{x}_{i} = 0.2$, $h_{i}^{z} = 0.4$.}
	\label{fig: XYZ}
\end{figure}

Symmetry manifests itself in the complex spectrum.
We obtain the complex spectrum of the dephasing XYZ chain, as shown in Fig.~\ref{fig: XYZ}\,(a).
Here, we consider the XYZ spin chain on the one-dimensional lattice with periodic boundaries that only has the nearest-neighbor coupling.
The complex spectrum is symmetric about the real axis, which originates from modular conjugation symmetry.
In addition, because of particle-hole symmetry$^{\dag}$ in Eq.~(\ref{eq: spin-PHS-dag}), it is also symmetric about the line $\mathrm{Re}\,\lambda = \mathrm{tr}\,\mathcal{L}/\mathrm{tr}\,I = - \sum_{i} \gamma_i$. 
As a combination of these two antiunitary symmetries, the complex spectrum forms a cross.
Owing to the nonintegrable nature of the XYZ model, the level repulsion around these symmetric lines is observed (see Appendix~\ref{asec: NH RMT TRS} for details)~\cite{Xiao-22}.
To clarify the role of particle-hole symmetry$^{\dag}$, we also investigate the complex spectrum of the same dephasing XYZ model with an additional magnetic field $-\sum_{i} h_{i}^{z} Z_{i}$ [Fig.~\ref{fig: XYZ}\,(b)].
Such a magnetic field along the $z$ direction breaks particle-hole symmetry$^{\dag}$ in Eq.~(\ref{eq: spin-PHS-dag}), 
and the model now belongs to class BDI$^{\dag}$ instead of class BDI$^{\dag}$ + $\mathcal{S}_{++}$.
While some complex eigenvalues remain in the symmetric line $\mathrm{Re}\,\lambda = - \sum_{i} \gamma_i$, the symmetry of the complex spectrum about this line is broken.
In addition, the level repulsion around the symmetric line disappears, which also indicates the absence of particle-hole symmetry$^{\dag}$.
It is also notable that similar cross-shaped complex spectra of Lindbladians due to parity-time symmetry were studied~\cite{Prosen-12, Huber-20, Shibata-20}.
Since the cross-shaped complex spectra in Refs.~\cite{Prosen-12, Huber-20, Shibata-20} rely on the combination of spatial-inversion symmetry and time-reversal symmetry, they are sensitive to disorder.
On the other hand, the cross-shaped complex spectrum in our model, as well as the model in Ref.~\cite{Prosen-12A}, relies only on internal (i.e., nonspatial) symmetry and is robust against disorder that preserves particle-hole symmetry$^{\dag}$.

\subsubsection{Dephasing XYZ model with Dzyaloshinskii-Moriya interactions}

In the previous dephasing XYZ model, time-reversal symmetry (or equivalently, particle-hole symmetry$^{\dag}$) in Eq.~(\ref{eq: spin-PHS-dag}) appears, the sign of which is $+1$.
In general, antiunitary symmetry with the sign $-1$ can also appear.
As an example, we consider the XYZ model with the Dzyaloshinskii-Moriya (DM) interactions
\begin{align}
    &H = - \sum_{i, j} \left( J_{ij}^{x} X_{i} X_{j} + J_{ij}^{y} Y_{i} Y_{j} + J_{ij}^{z} Z_{i} Z_{j}\right) \nonumber \\ 
    &\qquad\qquad+ \sum_{i, j} D_{ij} \left( X_{i} Z_{j} - Z_{i} X_{j} \right),
        \label{eq: XYZ-DM}
\end{align}
where $D_{ij} \in \mathbb{R}$ describes the strength of the DM interactions.
The DM interactions break the $\mathbb{Z}_2$ unitary symmetry in Eq.~(\ref{eq: XYZ-Z2-flip}).
Still, this spin model respects another $\mathbb{Z}_2$ unitary symmetry
\begin{align}
    U H U^{-1} = H,\quad U \coloneqq \prod_{i} Y_i.
\end{align}

Similarly to the previous case, let us add dephasing in Eq.~(\ref{eq: spin-dephasing}).
Modular conjugation symmetry in Eq.~(\ref{eq: modular conjugation}) and time-reversal symmetry$^{\dag}$ in Eq.~(\ref{eq: spin-TRS-dag}) remain to be respected.
On the other hand, the Lindbladian $\mathcal{L}$ no longer respects the $\mathbb{Z}_2$ unitary symmetry in Eq.~(\ref{eq: spin-total-Z2}) but rather respects
\begin{align}
    \mathcal{U} \mathcal{L} \mathcal{U}^{-1} = \mathcal{L},\quad \mathcal{U} \coloneqq \prod_{i} Y_{i}^{+} Y_{i}^{-}.
\end{align}
In addition, under the combination of the spin flip and time reversal, the Lindbladian respects Eq.~(\ref{eq: spin-PHS-dag}) with $U^{\pm} \coloneqq \prod_{i} Y_{i}^{\pm}$.
A distinctive feature of this antiunitary symmetry is that its sign can be either $+1$ and $-1$, depending on whether the number $N$ of sites is even or odd:
\begin{align}
    \left( U^{\pm} \mathcal{K} \right)^2 = \left( -1 \right)^{N}.
\end{align}
Consequently, this dephasing XYZ model with the DM interactions belongs to class BDI$^{\dag} + \mathcal{S}_{++}$ for even $N$  and class CI$^{\dag} + \mathcal{S}_{+-}$ for odd $N$ (or equivalently, class BDI + $\mathcal{S}_{++}$ and class BDI + $\mathcal{S}_{-+}$, respectively;
see Appendix~\ref{asec: 38-fold symmetry} for the definitions of the symmetry classes).
Owing to this antiunitary symmetry for odd $N$, complex eigenvalues on the symmetric line $\mathrm{Re}\,\lambda = \mathrm{tr}\,\mathcal{L}/\mathrm{tr}\,I$ exhibit the Kramers degeneracy while generic complex eigenvalues away from it do not.

Antiunitary symmetry with the sign $-1$ also appears in the SYK Lindbladians with the odd number $p$ of dissipators (see Table~\ref{tab: SYK Lindbladian p=1}).
We study the consequent complex-spectral statistics in Sec.~\ref{sec: dissipative quantum chaos - p=1, K in R}.

\section{Fermionic open quantum systems}
    \label{sec: general - fermion}

In this section, we generally analyze symmetry of open quantum fermionic systems.
In contrast to the bosonic case discussed in Sec.~\ref{sec: general - boson}, the operator-state mapping is intricate in fermionic open quantum systems because of the fermionic nature of the Hilbert space;
we clarify the operator-state mapping of open quantum fermionic systems in Sec.~\ref{subsec: operator-state mapping fermion}.
As a general feature of open quantum fermionic systems, we discuss the importance of fermion number parity in Sec.~\ref{subsec: fermion number parity}.
Similarly to open quantum bosonic systems, generic fermionic Lindbladians are invariant under modular conjugation, as shown in Sec.~\ref{subsec: KMS}.
Finally, in Sec.~\ref{subsec: PSA}, we develop the $\mathbb{Z}_4$ algebraic structure of fermion parity symmetry and antiunitary symmetry in the double Hilbert space of open quantum fermionic systems.

\subsection{Operator-state mapping}
    \label{subsec: operator-state mapping fermion}

In contrast to bosonic systems, the operator-state mapping is intricate for fermionic open quantum systems because we have to reproduce the fermionic anticommutation relations in the double Hilbert space.
Here, we follow the fermionic operator-state mapping in Refs.~\cite{Schmutz-78, Dzhioev-11, *Dzhioev-12}.
Let us consider a fermionic Lindbladian $\mathcal{L}$, which is generally described by Majorana fermions $\psi_{i}$'s in the Hilbert (Fock) space.
We can construct the fermionic operator-state mapping also by complex fermions in a similar manner.
To develop the operator-state mapping, we double the Hilbert space and introduce the two types of fermions $\psi^{+}_{i}$'s and $\psi^{-}_{i}$'s that belong to the ket and bra spaces, respectively.
These Majorana fermions $\psi^{+}_{i}$'s and $\psi^{-}_{i}$'s are required to satisfy the fermionic anticommutation relations
\begin{align}
    \{ \psi_{i}^{s}, \psi_{j}^{s'} \} = \delta_{ij} \delta_{ss'}
\end{align}
for $s, s' \in \{ +, - \}$.
While the fermion operators $\psi^{+}_{i}$'s implement the action on the density operator $\rho$ from the left, the fermion operators $\psi^{-}_{i}$'s implement the action on $\rho$ from the right.
Notably, $\psi^{+}_{i}$'s and $\psi^{-}_{i}$'s are required to anticommute with each other as a consequence of the fermionic nature of the Hilbert space.
For the introduction of $\psi^{+}_{i}$'s and $\psi^{-}_{i}$'s, we begin with mapping the identity operator: $I \rightarrow \ket{I}$.
The fermion operators $\psi^{+}_{i}$'s and $\psi^{-}_{i}$'s are connected to each other through this reference state $\ket{I}$,
\begin{equation}
    \psi_{i}^{+} \ket{I} = - \ii \psi_{i}^{-} \ket{I},
        \label{eq: reference state}
\end{equation}
which is compatible with $\psi_{i} I = I \psi_{i}$ in the original Hilbert space.
The factor ``$-\ii$" in Eq.~(\ref{eq: reference state}) ensures the fermionic anticommutation relation between $\psi_{i}^{+}$ and $\psi_{i}^{-}$.
The reference state $\ket{I}$ plays a role of a reflector that maps operators in the ket space to those in the bra space, and vice versa.
Using the reference state $\ket{I}$, we can map operators to states in the double Hilbert space. 
In particular, the density operator $\rho$ is mapped to the state
\begin{align}
    \ket{\rho} \coloneqq \rho^{+} \ket{I}.
\end{align}
Then, we multiply the Lindblad equation in Eqs.~(\ref{eq: Lindblad equation}) and (\ref{eq: Lindbladian}) from the right on $\ket{I}$ and convert the fermions acting on the density operator from the right to the fermions $\psi_{i}^{-}$'s that belong to the bra space.
In this manner, generic fermionic Lindbladians can also be effectively recast as the Schr\"odinger equation~(\ref{eq: imaginary Schrodinger equation}) with non-Hermitian fermionic operators.

For example, if we consider the one-body dissipators $L_{m} = \gamma \psi_{m}$ ($\gamma \in \mathbb{R}$; $m=1, 2, \cdots, M$), the Lindbladian $\mathcal{L}$ is written as the non-Hermitian operator
\begin{align}
    \mathcal{L}
    = - \ii \gamma^2 \sum_{m=1}^{M} \psi_{m}^{+} \psi_{m}^{-} - \frac{\gamma^2 M}{2} I^{+} I^{-},
\end{align}
where the fermion parity of the density operator is assumed to be even.
If we consider the two-body dissipators $L_{m} = \gamma \psi_{m} \psi_{m+1}$ ($\gamma \in \mathbb{R}$; $m =1, 2, \cdots, M$; $\psi_{M+1} \coloneqq \psi_{1}$), the Lindbladian $\mathcal{L}$ is written as 
\begin{align}
    \mathcal{L}
    = \gamma^2 \sum_{m=1}^{M} \psi_{m}^{+} \psi_{m+1}^{+} \psi_{m}^{-} \psi_{m+1}^{-} - \frac{\gamma^2 M}{4} I^{+} I^{-}.
\end{align}
Below, we focus on symmetry of open quantum many-body systems described by fermionic Lindbladians and develop their symmetry classification based on the 38-fold internal symmetry of non-Hermitian operators (see Appendix~\ref{asec: 38-fold symmetry} for details)~\cite{KSUS-19}.

\subsection{Fermion number parity}
    \label{subsec: fermion number parity}
    
In open quantum fermionic systems, the particle number can be preserved, for example, when dissipation only results in dephasing.
However, in typical situations, including the SYK Lindbladians discussed in the subsequent sections, dissipation breaks the conservation of the particle number.
Even in such a case, Lindbladians $\mathcal{L}$ always consist of the even number of fermion operators in the double Hilbert space, and hence total fermion parity $\left( -1 \right)^{\cal F}$ is preserved.
The symmetry operation acts as
\begin{align}
    \left( -1 \right)^{\cal F} \psi_{n}^{\pm} \left( -1 \right)^{\cal F} = - \psi_{n}^{\pm}
        \label{eq: fermion parity calF - psi}
\end{align}
for each fermion, and the Lindbladian $\mathcal{L}$ is invariant under this $\mathbb{Z}_2$ unitary operation:
\begin{equation}
    \left( -1 \right)^{\cal F} \mathcal{L} \left( -1 \right)^{\cal F} = \mathcal{L}.
        \label{eq: fermion parity symmetry}
\end{equation}
As discussed 
in Sec.\ \ref{subsec: unitary symmetry},
we may also define fermion parity $\left( -1 \right)^{F^{\pm}}$ in each ket or bra space by
\begin{align}
    \left( -1 \right)^{F^{\pm}} \psi_{n}^{\pm} \left( -1 \right)^{F^{\pm}} &= - \psi_{n}^{\pm}, \label{eq: Fpm - psi - 1} \\
    \left( -1 \right)^{F^{\pm}} \psi_{n}^{\mp} \left( -1 \right)^{F^{\pm}} &= \psi_{n}^{\mp}. \label{eq: Fpm - psi - 2}
\end{align}
However, such strong versions of fermion parity symmetry may not be necessarily respected in general while the weak version in Eq.~(\ref{eq: fermion parity symmetry}) is always respected.
In Sec.~\ref{sec: Lindblad SYK symmetry}, we explicitly construct the weak and strong fermion parity symmetry and show that it leads to the rich symmetry classification of open quantum many-body fermionic systems.

\subsection{Modular conjugation}
    \label{subsec: KMS}
    

For open quantum fermionic systems, we cannot straightforwardly introduce the antiunitary swap operation in Eq.~(\ref{eq: boson modular conjugation 1}) because of the fermionic anticommutation relation. 
We can still find modular conjugation
by demanding 
(i) it exchanges $\psi^+$ and $\psi^-$;
(ii) it is antiunitary;
and 
(iii) it leaves the reference state invariant.
We are thus led to introduce 
modular conjugation by 
\begin{align}
    \mathcal{J} \psi_{n}^{+} \mathcal{J}^{-1} 
    = \psi_{n}^{-},~~
    \mathcal{J} \psi_{n}^{-} {\cal J}^{-1} 
    = \psi_{n}^{+},~~
    \mathcal{J} z \mathcal{J}^{-1}
    = z^{*} 
    \label{eq: modular conjugation - fermion}
\end{align}
for each fermion $\psi_{n}^{\pm}$ and $z \in \mathbb{C}$.
Under this fermionic modular conjugation, we also have Eq.~(\ref{mod conj def}) for arbitrary bosonic operators $O^{+}$ (i.e., arbitrary operators that consist of the even number of fermionic operators).
Since arbitrary density operators consist of the even number of fermionic operators,
fermionic Lindbladians $\mathcal{L}$ are generally invariant under modular conjugation:
\begin{equation}
    \mathcal{J}\,\mathcal{L}\,\mathcal{J}^{-1} = \mathcal{L}. 
        \label{eq: modular conjugation symmetry}
\end{equation}

As discussed in Sec.~\ref{subsec: fermion number parity}, total fermion parity $\left( -1 \right)^{\cal F}$ in the double Hilbert space is always conserved. 
Thus, we can introduce another type of modular conjugation $\tilde{\cal J}$ by
\begin{align}
    \tilde{\cal J} \psi_{n}^{+} \tilde{\cal J}^{-1} 
    = - \psi_{n}^{-},~
    \tilde{\cal J} \psi_{n}^{-} \tilde{\cal J}^{-1} 
    = - \psi_{n}^{+},~
    \tilde{\cal J} z \tilde{\cal J}^{-1}
    = z^{*},
        \label{eq: modular conjugation - fermion - tilde}
\end{align}
satisfying $\tilde{\cal J} \propto \left( -1 \right)^{\cal F} {\cal J}$.
Generic fermionic Lindbladians are also invariant under this type of modular conjugation (i.e., $\tilde{\cal J}\,\mathcal{L}\,\tilde{\cal J}^{-1} = \mathcal{L}$).
In the subspace with fixed fermion parity $\left( -1 \right)^{\cal F}$, the two antiunitary operators $\mathcal{J}$ and $\tilde{\cal J}$ are equivalent to each other.
On the other hand, fermion parity $\left( -1 \right)^{F^{\pm}}$ in the individual ket or bra space is not necessarily conserved, as discussed in Sec.~\ref{subsec: fermion number parity}.
Consequently, the products of ${\cal J}$ and $\left( -1 \right)^{F^{\pm}}$ do not necessarily give rise to symmetry of Lindbladians in contrast to $\mathcal{J}$ and $\tilde{\cal J}$.

In Sec.~\ref{sec: Lindblad SYK symmetry}, we explicitly construct fermionic modular conjugation in Eqs.~(\ref{eq: modular conjugation - fermion}) and (\ref{eq: modular conjugation - fermion - tilde}).
Furthermore, in Sec.~\ref{sec: dissipative quantum chaos}, we demonstrate that modular conjugation symmetry is relevant to the complex-spectral statistics and dissipative quantum chaos of the SYK Lindbladians.
It is also notable that while all fermionic Lindbladians respect Eq.~(\ref{eq: modular conjugation symmetry}), modular conjugation $\mathcal{J}$ does not necessarily commute with fermion parity $\left( -1 \right)^{F^{\pm}}$ in the individual ket or bra space.
Hence, when $\mathcal{J}$ and $\left( -1 \right)^{F^{\pm}}$ anticommute with each other, modular conjugation symmetry $\mathcal{J}$ is not respected in the subspace with fixed fermion parity $\left( -1 \right)^{F^{\pm}}$.
We demonstrate this breaking of modular conjugation symmetry for the SYK Lindbladians with $N \equiv 2$ (mod $4$) and even $p$.


\begin{table}[t]
	\centering
	\caption{Four-fold algebraic structure of total fermion parity symmetry $\left( -1 \right)^{\cal F}$, modular conjugation symmetry $\mathcal{J}$, and antiunitary symmetry $\mathcal{R}$ in the double Hilbert space of open quantum fermionic systems. Here, $a \in \{ +1, -1 \}$ specifies the commutation or anticommutation relation between $\left( -1 \right)^{\cal F}$ and $\mathcal{R}$ [i.e., $\mathcal{R} \left( -1 \right)^{\cal F} = a \left( -1 \right)^{\cal F} \mathcal{R}$]; $b \in \{ +1, -1 \}$ specifies the commutation or anticommutation relation between $\mathcal{J}$ and $\mathcal{R}$ (i.e., $\mathcal{R} \mathcal{J} = b \mathcal{J} \mathcal{R}$).
    }
     \begin{tabular}{c|cccc} \hline \hline
     ~~$N$ (mod $4$)~~ & $0$ & $1$ & $2$ & $3$ \\ \hline
     $a$ & ~~$+1$~~ & ~~$-1$~~ & ~~$+1$~~ & ~~$-1$~~ \\
     $b$ & ~~$+1$~~ & ~~$+1$~~ & ~~$-1$~~ & ~~$-1$~~ \\
     $\mathcal{R}^2$ & $+1$ & $+1$ & $-1$ & $-1$ \\ \hline \hline
    \end{tabular}
        \label{tab: PSA}
\end{table}

\subsection{Projective symmetry analysis: \texorpdfstring{$\mathbb{Z}_4$}{Z4} periodicity} 
    \label{subsec: PSA}

We now discuss the algebraic properties 
of total fermion parity and 
modular conjugation in more detail. 
For Hermitian Majorana fermion systems,
which can be thought of as being realized as 
a boundary of (1+1)-dimensional topological superconductors,
it is known that symmetry actions are realized projectively
\cite{Fidkowski-Kitaev-10, *Fidkowski-Kitaev-11, Turner-11}.
Specifically, there is an eight-fold algebraic structure 
of fermion parity and time-reversal symmetry (see also Appendix~\ref{asec: symmetry classification - SYK Hamiltonian}).

In the double Hilbert space, 
modular conjugation $\mathcal{J}$
and total fermion parity $(-1)^{\mathcal{F}}$
realize a linear representation, 
i.e., 
modular conjugation $\mathcal{J}$ always satisfies $\mathcal{J}^2 = +1$ and commutes with $\left( -1 \right)^{\cal F}$
for any $N$. 
This can be seen by 
explicitly constructing 
these operators 
in the double Hilbert space.
For even $N$, 
they are constructed explicitly
as
\begin{align}
   & \left( -1 \right)^{\cal F} = \prod_{i=1}^{N/2} \left( 2\ii \psi_{2i-1}^{+} \psi_{2i}^{+} \right) \left( 2\ii \psi_{2i-1}^{-} \psi_{2i}^{-} \right),
    \label{eq: total fermion parity - even N} \\
    &\mathcal{J} = \prod_{i=1}^{N/2} 
    \left( \psi_{2i-1}^{+} - \psi_{2i-1}^{-} \right)
    \left( \ii \psi_{2i}^{+} +
    \ii \psi_{2i}^{-} \right) \mathcal{K}. \label{eq: J - even N}
\end{align}
Similarly, for odd $N$, 
\begin{align}
    &\left( -1 \right)^{\cal F} = \left[ \prod_{i=1}^{\left(N-1\right)/2} \left( 2\ii \psi_{2i-1}^{+} \psi_{2i}^{+} \right) \left( 2\ii \psi_{2i-1}^{-} \psi_{2i}^{-} \right) \right] \nonumber \\
    &\qquad\qquad\qquad\qquad\qquad\qquad\qquad 
    \times ( 2\ii \psi_{N}^{+} \psi_{N}^{-} ), \label{eq: total fermion parity - odd N} \\
    \nonumber \\
        &\mathcal{J} = \left[ \prod_{i=1}^{(N-1)/2} 
    \left( \psi_{2i-1}^{+} +
    \psi_{2i-1}^{-} \right)
    \left( \ii\psi_{2i}^{+} -
    \ii \psi_{2i}^{-} \right) \right] \nonumber \\ 
    &\qquad\qquad\qquad\qquad\qquad\qquad \times 
    \left( \psi_{N}^{+} 
    +
    \psi_{N}^{-} \right) \mathcal{K}, \label{eq: J - odd N}
\end{align}
Here, $N$ is the total number of fermion operators, and we choose $\psi_i$ with odd $i$ to be real and symmetric and $\psi_i$ with even $i$ to be pure imaginary and antisymmetric,
\begin{equation}
    \mathcal{K} \psi_{2i-1} \mathcal{K} = \psi_{2i-1},\quad
    \mathcal{K} \psi_{2i} \mathcal{K} = - \psi_{2i}
        \label{eq: Majorana - complex conjugation}
\end{equation}
with complex conjugation $\mathcal{K}$, 
so that the corresponding complex fermion operators will be real.
From these representations, 
we verify the relations
\begin{align}
    \mathcal{J}^2 = +1,\quad
    \mathcal{J} \left( -1 \right)^{\cal F} = \left( -1 \right)^{\cal F} \mathcal{J}.
        \label{eq: J-basic}
\end{align}

On the other hand, once we add an antiunitary symmetry, a four-fold algebraic structure emerges.
In addition to total fermion parity and modular 
conjugation, 
let us consider, similarly to time reversal, an antiunitary operation $\mathcal{R}$ that acts as
\begin{align}
    \mathcal{R} \psi_{n}^{\pm} \mathcal{R}^{-1} = \psi_{n}^{\pm},\quad \mathcal{R} z \mathcal{R}^{-1} = z^{*}
        \label{eq: OQS-R}
\end{align}
for $z \in \mathbb{C}$.
Notably, while the action of this antiunitary operation $\mathcal{R}$ is similar to time reversal $T$ in closed quantum systems, $\mathcal{R}$ does not necessarily describe physical time reversal in open quantum systems.
In fact, while $T e^{-\ii H t} T^{-1} = e^{+\ii\,(THT^{-1})\,t}$ gives the time-reversed dynamics of the unitary dynamics $e^{-\ii Ht}$, the open-quantum counterpart $\mathcal{R} e^{\mathcal{L} t} \mathcal{R}^{-1} = e^{(\mathcal{R}\mathcal{L}\mathcal{R}^{-1})\,t}$ does not give the time-reversed dynamics of the Lindbladian dynamics $e^{\mathcal{L} t}$.
Still, the antiunitary symmetry $\mathcal{R}$ gives constraints on open quantum systems.
Total fermion parity symmetry $\left( -1 \right)^{\cal F}$ and antiunitary symmetry $\mathcal{R}$, as well as modular conjugation symmetry $\mathcal{J}$, also play a central role in the symmetry classification of the SYK Lindbladians, as discussed in Sec.~\ref{sec: Lindblad SYK symmetry}.
There, fermion parity $\left( -1 \right)^{F^{\pm}}$ in the individual ket and bra spaces, and other types of antiunitary symmetry $\mathcal{P}$, $\mathcal{Q}$, $\mathcal{S}$ are also relevant.

As before,
we explicitly construct 
the symmetry operation
$\mathcal{R}$
and obtain their algebraic structure (Table~\ref{tab: PSA}).
For even $N$, 
the antiunitary operator $\mathcal{R}$ satisfying Eq.~(\ref{eq: OQS-R}) is
\begin{align}
    {\cal R} = \left( \prod_{i=1}^{N/2} 2\,( \ii {\psi}_{2i}^{+} ) ( \ii {\psi}_{2i}^{-} ) \right) \mathcal{K}.
\end{align}
From this representation, we verify 
\begin{align}
    \mathcal{R}^2 =
    \prod_{i=1}^{N/2} 4\,(\ii {\psi}_{2i}^{+}) (\ii {\psi}_{2i}^{-}) (\ii {\psi}_{2i}^{+}) (\ii {\psi}_{2i}^{-})
    = \left( -1 \right)^{N/2}
\end{align}
and the relations
\begin{align}
&
    \mathcal{R} \left( -1 \right)^{\cal F} =  \left( -1 \right)^{\cal F} \mathcal{R}, 
    \\
    &
    \mathcal{R} \mathcal{J} = \left( -1 \right)^{N/2} \mathcal{J} \mathcal{R}.
\end{align}
Similarly, for odd $N$, 
$\mathcal{R}$ is constructed as
\begin{align}
    {\cal R} = \left( \prod_{i=1}^{(N-1)/2} 2\,( \ii {\psi}_{2i}^{+} ) ( \ii {\psi}_{2i}^{-} ) \right) \mathcal{K},
\end{align}
leading to
\begin{align}
&
    \mathcal{R}^2 = \left( -1 \right)^{(N-1)/2},
    \\
    &
    \mathcal{R} \left( -1 \right)^{\cal F} = -\left( -1 \right)^{\cal F} \mathcal{R}, 
    \\
    &
    \mathcal{R} \mathcal{J} = \left( -1 \right)^{(N-1)/2} \mathcal{J} \mathcal{R}.
\end{align}
These algebraic structures are summarized as Table~\ref{tab: PSA}.
The algebraic relation between $\mathcal{J}$ and $\mathcal{R}$ (i.e., $b \in \{ +1, -1 \}$ with $\mathcal{R} \mathcal{J} = b \mathcal{J} \mathcal{R}$) is equivalent to the sign $\mathcal{R}^2$ of the antiunitary symmetry for arbitrary $N$.

Importantly, the algebraic structure of total fermion parity $\left( -1 \right)^{\cal F}$, modular conjugation $\mathcal{J}$, and antiunitary symmetry $\mathcal{R}$ is four-fold periodic in $N$
(Table~\ref{tab: PSA}), which contrasts with the eight-fold periodicity for closed quantum systems~\cite{Fidkowski-Kitaev-10, *Fidkowski-Kitaev-11, Turner-11}. 
Consequently, the classification of the SYK Lindbladians is also four-fold periodic in terms of $N$, as shown in Sec.~\ref{sec: Lindblad SYK symmetry}.
The different periodicity is a direct consequence of the operator-state mapping.
As mentioned above, the eight-fold periodicity of the algebraic structure of symmetry is closely related to the $\mathbb{Z}_8$ topological classification of interacting Majorana fermions with time-reversal symmetry~\cite{Fidkowski-Kitaev-10, *Fidkowski-Kitaev-11, Turner-11}. 
Similarly, our four-fold periodicity of the algebraic structure of symmetry should be relevant to the topological classification of open quantum fermionic systems.
Finally, 
it is worthwhile noticing that 
the real structures on 
spectral triplets in noncommutative geometry
are similarly classified with eight-fold periodicity~\cite{Connes:1995tu, *Connes:1996gi}.
In these works, the case of modular conjugation
that squares to $-1$ is also considered.

\section{Symmetry classification of the Sachdev-Ye-Kitaev Lindbladians}
    \label{sec: Lindblad SYK symmetry}

\begin{table*}[t]
	\centering
	\caption{Periodic table of the Sachdev-Ye-Kitaev (SYK) Lindbladians with the linear dissipators $p=1$ for $q \equiv 0, 2$ (mod $4$) and the number $N$ (mod $4$) of Majorana fermions. See Appendix~\ref{asec: 38-fold symmetry} for the detailed definitions of the symmetry classes. For the antiunitary symmetry $\mathcal{J}$, $\mathcal{P}$, $\mathcal{Q}$, $\mathcal{R}$, $\mathcal{S}$, the entries $\pm 1$ mean the presence of the symmetry and its sign, and the entries $0$ mean the absence of the symmetry. Additional symmetry can be present for $K_{m; i} K_{m; j}^{*} \in \mathbb{R}$. For odd $p \geq 3$, the antiunitary symmetry ${\cal P}$, ${\cal R}$ is no longer respected while the antiunitary symmetry ${\cal Q}$, ${\cal S}$ remains to be respected; for arbitrary odd $p \geq 3$ and $N$, we have class AI (or equivalently class D$^{\dag}$) for $q \equiv 0$ with $K_{m; i} K_{m; j}^{*} \notin \mathbb{R}$ and $q \equiv 2$, and class BDI$^{\dag}$ for $q \equiv 0$ with $K_{m; i} K_{m; j}^{*} \in \mathbb{R}$.}
     \begin{tabular}{c|cccc} \hline \hline
     ~~$N$ (mod $4$)~~ & ~~$0$~~ & ~~$1$~~ & ~~$2$~~ & ~~$3$~~ \\ \hline
     fermion parity $\left( -1 \right)^{\cal F}$ & $\mathbb{Z}_2$ & $\mathbb{Z}_2$ & $\mathbb{Z}_2$ & $\mathbb{Z}_2$ \\
     modular conjugation $\mathcal{J}$ & $+1$ & $+1$ & $+1$ & $+1$ \\
     $\mathcal{P}$ & $+1$ & $0$ & $-1$ & $0$ \\
     $\mathcal{Q}$ & $+1$ & $+1$ & $+1$ & $+1$ \\
     $\mathcal{R}$ & $+1$ & $0$ & $-1$ & $0$ \\
     $\mathcal{S}$ & $+1$ & $+1$ & $+1$ & $+1$ \\ \hline
     ~~$q \equiv 0$ (mod $4$) ($K_{m; i} K_{m; j}^{*} \notin \mathbb{R}$)~~ & ~~AI = D$^{\dag}$~~ & ~~AI = D$^{\dag}$~~ & ~~AI = D$^{\dag}$~~ & ~~AI = D$^{\dag}$~~ \\
     ~~$q \equiv 0$ (mod $4$) ($K_{m; i} K_{m; j}^{*} \in \mathbb{R}$)~~ & ~~BDI + $\mathcal{S}_{++}$ = BDI$^{\dag}$ + $\mathcal{S}_{++}$~~ & BDI$^{\dag}$ & ~~BDI + $\mathcal{S}_{-+}$ = CI$^{\dag}$ + $\mathcal{S}_{+-}$~~ & BDI$^{\dag}$ \\ 
     $q \equiv 2$ (mod $4$) & BDI & AI = D$^{\dag}$ & CI & AI = D$^{\dag}$ \\ \hline \hline 
    \end{tabular}
	\label{tab: SYK Lindbladian p=1}
\end{table*}

\begin{table*}[t]
	\centering
	\caption{Periodic table of the Sachdev-Ye-Kitaev (SYK) Lindbladians with the even number $p$ of dissipators for $q \equiv 0, 2$ (mod $4$) and the number $N$ (mod $4$) of Majorana fermions. See Appendix~\ref{asec: 38-fold symmetry} for the detailed definitions of the symmetry classes. For the antiunitary symmetry $\mathcal{J}$, $\mathcal{P}$, $\mathcal{Q}$, $\mathcal{R}$, $\mathcal{S}$, the entries $\pm 1$ mean the presence of the symmetry and its sign, and the entries $0$ mean the absence of the symmetry. Additional symmetry can be present for $K_{m; i} K_{m; j}^{*} \in \mathbb{R}$. For even $N$, we assume even fermion parity $\left( -1 \right)^{\cal F} = +1$ in the double Hilbert space, which is relevant to the presence of modular conjugation symmetry $\mathcal{J}$ in the subspace with fixed fermion parity $\left( -1 \right)^{F^{\pm}}$.
	}
     \begin{tabular}{c|cccc} \hline \hline
     ~~~$N$ (mod $4$)~~~ & ~~~$0$~~~ & ~~~$1$~~~ & ~~~$2$~~~ & ~~~$3$~~~ \\ \hline
     fermion parity $\left( -1 \right)^{\cal F}$, $\left( -1 \right)^{F^{\pm}}$ & ~~$\mathbb{Z}_2 \times \mathbb{Z}_2$~~ & $\mathbb{Z}_2$ & $\mathbb{Z}_2 \times \mathbb{Z}_2$ & $\mathbb{Z}_2$ \\
     modular conjugation $\mathcal{J}$ & $+1$ & $+1$ & $0$ & $+1$ \\
     $\mathcal{P}$ & $+1$ & $0$ & $0$ & $0$ \\
     $\mathcal{Q}$ & $+1$ & $+1$ & $0$  & $+1$ \\
     $\mathcal{R}$ & $+1$ & $0$ & $0$ & $0$ \\
     $\mathcal{S}$ & $+1$ & $+1$ & $0$ & $+1$ \\ \hline
     ~~$q \equiv 0$ (mod $4$) ($K_{m; i} K_{m; j}^{*} \notin \mathbb{R}$), $q \equiv 2$ (mod $4$)~~ & ~~AI=D$^{\dag}$~~ & ~~AI=D$^{\dag}$~~ & ~~A~~ & ~~AI=D$^{\dag}$~~ \\ 
     $q \equiv 0$ (mod $4$) ($K_{m; i} K_{m; j}^{*} \in \mathbb{R}$) & BDI$^{\dag}$ & BDI$^{\dag}$  & A + $\eta$ = AIII & BDI$^{\dag}$ \\ \hline \hline
    \end{tabular}
	\label{tab: SYK Lindbladian p=2}
\end{table*}

As a prototypical example of open quantum many-body fermionic systems, 
we now develop the symmetry classification of the dissipative SYK model described by the Lindblad master equation~\cite{Sa-22, Kulkarni-22}.
The results are summarized 
in the periodic tables~\ref{tab: SYK Lindbladian p=1} and \ref{tab: SYK Lindbladian p=2}.
In the absence of dissipation, the SYK model is classified by two types of antiunitary symmetry 
(see Appendix~\ref{asec: symmetry classification - SYK Hamiltonian} for details)~\cite{You-17, Fu-16, GarciaGarcia-16, Cotler-17, Li-17, Kanazawa-17, Behrends-19, Sun-20}.
We generalize 
them 
to the SYK Lindbladians and investigate whether 
they remain
symmetry even in the presence of dissipation (Fig.~\ref{fig: PQRS}).
In addition to the antiunitary symmetry that is reminiscent of the symmetry of the SYK Hamiltonians, the SYK Lindbladians respect modular conjugation symmetry, as discussed in Sec.~\ref{subsec: KMS}.
Furthermore, as also discussed in Sec.~\ref{subsec: fermion number parity}, we need to consider the subspace with fixed fermion parity to study symmetry and spectral statistics.
Taking everything into consideration, we obtain the periodic tables of the SYK Lindbladians 
having dissipators with the odd/even number of fermion operators 
(Table~\ref{tab: SYK Lindbladian p=1}/\ref{tab: SYK Lindbladian p=2}).
The period of this symmetry classification is four in terms of the number $N$ of Majorana fermions while that of the SYK Hamiltonians is eight.
This is a consequence of the operator-state mapping and also consistent with the symmetry analysis in Sec.~\ref{subsec: PSA}.
The SYK Lindbladians exhibit several unique symmetry classes. 
Some of such symmetry classes were not argued to appear in noninteracting quadratic Lindbladians~\cite{Lieu-20} and arise from the many-body nature of the SYK Lindbladians.

\subsection{Model}

The Hamiltonian $H$ of the SYK Lindbladian~\cite{Sa-22, Kulkarni-22} is defined as the $q$-body SYK model
\begin{align}
    H = \ii^{q/2} \sum_{1 \leq i_1 < \cdots < i_q \leq N} J_{i_1, \cdots, i_q} \psi_{i_1} \cdots \psi_{i_q},
        \label{eq: SYK Hamiltonian}
\end{align}
where $q$ is even, and $J_{i_1, \cdots, i_q}$'s are the real-valued random coupling drawn from the Gaussian distribution satisfying
\begin{equation}
    \overline{J_{i_1, \cdots, i_q}} = 0,\quad
    \overline{J_{i_1, \cdots, i_q}^2} = \frac{(q-1)!}{N^{q-1}}J^2,
\end{equation}
where the overlines denote the disorder average.
This Hamiltonian includes $N$ Majorana fermions that are defined to satisfy
\begin{align}
    \{ \psi_{i}, \psi_{j} \} = \delta_{ij},\quad
    \psi_{i}^{\dag} = \psi_{i}.
\end{align}
We add the Markovian dissipation described by Eqs.~(\ref{eq: Lindblad equation}) and (\ref{eq: Lindbladian}) to the SYK Hamiltonian $H$.
The dissipators $L_m$'s are chosen to be 
generic 
all-to-all $p$-body Majorana fermion 
operators 
\begin{align}
    L_m = \sum_{1 \leq i_1 < \cdots < i_p \leq N} K_{m; i_1, \cdots, i_p} \psi_{i_1} \cdots \psi_{i_p}
\end{align}
for $m=1, 2, \cdots, M$.
Here, $K_{m; i_1, \cdots, i_p}$'s are the complex-valued random coupling drawn from the Gaussian distribution satisfying
\begin{equation}
    \overline{K_{m; i_1, \cdots, i_p}} = 0,\quad
    \overline{\left|K_{m; i_1, \cdots, i_p}\right|^2} = \frac{p!}{N^p}K^2
\end{equation}
with a constant $K \geq 0$.
Previously, the complex-spectral properties of the SYK Lindbladian were investigated for the linear dissipators $p=1$~\cite{Kulkarni-22} and the quadratic dissipators $p=2$~\cite{Sa-22, Kulkarni-22}, mainly for large $N$ and $M$.
It is also notable that Ref.~\cite{Xu-21} investigated the stochastic dynamics of the four-body SYK Hamiltonian, which is equivalent to the SYK Lindbladian with $p=q=4$.
Here, we study the symmetry classification and dissipative quantum chaos for finite $N$ and $M$.

As discussed in Sec.~\ref{subsec: operator-state mapping}, it is useful for the symmetry analysis to vectorize the Lindblad equation on the basis of the operator-state mapping.
The SYK Lindbladian is mapped to the non-Hermitian many-body operator
\begin{align}
    \mathcal{L} = - \ii H^{+} + \ii \left( -\ii \right)^{q} H^{-} + \mathcal{D}.
\end{align}
Here, the Hamiltonians $H^{+}$ and $H^{-}$ are the SYK Hamiltonians described by the Majorana fermions $\psi^{+}$ and $\psi^{-}$ that belong to the ket and bra spaces, respectively.
The dissipation term $\mathcal{D}$ that couples $\psi^{+}$ and $\psi^{-}$ is given as
\begin{align}
    &\mathcal{D} = \sum_{m, i, j} K_{m; i} K_{m; j}^{*} \left( -\ii \psi_{i_1}^{+} \cdots \psi_{i_{p}}^{+} \psi_{j_1}^{-} \cdots \psi_{j_{p}}^{-} \right. \nonumber \\
    &\qquad\qquad\quad- \frac{1}{2} \psi_{j_p}^{+} \cdots \psi_{j_{1}}^{+} \psi_{i_1}^{+} \cdots \psi_{i_{p}}^{+} \nonumber \\
    &\qquad\qquad\quad\left.- \frac{1}{2} \psi_{i_p}^{-} \cdots \psi_{i_{1}}^{-} \psi_{j_1}^{-} \cdots \psi_{j_{p}}^{-} \right)
        \label{eq: dissipator odd p}
\end{align}
for odd $p$ and
\begin{align}
    &\mathcal{D} = \sum_{m, i, j} K_{m; i} K_{m; j}^{*} \left( \psi_{i_1}^{+} \cdots \psi_{i_{p}}^{+} \psi_{j_1}^{-} \cdots \psi_{j_{p}}^{-} \right. \nonumber \\
    &\qquad\qquad\quad- \frac{1}{2} \psi_{j_p}^{+} \cdots \psi_{j_{1}}^{+} \psi_{i_1}^{+} \cdots \psi_{i_{p}}^{+} \nonumber \\
    &\qquad\qquad\quad\left. - \frac{1}{2} \psi_{i_p}^{-} \cdots \psi_{i_{1}}^{-} \psi_{j_1}^{-} \cdots \psi_{j_{p}}^{-} \right)
        \label{eq: dissipator even p}
\end{align}
for even $p$, with $\sum_{i} \coloneqq \sum_{1 \leq i_1 < \cdots < i_p \leq N}$ and $K_{m; i} \coloneqq K_{m; i_1, \cdots, i_{p}}$.
The trace of the SYK Lindbladian is obtained as
\begin{equation}
    \mathrm{tr}\,\mathcal{L} = - \frac{1}{2^p} \sum_{m, i} \left| K_{m; i} \right|^2 \leq 0.
\end{equation}
Furthermore, the Hermitian conjugate of $\mathcal{D}$ is given as
\begin{align}
    &\mathcal{D}^{\dag} = \sum_{m, i, j} K_{m; i}^{*} K_{m; j} \left( -\ii \psi_{i_1}^{+} \cdots \psi_{i_{p}}^{+} \psi_{j_1}^{-} \cdots \psi_{j_{p}}^{-} \right.\nonumber \\
    &\qquad\qquad\quad- \frac{1}{2} \psi_{i_p}^{+} \cdots \psi_{i_{1}}^{+} \psi_{j_1}^{+} \cdots \psi_{j_{p}}^{+} \nonumber \\
    &\qquad\qquad\quad\left. - \frac{1}{2} \psi_{j_p}^{-} \cdots \psi_{j_{1}}^{-} \psi_{i_1}^{-} \cdots \psi_{i_{p}}^{-} \right)
\end{align}
for odd $p$ and
\begin{align}
    &\mathcal{D}^{\dag} = \sum_{m, i, j} K_{m; i}^{*} K_{m; j} \left( \psi_{i_1}^{+} \cdots \psi_{i_{p}}^{+} \psi_{j_1}^{-} \cdots \psi_{j_{p}}^{-} \right. \nonumber \\
    &\qquad\qquad\quad- \frac{1}{2} \psi_{i_p}^{+} \cdots {\psi}_{i_{1}}^{+} \psi_{j_1}^{+} \cdots \psi_{j_{p}}^{+} \nonumber \\
    &\qquad\qquad\quad\left.- \frac{1}{2} \psi_{j_p}^{-} \cdots \psi_{j_{1}}^{-} \psi_{i_1}^{-} \cdots \psi_{i_{p}}^{-} \right)
\end{align}
for even $p$.
Notably, for $K_{m; i} K_{m; j}^{*} = K_{m; i}^{*} K_{m; j}$ (i.e., $K_{m; i} K_{m; j}^{*} \in \mathbb{R}$), including the real-valued coupling $K_{m; i} \in \mathbb{R}$ and the pure-imaginary coupling $K_{m; i} \in \ii \mathbb{R}$, the dissipation term $\mathcal{D}$ is Hermitian for arbitrary $p$:
\begin{equation}
    \mathcal{D}^{\dag} = \mathcal{D}.
\end{equation}
This additional property is relevant to the symmetry classification, as discussed in the following.
We also note that Eq.~(\ref{eq: dissipator odd p}) assumes even fermion parity $\left( -1 \right)^{\cal F} = +1$ of the density operator $\rho$. 
When fermion parity of the density operator $\rho$ is odd $\left( -1 \right)^{\cal F} = -1$ and $p$ is odd, we instead have
\begin{align}
    &\mathcal{D} = \sum_{m, i, j} K_{m; i} K_{m; j}^{*} \left( +\ii \psi_{i_1}^{+} \cdots \psi_{i_{p}}^{+} \psi_{j_1}^{-} \cdots \psi_{j_{p}}^{-} \right.\nonumber \\
    &\qquad\qquad\quad- \frac{1}{2} \psi_{j_p}^{+} \cdots \psi_{j_{1}}^{+} \psi_{i_1}^{+} \cdots \psi_{i_{p}}^{+} \nonumber \\
    &\qquad\qquad\quad\left.- \frac{1}{2} \psi_{i_p}^{-} \cdots \psi_{i_{1}}^{-} \psi_{j_1}^{-} \cdots \psi_{j_{p}}^{-} \right).
\end{align}
Since this difference does not change the symmetry classification, we below study Eq.~(\ref{eq: dissipator odd p}) for even fermion parity.

\subsection{Fermion parity symmetry}
    \label{subsec: SYK - fermion parity}

\begin{table}[t]
	\centering
	\caption{Fermion parity symmetry for the Sachdev-Ye-Kitaev (SYK) Lindbladians. Total fermion parity $\left( -1 \right)^{\cal F}$ is always conserved and gives rise to $\mathbb{Z}_2$ unitary symmetry in the double Hilbert space. For even $N$ and even $p$, fermion parity $\left( -1 \right)^{F^{\pm}}$ is conserved even in each ket or bra space. While total fermion parity commutes with complex conjugation $\mathcal{K}$ for even $N$, it anticommutes with $\mathcal{K}$ for odd $N$.}
     \begin{tabular}{c|cc} \hline \hline
     ~~~~ & ~~Even $N$~~ & ~~Odd $N$~~ \\ \hline
     Odd $p$ & $\mathbb{Z}_2$; $[ \left( -1 \right)^{\cal F},  \mathcal{K} ] = 0$ & $\mathbb{Z}_2$; $\{ \left( -1 \right)^{\cal F},  \mathcal{K} \} = 0$ \\
     ~~Even $p$~~ & ~~$\mathbb{Z}_2 \times \mathbb{Z}_2$; $[ \left( -1 \right)^{F^{\pm}},  \mathcal{K} ] = 0$~~ & $\mathbb{Z}_2$; $\{ \left( -1 \right)^{\cal F}, \mathcal{K} \} = 0$ \\ \hline \hline
    \end{tabular}
        \label{tab: fermion parity}
\end{table}

As also discussed in Sec.~\ref{subsec: fermion number parity}, total fermion parity $\left( -1 \right)^{\cal F}$ defined by Eq.~(\ref{eq: fermion parity calF - psi}) is always conserved in the double Hilbert space because of the operator-state mapping.
Here, we explicitly construct total fermion parity $\left( -1 \right)^{\cal F}$ for both even and odd $N$.
For even $p$, fermion parity $\left( -1 \right)^{F^{\pm}}$ in the individual ket and bra spaces defined by Eqs.~(\ref{eq: Fpm - psi - 1}) and (\ref{eq: Fpm - psi - 2}) is also relevant.
We also explicitly construct $\left( -1 \right)^{F^{\pm}}$ for even and odd $N$ and discuss their relationship.
Relevant fermion parity symmetry for each $N$ and $p$ is summarized in Table~\ref{tab: fermion parity}.

\subsubsection{Even \texorpdfstring{$N$}{N}}

For even $N$, total fermion parity $\left( -1 \right)^{\cal F}$ in Eq.~(\ref{eq: fermion parity calF - psi}) is given as [i.e., Eq.~(\ref{eq: total fermion parity - even N})]
\begin{align}
    \left( -1 \right)^{\cal F} = \prod_{i=1}^{N/2} \left( 2\ii \psi_{2i-1}^{+} \psi_{2i}^{+} \right) \left( 2\ii \psi_{2i-1}^{-} \psi_{2i}^{-} \right).
\end{align}
Generic fermionic Lindbladians, including the SYK Lindbladian, respect total fermion parity symmetry in Eq.~(\ref{eq: fermion parity symmetry}) for any case because it consists of the even number of fermion operators. 
To study the symmetry and spectral statistics, we consider the symmetry operations in the subspace with fixed total fermion parity $\left( -1 \right)^{\cal F}$.

Fermion parity operators $\left( -1 \right)^{F^{\pm}}$ in each ket or bra space defined by Eqs.~(\ref{eq: Fpm - psi - 1}) and (\ref{eq: Fpm - psi - 2}) are also constructed as
\begin{align}
    \left( -1 \right)^{F^{+}} &= \prod_{i=1}^{N/2} \left( 2\ii \psi_{2i-1}^{+} \psi_{2i}^{+} \right), 
        \label{eq: fermion parity even N +} \\
    \left( -1 \right)^{F^{-}} &= \prod_{i=1}^{N/2} \left( 2\ii \psi_{2i-1}^{-} \psi_{2i}^{-} \right).
        \label{eq: fermion parity even N -}
\end{align}
The product of $\left( -1 \right)^{F^{+}}$ and $\left( -1 \right)^{F^{-}}$ gives total fermion parity $\left( -1 \right)^{\cal F}$ in the double Hilbert space:
\begin{equation}
    \left( -1 \right)^{F^{+}} \left( -1 \right)^{F^{-}} = \left( -1 \right)^{\cal F}.
\end{equation}
For odd $p$, these unitary operators do not give rise to symmetry of the SYK Lindbladians since the dissipation term $\mathcal{D}$ contains the odd number of Majorana fermions in each space.
For even $p$, on the other hand, the dissipation term $\mathcal{D}$ contains the even number of Majorana fermions, and these operators individually give $\mathbb{Z}_2$ unitary symmetry of the SYK Lindbladian.

\subsubsection{Odd \texorpdfstring{$N$}{N}}

For odd $N$, total fermion parity defined by Eq.~(\ref{eq: fermion parity calF - psi}) is constructed as [i.e., Eq.~(\ref{eq: total fermion parity - odd N})]
\begin{align}
    &\left( -1 \right)^{\cal F} = \left[ \prod_{i=1}^{\left(N-1\right)/2} \left( 2\ii \psi_{2i-1}^{+} \psi_{2i}^{+} \right) \left( 2\ii \psi_{2i-1}^{-} \psi_{2i}^{-} \right) \right] \nonumber \\
    &\qquad\qquad\qquad\qquad\qquad\qquad\qquad \times ( 2\ii \psi_{N}^{+} \psi_{N}^{-} ).
\end{align}
Even for odd $N$, the SYK Lindbladian respects total fermion parity symmetry in Eq.~(\ref{eq: fermion parity symmetry}).
Notably, the additional term $2\ii \psi_{N}^{+} \psi_{N}^{-}$ is not invariant under complex conjugation with our choice of Majorana fermion operators in Eq.~(\ref{eq: Majorana - complex conjugation}).
Consequently, total fermion parity symmetry for odd $N$ anticommutes with complex conjugation:
\begin{align}
    \mathcal{K} \left( -1 \right)^{\cal F} \mathcal{K} = - \left( -1 \right)^{\cal F}.
\end{align}
This property changes the algebra between the symmetry operations and hence leads to the different symmetry classification.

For odd $N$, fermion parity $\left( -1 \right)^{F^{\pm}}$ in the individual ket and bra spaces defined by Eqs.~(\ref{eq: Fpm - psi - 1}) and (\ref{eq: Fpm - psi - 2}) is constructed as
\begin{align}
    \left( -1 \right)^{F^{+}} &= \left[ \prod_{i=1}^{\left( N-1 \right)/2} \left( 2\ii \psi_{2i-1}^{-} \psi_{2i}^{-} \right) \right]\,( \sqrt{2} \psi_{N}^{-} ), \\
    \left( -1 \right)^{F^{-}} &= \left[ \prod_{i=1}^{\left( N-1 \right)/2} \left( 2\ii \psi_{2i-1}^{+} \psi_{2i}^{+} \right) \right]\,( \sqrt{2} \psi_{N}^{+} ).
\end{align}
In contrast with the previous case for even $N$ [i.e., Eqs.~(\ref{eq: fermion parity even N +}) and (\ref{eq: fermion parity even N -})], fermion parity $\left( -1 \right)^{F^{\pm}}$ for odd $N$ consists of Majorana fermions $\psi_{i}^{\mp}$'s in the opposite Hilbert space.
In closed quantum systems with the odd number of Majorana fermions, fermion parity is constructed by adding a Majorana fermion in different Hilbert space~\cite{Fidkowski-Kitaev-11, Turner-11, You-17}.
In the double Hilbert space of open quantum systems, Majorana fermions in the opposite Hilbert space play a similar role of the additional Majorana fermion in the different Hilbert space.
Notably, these two unitary operators anticommute with each other:
\begin{align}
    \{ \left( -1 \right)^{F^{+}}, \left( -1 \right)^{F^{-}}\} = 0.
\end{align}
Thus, the two unitary operators cannot be diagonalized simultaneously in contrast with the previous case for even $N$.
Consequently, only total fermion parity $\left( -1 \right)^{\cal F}$ in the double Hilbert space is relevant for odd $N$.

\subsection{Modular conjugation symmetry}
    \label{subsec: SYK modular conjugation}

As discussed in Sec.~\ref{subsec: KMS}, Lindbladians generally respect additional antiunitary symmetry defined by modular conjugation.
Here, we explicitly construct modular conjugation $\mathcal{J}$ in Eq.~(\ref{eq: modular conjugation - fermion}) and $\tilde{\cal J}$ in Eq.~(\ref{eq: modular conjugation - fermion - tilde}) for even and odd $N$ and discuss their relationship with fermion parity symmetry.

\subsubsection{Even \texorpdfstring{$N$}{N}}

For even $N$, the antiunitary operators $\mathcal{J}$ and $\tilde{\cal J}$ defined by Eqs.~(\ref{eq: modular conjugation - fermion}) and (\ref{eq: modular conjugation - fermion - tilde}) are explicitly constructed as [i.e., Eq.~(\ref{eq: J - even N})]
\begin{align}
    \mathcal{J} &= \prod_{i=1}^{N/2} 
    \left( \psi_{2i-1}^{+} - \psi_{2i-1}^{-} \right)
    \left( \ii \psi_{2i}^{+} +
    \ii \psi_{2i}^{-} \right) \mathcal{K}, \\
    \tilde{\cal J} &= \prod_{i=1}^{N/2} 
    \left( \psi_{2i-1}^{+} + \psi_{2i-1}^{-} \right)
    \left( \ii \psi_{2i}^{+} -
    \ii \psi_{2i}^{-} \right) \mathcal{K},
\end{align}
satisfying
\begin{equation}
    \mathcal{J}^2 = \tilde{\cal J}^2 = +1.
        \label{eq: modular conjugation sign}
\end{equation}
Generic Lindbladians, including the SYK Lindbladians for arbitrary $N$ and $p$, are invariant under modular conjugation $\mathcal{J}$ and $\tilde{\cal J}$.
In addition, $\mathcal{J}$ and $\tilde{\cal J}$ commute with total fermion parity $\left( -1 \right)^{\cal F}$ in the double Hilbert space:
\begin{equation}
    \left( -1 \right)^{\cal F} \mathcal{J} \left( -1 \right)^{\cal F} = \mathcal{J},~\left( -1 \right)^{\cal F} \tilde{\cal J} \left( -1 \right)^{\cal F} = \tilde{\cal J}.
        \label{eq: modular conjugation & fermion parity}
\end{equation}
Thus, modular conjugation remains to give symmetry even in the subspace with fixed total fermion parity $\left( -1 \right)^{\cal F}$.

While only total fermion parity $\left( -1 \right)^{\cal F}$ is relevant in general, fermion parity $\left( -1 \right)^{F^{\pm}}$ in the individual ket or bra space is also relevant for even $p$, as described in Sec.~\ref{subsec: SYK - fermion parity}.
Even if ${\cal J}$ and $\tilde{\cal J}$ remain symmetry in the subspace with fixed $\left( -1 \right)^{\cal F}$, it is nontrivial whether they are still symmetry in the subspace with fixed fermion parity $\left( -1 \right)^{F^{\pm}}$.
To see this, we have
\begin{align}
    &\left( -1 \right)^{F^{+}} \mathcal{J} \left( -1 \right)^{F^{+}} = \tilde{\cal J},
\end{align}
and 
\begin{align}
    \mathcal{J} \tilde{\cal J} = \left( -1 \right)^{N/2} \left( -1 \right)^{\cal F},
\end{align}
leading to
\begin{equation}
    \left( -1 \right)^{F^{+}} \mathcal{J} \left( -1 \right)^{F^{+}}
    = \left( -1 \right)^{N/2} \left( -1 \right)^{\cal F} \mathcal{J}.
        \label{eq: modular conjugation & sub-fermion parity}
\end{equation}
Modular conjugation $\tilde{\cal J}$ also satisfies the same algebra.
Thus, for $\left( -1 \right)^{N/2} \left( -1 \right)^{\cal F} = +1$, modular conjugation $\mathcal{J}$ commutes with $\left( -1 \right)^{F^{\pm}}$ and hence remains symmetry even in the subspace with fixed $\left( -1 \right)^{F^{\pm}}$;
for $\left( -1 \right)^{N/2} \left( -1 \right)^{\cal F} = -1$, by contrast, $\mathcal{J}$ anticommutes with $\left( -1 \right)^{F^{\pm}}$ and is no longer symmetry in the subspace with fixed $\left( -1 \right)^{F^{\pm}}$.
We numerically confirm the breaking of modular conjugation symmetry in Sec.~\ref{sec: dissipative quantum chaos}.
It is also notable that even if modular conjugation no longer gives symmetry, the combination of modular conjugation and another operation can give rise to symmetry even in the subspace with fixed $\left( -1 \right)^{F^{\pm}}$.

\subsubsection{Odd \texorpdfstring{$N$}{N}}

For odd $N$, the antiunitary operators $\mathcal{J}$ and $\tilde{\cal J}$ defined by Eqs.~(\ref{eq: modular conjugation - fermion}) and (\ref{eq: modular conjugation - fermion - tilde})
are constructed as [i.e., Eq.~(\ref{eq: J - odd N})]
\begin{align}
    &\mathcal{J} = \left[ \prod_{i=1}^{(N-1)/2} 
    \left( \psi_{2i-1}^{+} +
    \psi_{2i-1}^{-} \right)
    \left( \ii\psi_{2i}^{+} -
    \ii \psi_{2i}^{-} \right) \right] \nonumber \\ 
    &\qquad\qquad\qquad\qquad\qquad\qquad \times \left( \psi_{N}^{+} 
    +
    \psi_{N}^{-} \right) \mathcal{K}, \\
    &\tilde{\cal J} = \left[ \prod_{i=1}^{(N-1)/2} 
    \left( \psi_{2i-1}^{+} -
    \psi_{2i-1}^{-} \right)
    \left( \ii\psi_{2i}^{+} +
    \ii \psi_{2i}^{-} \right) \right] \nonumber \\ 
    &\qquad\qquad\qquad\qquad\qquad\qquad \times \left( \psi_{N}^{+} 
    -
    \psi_{N}^{-} \right) \mathcal{K}.
\end{align}
With this definition, we have Eqs.~(\ref{eq: modular conjugation sign}) and (\ref{eq: modular conjugation & fermion parity}), and hence modular conjugation symmetry still gives symmetry in the subspace with fixed total fermion parity $\left( -1 \right)^{\cal F}$.

\subsection{Antiunitary symmetry}
    \label{subsec: PQRS}

\begin{table}[t]
	\centering
	\caption{Antiunitary symmetry for the Sachdev-Ye-Kitaev (SYK) Lindbladians. ${\cal P}$ and ${\cal Q}$ can be respectively replaced with ${\cal R}$ and ${\cal S}$ in each subspace with fixed fermion parity.}
     \begin{tabular}{c|cccc} \hline \hline
     ~~~$N$ (mod $4$)~~~ & ~~~$0$~~~ & ~~~$1$~~~ & ~~~$2$~~~ & ~~~$3$~~~ \\ \hline
     Odd $p$ & ~${\cal P}$ \& ${\cal Q}$~ & ${\cal Q}$ & ~${\cal P}$ \& ${\cal Q}$~ & ${\cal Q}$ \\ 
     Even $p$ & ${\cal P}$ & ${\cal Q}$ & none & ${\cal Q}$ \\ \hline \hline
    \end{tabular}
        \label{tab: PQRS}
\end{table}

\begin{figure}[t]
\centering
\includegraphics[width=54mm]{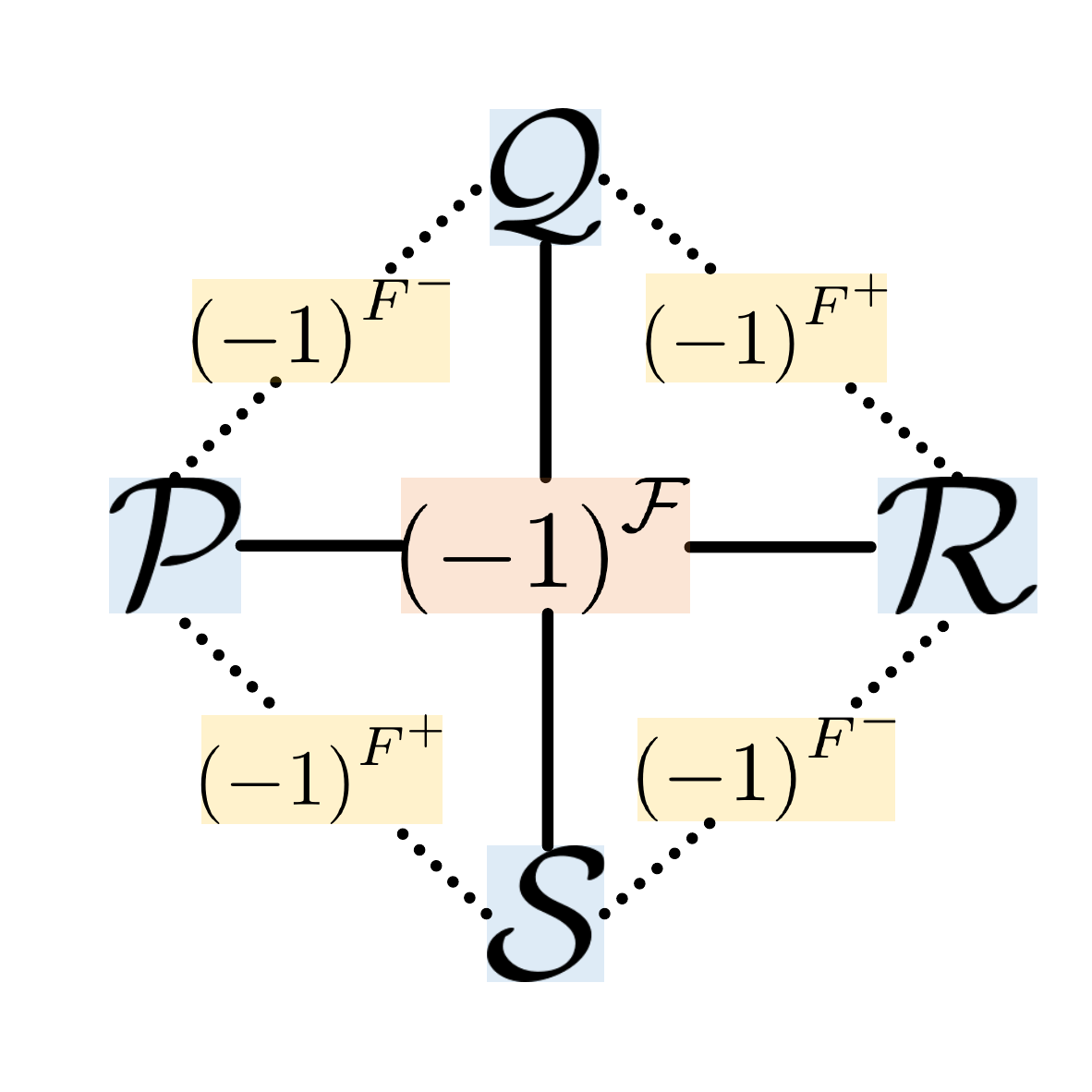} 
\caption{Relationship between the four types of antiunitary symmetry $\mathcal{P}$, $\mathcal{Q}$, $\mathcal{R}$, $\mathcal{S}$ and fermion parity symmetry $\left( -1 \right)^{\cal F}$, $\left( - 1\right)^{F^{\pm}}$. 
The product of $\mathcal{P}$ and $\mathcal{R}$ and that of $\mathcal{Q}$ and $\mathcal{S}$ are proportional to total fermion parity $\left( -1 \right)^{\cal F}$ in the double Hilbert space. 
The product of $\mathcal{P}$ and $\mathcal{S}$ and that of $\mathcal{Q}$ and $\mathcal{R}$ are proportional to fermion parity $\left( -1 \right)^{F^{+}}$ in the ket space. 
The product of $\mathcal{P}$ and $\mathcal{Q}$ and that of $\mathcal{R}$ and $\mathcal{S}$ are proportional to fermion parity $\left( -1 \right)^{F^{-}}$ in the bra space.}
	\label{fig: PQRS}
\end{figure}

The SYK Hamiltonians $H$ are generally classified according to two types of antiunitary symmetry (see Appendix~\ref{asec: symmetry classification - SYK Hamiltonian} for details)~\cite{You-17, Fu-16, GarciaGarcia-16, Cotler-17, Li-17, Kanazawa-17, Behrends-19, Sun-20}.
We study their counterparts in the SYK Lindbladians.
As a generalization of the antiunitary operators for the SYK Lindbladians, let us introduce the four antiunitary operators $\mathcal{P}$, $\mathcal{Q}$, $\mathcal{R}$, $\mathcal{S}$ in the double Hilbert space.
Each antiunitary operation is defined to respect
\begin{align}
    {\cal P} {\psi}_{n}^{\pm} {\cal P}^{-1} &= - {\psi}_{n}^{\pm}, \label{eq: P} \\
    {\cal Q} {\psi}_{n}^{\pm} {\cal Q}^{-1} &= \mp {\psi}_{n}^{\pm}, \label{eq: Q} \\
    {\cal R} {\psi}_{n}^{\pm} {\cal R}^{-1} &= + {\psi}_{n}^{\pm}, \label{eq: R} \\
    {\cal S} {\psi}_{n}^{\pm} {\cal S}^{-1} &= \pm {\psi}_{n}^{\pm}. \label{eq: S}
\end{align}
Similarly to closed quantum systems, the four antiunitary operators $\mathcal{P}$, $\mathcal{Q}$, $\mathcal{R}$, $\mathcal{S}$ are related with each other via fermion parity $\left( -1 \right)^{\cal F}$ and $\left( -1 \right)^{F^{\pm}}$ (Fig.~\ref{fig: PQRS}).

From these definitions of the antiunitary operations in Eqs.~(\ref{eq: P})-(\ref{eq: S}), we have
\begin{align}
    {\cal A} {H}^{\pm} {\cal A}^{-1} &= \left( -1 \right)^{q/2} {H}^{\pm},
\end{align}
for ${\cal A} = {\cal P}, {\cal Q}, {\cal R}, {\cal S}$, where $H^{+}$ and $H^{-}$ are the SYK Hamiltonians described by the Majorana fermions $\psi_{n}^{+}$ and $\psi_{n}^{-}$ in the ket and bra spaces, respectively.
Hence, we have
\begin{align}
    &{\cal A}\,(-\ii {H}^{+} + \ii \left( -\ii \right)^{q} {H}^{-})\,{\cal A}^{-1} \nonumber \\
    &\qquad\qquad= \left( -1 \right)^{q/2 + 1}\,(-\ii {H}^{+} + \ii \left( -\ii \right)^{q} {H}^{-}), \label{eq: AH1} \\
    &{\cal A}\,(-\ii {H}^{+} + \ii \left( -\ii \right)^{q} {H}^{-})^{\dag}\,{\cal A}^{-1} \nonumber \\
    &\qquad\qquad= \left( -1 \right)^{q/2}\,(-\ii {H}^{+} + \ii \left( -\ii \right)^{q} {H}^{-}). \label{eq: AH2}
\end{align}
Thus, the original symmetry of the SYK Hamiltonian can remain to be respected even in the double Hilbert space.
Whether it survives even in the presence of dissipation depends on $p$, which we study for $p=1$ (Sec.~\ref{subsec: SYK Lindbladian p1 classification}), $p=2$ (Sec.~\ref{subsec: SYK Lindbladian p2 classification}), and $p \geq 3$ (Sec.~\ref{subsec: generic dissipator}).
Before studying whether the dissipation term $\mathcal{D}$ is invariant under the antiunitary operations, we here generalize the results in Sec.~\ref{subsec: PSA} and discuss their algebraic structures with fermion parity in the double Hilbert space, clarifying which antiunitary operations are relevant in the subspace with fixed fermion parity.

\subsubsection{Even \texorpdfstring{$N$}{N}}

For even $N$, the antiunitary operators defined by Eqs.~(\ref{eq: P})-(\ref{eq: S}) are explicitly constructed as
\begin{align}
    {\cal P} &\coloneqq \left( \prod_{i=1}^{N/2} 2 {\psi}_{2i-1}^{+} {\psi}_{2i-1}^{-} \right) \mathcal{K}, \\
    {\cal Q} &\coloneqq \left( \prod_{i=1}^{N/2} 2 {\psi}_{2i-1}^{+}\,( \ii {\psi}_{2i}^{-} ) \right) \mathcal{K}, \\
    {\cal R} &\coloneqq \left( \prod_{i=1}^{N/2} 2\,( \ii {\psi}_{2i}^{+} ) ( \ii {\psi}_{2i}^{-} ) \right) \mathcal{K}, \\
    {\cal S} &\coloneqq \left( \prod_{i=1}^{N/2} 2\,( \ii {\psi}_{2i}^{+} )\,{\psi}_{2i-1}^{-} \right) \mathcal{K}.
\end{align}
The antiunitary operators $\mathcal{P}$ and $\mathcal{R}$ contain all the Majorana fermions $\psi_{n}^{\pm}$'s with odd $n$ and $\psi_{n}^{\pm}$'s with even $n$ in the double Hilbert space, respectively.
On the other hand, the antiunitary operator $\mathcal{Q}$ contains $\psi_{n}^{+}$'s with odd $n$ in the ket space and $\psi_{n}^{-}$'s with even $n$ in the bra space, and $\mathcal{S}$ contains $\psi_{n}^{+}$'s with even $n$ in the ket space and $\psi_{n}^{-}$'s with odd $n$ in the bra space.

Even though the antiunitary operators $\mathcal{P}$, $\mathcal{Q}$, $\mathcal{R}$, $\mathcal{S}$ respect Eqs.~(\ref{eq: AH1}) and (\ref{eq: AH2}), they do not necessarily give rise to symmetry in the subspace with fixed fermion parity $\left( -1 \right)^{\cal F}$ or $\left( -1 \right)^{F^{\pm}}$.
Importantly, these antiunitary operators form the algebra different from that of the SYK Hamiltonians, which we partially studied in Sec.~\ref{subsec: PSA}.
First, we have
\begin{align}
    {\cal P}^2 &= \prod_{i=1}^{N/2} 4 {\psi}_{2i-1}^{+} {\psi}_{2i-1}^{-} {\psi}_{2i-1}^{+} {\psi}_{2i-1}^{-}
    = \left( -1 \right)^{N/2}, \label{eq: P - sign} \\
    {\cal Q}^2 &= \prod_{i=1}^{N/2} 4 {\psi}_{2i-1}^{+} (\ii {\psi}_{2i}^{-}) {\psi}_{2i-1}^{+} (\ii {\psi}_{2i}^{-})
    = +1, \\
    {\cal R}^2 &=
    \prod_{i=1}^{N/2} 4\,(\ii {\psi}_{2i}^{+}) (\ii {\psi}_{2i}^{-}) (\ii {\psi}_{2i}^{+}) (\ii {\psi}_{2i}^{-})
    = \left( -1 \right)^{N/2}, \\
    {\cal S}^2 &=
    \prod_{i=1}^{N/2} 4\,(\ii {\psi}_{2i}^{+}) {\psi}_{2i-1}^{-} (\ii {\psi}_{2i}^{+}) {\psi}_{2i-1}^{-}
    = +1.
\end{align}
Notably, the signs of these antiunitary operators are cyclic for $N$ modulo $4$, and the period of the symmetry classification is $4$ in terms of $N$.
This contrasts with the symmetry classification of the SYK Hamiltonians, for which the signs of the antiunitary operators are cyclic for $N$ modulo $8$.
In addition, these antiunitary operators commute with fermion parity $\left( -1 \right)^{\cal F}$ in the double Hilbert space:
\begin{align}
    \left( -1 \right)^{\cal F} \mathcal{A} \left( -1 \right)^{\cal F} = \mathcal{A}
        \label{eq: total fermion parity & A - even N}
\end{align}
for $\mathcal{A} = \mathcal{P}, \mathcal{Q}, \mathcal{R}, \mathcal{S}$.
On the other hand, they do not necessarily commute with fermion parity $\left( -1 \right)^{F^{\pm}}$ in each ket or bra space:
\begin{equation}
    \left( -1 \right)^{F^{\pm}} \mathcal{A} \left( -1 \right)^{F^{\pm}} = \left( -1 \right)^{N/2} \mathcal{A}.
        \label{eq: fermion parity braket PQRS}
\end{equation}
We also have the following relationship between the antiunitary operators and fermion parity (Fig.~\ref{fig: PQRS}):
\begin{align}
    \mathcal{PR}, \mathcal{QS} &\propto \left( -1 \right)^{\cal F}, \\
    \mathcal{PS}, \mathcal{QR} &\propto \left( -1 \right)^{F^{+}}, \\
    \mathcal{PQ}, \mathcal{RS} &\propto \left( -1 \right)^{F^{-}}.
\end{align}
Therefore, for odd $p$, in the subspace with fixed total fermion parity $\left( -1 \right)^{\cal F}$, two of the antiunitary symmetry---for example, $\mathcal{P}$ and $\mathcal{Q}$---are relevant.
For even $p$, in the subspace with fixed fermion parity $\left( -1 \right)^{F^{+}}$ and $\left( -1 \right)^{F^{-}}$, only one of the antiunitary symmetry---for example, $\mathcal{P}$---is relevant for $N \equiv 0$ (mod $4$) and none of the antiunitary symmetry is relevant for $N \equiv 2$ (mod $4$).
We summarize the relevant symmetry in Table~\ref{tab: PQRS}.
Notably, for even $p$ and $N \equiv 2$ (mod $4$), although each antiunitary operation is no longer symmetry, the combination with modular conjugation can give rise to symmetry even in the subspace with fixed $\left( -1 \right)^{F^{+}}$ and $\left( -1 \right)^{F^{-}}$.

\subsubsection{Odd \texorpdfstring{$N$}{N}}

For odd $N$, the antiunitary symmetry operators defined by Eqs.~(\ref{eq: P})-(\ref{eq: S}) are constructed as
\begin{align}
    {\cal P} &\coloneqq \left( \prod_{i=1}^{(N+1)/2} 2 {\psi}_{2i-1}^{+} {\psi}_{2i-1}^{-} \right) \mathcal{K}, \\
    {\cal Q} &\coloneqq \left( \prod_{i=1}^{(N-1)/2} 2 {\psi}_{2i-1}^{+}\,( \ii {\psi}_{2i}^{-} ) \right) \left( \sqrt{2} {\psi}^{+}_{N} \right) \mathcal{K}, \\
    {\cal R} &\coloneqq \left( \prod_{i=1}^{(N-1)/2} 2\,( \ii {\psi}_{2i}^{+} ) ( \ii {\psi}_{2i}^{-} ) \right) \mathcal{K}, \\
    {\cal S} &\coloneqq \left( \prod_{i=1}^{(N-1)/2} 2\,( \ii {\psi}_{2i}^{+} )\,{\psi}_{2i-1}^{-} \right)  \left( \sqrt{2} {\psi}^{-}_{N} \right) \mathcal{K}.
\end{align}
The antiunitary operators satisfy 
\begin{align}
    \mathcal{P}^2 
    &= \left( -1 \right)^{\left( N+1 \right)/2}, \\
    \mathcal{Q}^2 
    &= +1, \\
    \mathcal{R}^2 
    &= \left( -1 \right)^{\left( N-1 \right)/2}, \\
    \mathcal{S}^2 &= +1.
\end{align}
Furthermore, while the antiunitary operators $\mathcal{P}$ and $\mathcal{R}$ anticommute with fermion parity symmetry $\left( -1 \right)^{\cal F}$ in the double Hilbert space, the antiunitary operators $\mathcal{Q}$ and $\mathcal{S}$ commute with $\left( -1 \right)^{\cal F}$:
\begin{align}
    \left( -1 \right)^{\cal F} \mathcal{A} \left( -1 \right)^{\cal F} = \begin{cases}
    - \mathcal{A} & \left( \mathcal{A} = \mathcal{P}, \mathcal{R} \right); \\
    + \mathcal{A} & \left( \mathcal{A} = \mathcal{Q}, \mathcal{S} \right).
    \end{cases}
        \label{eq: total fermion parity & A - odd N}
\end{align}
This means that the antiunitary operations $\mathcal{P}$ and $\mathcal{R}$ switch fermion parity $\left( -1 \right)^{\cal F}$ and are no longer symmetry in the subspace with fixed fermion parity.
As a result, for odd $N$ and arbitrary $p$, in the subspace with fixed total fermion parity $\left( -1 \right)^{\cal F}$, only one of the antiunitary symmetry---$\mathcal{Q}$ or $\mathcal{S}$---is relevant (Table~\ref{tab: PQRS}).

\subsection{Linear dissipator (\texorpdfstring{$p=1$}{p=1})}
    \label{subsec: SYK Lindbladian p1 classification}

For the linear dissipators $p=1$, the dissipation term $\mathcal{D}$ in Eq.~(\ref{eq: dissipator odd p}) reads
\begin{align}
    \mathcal{D} = \sum_{m, i, j} K_{m; i} K_{m; j}^{*} \left( -\ii \psi_{i}^{+} \psi_{j}^{-} - \frac{1}{2} \psi_{j}^{+} \psi_{i}^{+} - \frac{1}{2} \psi_{i}^{-} \psi_{j}^{-} \right),
\end{align}
and the shifted dissipation term $\tilde{\cal D}$ reads
\begin{align}
    \tilde{\cal D}
    &\coloneqq \mathcal{D} - (\mathrm{tr}\,\mathcal{L}/\mathrm{tr}\,I)\,I \nonumber \\
    &= \sum_{m, i, j} K_{m; i} K_{m; j}^{*} \left( -\ii \psi_{i}^{+} \psi_{j}^{-} \right) \nonumber \\
    &\qquad- \frac{1}{2} \sum_{m; i \neq j} K_{m; i} K_{m; j}^{*} \left( \psi_{j}^{+} \psi^{+}_{i} + \psi_{i}^{-} \psi_{j}^{-} \right).
\end{align}
While the linear dissipators were chosen to be $K_{m; i} = \sqrt{\mu} \delta_{m, i}$ ($\mu \geq 0$) in Ref.~\cite{Kulkarni-22}, we here consider the symmetry classification for generic dissipative coupling $K_{m; i} \in \mathbb{C}$.
As discussed below, the reality of $K_{m; i}$, or more generally the condition $K_{m; i} K_{m; j}^{*} \in \mathbb{R}$, is relevant to the symmetry classification.

In Sec.~\ref{subsec: PQRS}, we confirm that the antiunitary operations $\mathcal{P}$ and $\mathcal{Q}$ still give rise to symmetry of the Hamiltonian part in the double Hilbert space [i.e., Eqs.~(\ref{eq: AH1}) and (\ref{eq: AH2})].
Let us study the operations of $\mathcal{P}$ and $\mathcal{Q}$ on the dissipation term $\mathcal{D}$. 
First, we have
\begin{align}
    &{\cal P}\,\tilde{\cal D}\,{\cal P}^{-1}
    = \sum_{m, i, j} K_{m; i}^{*} K_{m; j} \left( +\ii \psi_{i}^{+} \psi_{j}^{-} \right) \nonumber \\
    &\qquad\qquad- \frac{1}{2} \sum_{m; i \neq j} K_{m; i}^{*} K_{m; j} \left( \psi_{j}^{+} \psi_{i}^{+} + \psi_{i}^{-} \psi_{j}^{-} \right) \nonumber \\
    &\qquad\quad
    \begin{cases}
    = - \tilde{\cal D} & \left( K_{m; i} K_{m; j}^{*} \in \mathbb{R} \right); \\
    \neq - \tilde{\cal D} & \left( \text{otherwise} \right).
    \end{cases}
\end{align}
Furthermore, we have
\begin{align}
    &\mathcal{P}\,\tilde{\cal D}^{\dag}\,\mathcal{P}^{-1}
    = \sum_{m, i, j} K_{m; i} K_{m; j}^{*} \left( +\ii \psi_{i}^{+} \psi_{j}^{-} \right) \nonumber \\
    &\qquad\qquad\quad - \frac{1}{2} \sum_{m; i \neq j} K_{m; i}^{*} K_{m; j} \left( \psi_{j}^{+} \psi_{i}^{+} + \psi_{i}^{-} \psi_{j}^{-} \right) \nonumber \\
    &\qquad\qquad= - \tilde{\cal D}
\end{align}
for arbitrary $K_{m; i} \in \mathbb{C}$.
Combining these equations with Eqs.~(\ref{eq: AH1}) and (\ref{eq: AH2}), we have
\begin{align}
    {\cal P}\,\tilde{\cal L}\,{\cal P}^{-1} 
    &= - \tilde{\cal L}
\end{align}
for $q \equiv 0$ (mod $4$) and $K_{m; i} K_{m; j}^{*} \in \mathbb{R}$, and
\begin{align}
    {\cal P}\,\tilde{\cal L}^{\dag}\,{\cal P}^{-1} 
    &= - \tilde{\cal L}
\end{align}
for $q \equiv 2$ (mod $4$).
Here, $\tilde{\cal L} \coloneqq {\cal L} - (\mathrm{tr}\,\mathcal{L}/\mathrm{tr}\,I)\,I$ is the shifted Lindbladian (see also Sec.~\ref{subsec: shift}).

Similarly, we have
\begin{align}
    &{\cal Q}\,{\cal D}\,{\cal Q}^{-1} \nonumber \\
    &\quad= \sum_{m, i, j} K_{m; i}^{*} K_{m; j} \left( -\ii {\psi}_{i}^{+} {\psi}_{j}^{-} \right) \nonumber \\
    &\qquad\qquad- \frac{1}{2} \sum_{m; i, j} K_{m; i}^{*} K_{m; j} \left( {\psi}_{j}^{+} {\psi}_{i}^{+} + {\psi}_{i}^{-} {\psi}_{j}^{-} \right) \nonumber \\
    &\quad\begin{cases}
    = {\cal D} & \left( K_{m; i} K_{m; j}^{*} \in \mathbb{R} \right); \\
    \neq {\cal D} & \left( \text{otherwise} \right).
    \end{cases}
\end{align}
Furthermore, we have
\begin{align}
    &{\cal Q}\,{\cal D}^{\dag}\,{\cal Q}^{-1} \nonumber \\
    &\quad= \sum_{m, i, j} K_{m; i} K_{m; j}^{*} \left( -\ii {\psi}_{i}^{+} {\psi}_{j}^{-} \right) \nonumber \\
    &\qquad\qquad- \frac{1}{2} \sum_{m; i, j} K_{m; i}^{*} K_{m; j} \left( {\psi}_{j}^{+} {\psi}_{i}^{+} + {\psi}_{i}^{-} {\psi}_{j}^{-} \right) \nonumber \\
    &\quad\begin{cases}
    = {\cal D} & \left( K_{m; i} K_{m; j}^{*} \in \mathbb{R} \right); \\
    \neq {\cal D} & \left( \text{otherwise} \right).
    \end{cases}
\end{align}
Consequently, we have
\begin{align}
    {\cal Q}\,{\cal L}\,{\cal Q}^{-1} 
    &= + {\cal L},
        \label{eq: p1 Q}
\end{align}
for $q \equiv 2$ (mod $4$) and $K_{m; i} K_{m; j}^{*} \in \mathbb{R}$, and 
\begin{align}
    {\cal Q}\,{\cal L}^{\dag}\,{\cal Q}^{-1} 
    &= + {\cal L}
\end{align}
for $q \equiv 0$ (mod $4$) and $K_{m; i} K_{m; j}^{*} \in \mathbb{R}$.

Considering the two types of antiunitary symmetry $\mathcal{P}$ and $\mathcal{Q}$, as well as modular conjugation symmetry $\mathcal{J}$, we have the periodic table~\ref{tab: SYK Lindbladian p=1}.
Here, for each $N$ (mod $4$), we need to consider relevant antiunitary symmetry in the subspace with fixed fermion parity, as summarized in Table~\ref{tab: PQRS}.
In the certain cases, the antiunitary symmetry $\mathcal{P}$ and $\mathcal{Q}$ can be respected only for $K_{m; i} K_{m; j}^{*} \in \mathbb{R}$, which changes the symmetry classification.
Previously, the quadratic Lindbladians were argued to fit into one of the tenfold Altland-Zirnbauer$^{\dag}$ symmetry class~\cite{Lieu-20}.
While the SYK Lindbladians with $p=1$ contain the linear dissipators similarly to the quadratic Lindbladians in Ref.~\cite{Lieu-20}, they also contain the $q$-body Hamiltonian and hence cannot be reduced to the noninteracting non-Hermitian operators~\cite{Prosen-08}.
As a consequence of the many-body nature, the SYK Lindbladians exhibit the unique symmetry classes that were not predicted in the symmetry classification of the quadratic Lindbladians.
Such examples include classes BDI$^{\dag} + \mathcal{S}_{++}$ and CI$^{\dag} + \mathcal{S}_{+-}$ for $q \equiv 0$ (mod $4$), as well as classes BDI and CI for $q \equiv 2$ (mod $4$), all of which do not fit into the tenfold Altland-Zirnbauer$^{\dag}$ symmetry class (see Appendix~\ref{asec: 38-fold symmetry} for the definitions of the symmetry classes).

Another notable feature of our symmetry classification is the presence of symplectic antiunitary symmetry.
In fact, time-reversal symmetry (or equivalently particle-hole symmetry$^{\dag}$) with the sign $\mathcal{P}^2 = -1$ is respected for $q \equiv 0$ (mod $4$) and $K_{m; i} K_{m; j}^{*} \in \mathbb{R}$.
The different sign of antiunitary symmetry leads to the different spectral correlations (see Appendix~\ref{asec: NH RMT TRS} for details).
We show such symmetry-enriched dissipative quantum chaos in Sec.~\ref{sec: dissipative quantum chaos - p=1, K in R}.
By contrast, while time-reversal symmetry$^{\dag}$ with the sign $+1$ appears, time-reversal symmetry$^{\dag}$ with the sign $-1$ does not appear in our classification.
Time-reversal symmetry$^{\dag}$ with the sign $-1$ inevitably leads to the Kramers degeneracy for all complex eigenvalues~\cite{KSUS-19}, which may be incompatible with the double Hilbert space (see also Ref.~\cite{Lieu-22}).
On the other hand,
time-reversal symmetry with the sign $-1$ results in the Kramers degeneracy only on the symmetric line~\cite{Kawabata-19}. 
Nevertheless, it is also possible that time-reversal symmetry$^{\dag}$ with the sign $-1$ may appear in other models of Lindbladians.
It merits further study to investigate whether or not time-reversal symmetry$^{\dag}$ with the sign $-1$ can appear in open quantum systems.

We note in passing that the combination of the antiunitary symmetry $\mathcal{Q}$ in Eq.~(\ref{eq: p1 Q}) and modular conjugation symmetry $\mathcal{J}$ in Eq.~(\ref{eq: modular conjugation symmetry}) gives rise to additional unitary symmetry that commutes with the Lindbladian $\mathcal{L}$: 
\begin{equation}
    (\mathcal{J} \mathcal{Q})\,\mathcal{L}\,(\mathcal{J} \mathcal{Q})^{-1} = \mathcal{L}.
\end{equation}
Thus, we have to study the symmetry within the subspace that diagonalizes the unitary symmetry $\mathcal{J} \mathcal{Q}$.
Since $\mathcal{J}$ and $\mathcal{Q}$ have the same sign $\mathcal{J}^2 = \mathcal{Q}^2 = +1$, they can commute with each other.
In this subspace, the antiunitary symmetry $\mathcal{P}$ also remains symmetry for $\left( -1 \right)^{\cal F} = +1$.
It is also notable that since the unitary operator $\mathcal{JQ}$ exchanges the ket and bra degrees of freedom, the above unitary symmetry can be considered as KMS symmetry.

\subsection{Quadratic dissipator (\texorpdfstring{$p=2$}{p=2})}
    \label{subsec: SYK Lindbladian p2 classification}

For the quadratic dissipators $p=2$, the dissipation term $\mathcal{D}$ in Eq.~(\ref{eq: dissipator odd p}) reads
\begin{align}
    {\cal D} &= \sum_{m, i, j} K_{m; i} K_{m; j}^{*} \Bigl( {\psi}_{i_1}^{+} {\psi}_{i_2}^{+} {\psi}_{j_1}^{-} {\psi}_{j_2}^{-} \nonumber \\
    &\qquad\qquad- \frac{1}{2} {\psi}_{j_2}^{+} {\psi}_{j_1}^{+} {\psi}_{i_1}^{+} {\psi}_{i_2}^{+} - \frac{1}{2} {\psi}_{i_2}^{-} {\psi}_{i_1}^{-} {\psi}_{j_1}^{-} {\psi}_{j_2}^{-} \Bigr),
\end{align} 
and the shifted dissipation term $\tilde{\cal D}$ reads
\begin{align}
    &\tilde{\cal D}
    \coloneqq {\cal D} - \mathrm{tr}\,{\cal L}/\mathrm{tr}\,I \nonumber \\ 
    &= \sum_{m, i, j} K_{m; i} K_{m; j}^{*} \left( {\psi}_{i_1}^{+} {\psi}_{i_2}^{+} {\psi}_{j_1}^{-} {\psi}_{j_2}^{-} \right) \nonumber \\
    &- \frac{1}{2} \sum_{m; i_1 \neq j_1, i_2 \neq j_2} K_{m; i} K_{m; j}^{*} \left( {\psi}_{j_2}^{+} {\psi}_{j_1}^{+} {\psi}_{i_1}^{+} {\psi}_{i_2}^{+} + {\psi}_{i_2}^{-} {\psi}_{i_1}^{-} {\psi}_{j_1}^{-} {\psi}_{j_2}^{-} \right).
\end{align}
We study the action of the antiunitary operations ${\cal P}$ and ${\cal Q}$ on the dissipation term $\mathcal{D}$.
First, we have
\begin{align}
    &{\cal P}\,{\cal D}\,{\cal P}^{-1} \nonumber \\
    &\quad= \sum_{m, i, j} K_{m; i}^{*} K_{m; j} \left( {\psi}_{i_1}^{+} {\psi}_{i_2}^{+} {\psi}_{j_1}^{-} {\psi}_{j_2}^{-} \right) \nonumber \\
    &\qquad- \frac{1}{2} \sum_{m; i, j} K_{m; i}^{*} K_{m; j} \left( {\psi}_{j_2}^{+} {\psi}_{j_1}^{+} {\psi}_{i_1}^{+} {\psi}_{i_2}^{+} + {\psi}_{i_2}^{-} {\psi}_{i_1}^{-} {\psi}_{j_1}^{-} {\psi}_{j_2}^{-} \right) \nonumber \\
    &\quad\begin{cases}
    = + {\cal D} & \left( K_{m; i} K_{m; j}^{*} \in \mathbb{R} \right); \\
    \neq + {\cal D}\ & \left( \text{otherwise} \right).
    \end{cases}
\end{align}
We also have
\begin{align}
    &{\cal P}\,{\cal D}^{\dag}\,{\cal P}^{-1} \nonumber \\
    &\quad = \sum_{m, i, j} K_{m; i} K_{m; j}^{*} \left( {\psi}_{i_1}^{+} {\psi}_{i_2}^{+} {\psi}_{j_1}^{-} {\psi}_{j_2}^{-} \right) \nonumber \\  
    &\qquad- \frac{1}{2} \sum_{m; i, j} K_{m; i}^{*} K_{m; j} \left( {\psi}_{j_2}^{+} {\psi}_{j_1}^{+} {\psi}_{i_1}^{+} {\psi}_{i_2}^{+} + {\psi}_{i_2}^{-} {\psi}_{i_1}^{-} {\psi}_{j_1}^{-} {\psi}_{j_2}^{-} \right) \nonumber \\
    &\quad\begin{cases}
    = + {\cal D} & \left( K_{m; i} K_{m; j}^{*} \in \mathbb{R} \right); \\
    \neq + {\cal D}\ & \left( \text{otherwise} \right).
    \end{cases}
\end{align}
Combining these equations with Eqs.~(\ref{eq: AH1}) and (\ref{eq: AH2}), we have
\begin{align}
    {\cal P}\,{\cal L}\,{\cal P}^{-1} 
    &= + {\cal L}
\end{align}
for $q \equiv 2$ (mod $4$) and $K_{m; i} K_{m; j}^{*} \in \mathbb{R}$, and
\begin{align}
    {\cal P}\,{\cal L}^{\dag}\,{\cal P}^{-1} 
    &= + {\cal L}
        \label{eq: p=2 q=0 P}
\end{align}
for $q \equiv 0$ (mod $4$) and $K_{m; i} K_{m; j}^{*} \in \mathbb{R}$.

Similarly, for ${\cal Q}$, we have
\begin{align}
    &{\cal Q}\,{\cal D}\,{\cal Q}^{-1} \nonumber \\
    &\quad= \sum_{m, i, j} K_{m; i}^{*} K_{m; j} \left( {\psi}_{i_1}^{+} {\psi}_{i_2}^{+} {\psi}_{j_1}^{-} {\psi}_{j_2}^{-} \right) \nonumber \\
    &\qquad- \frac{1}{2} \sum_{m; i, j} K_{m; i}^{*} K_{m; j} \left( {\psi}_{j_2}^{+} {\psi}_{j_1}^{+} {\psi}_{i_1}^{+} {\psi}_{i_2}^{+} + {\psi}_{i_2}^{-} {\psi}_{i_1}^{-} {\psi}_{j_1}^{-} {\psi}_{j_2}^{-} \right) \nonumber \\
    &\quad\begin{cases}
    = + {\cal D} & \left( K_{m; i} K_{m; j}^{*} \in \mathbb{R} \right); \\
    \neq + {\cal D}\ & \left( \text{otherwise} \right).
    \end{cases}
\end{align}
Furthermore, we have
\begin{align}
    &{\cal Q}\,{\cal D}^{\dag}\,{\cal Q}^{-1} \nonumber \\
    &\quad= \sum_{m, i, j} K_{m; i} K_{m; j}^{*} \left( {\psi}_{i_1}^{+} {\psi}_{i_2}^{+} {\psi}_{j_1}^{-} {\psi}_{j_2}^{-} \right) \nonumber \\
    &\qquad- \frac{1}{2} \sum_{m; i, j} K_{m; i}^{*} K_{m; j} \left( {\psi}_{j_2}^{+} {\psi}_{j_1}^{+} {\psi}_{i_1}^{+} {\psi}_{i_2}^{+} + {\psi}_{i_2}^{-} {\psi}_{i_1}^{-} {\psi}_{j_1}^{-} {\psi}_{j_2}^{-} \right) \nonumber \\
    &\quad\begin{cases}
    = + {\cal D} & \left( K_{m; i} K_{m; j}^{*} \in \mathbb{R} \right); \\
    \neq + {\cal D}\ & \left( \text{otherwise} \right).
    \end{cases}
\end{align}
Consequently, we have
\begin{align}
    {\cal Q}\,{\cal L}\,{\cal Q}^{-1} 
    &= + {\cal L}
\end{align}
for $q \equiv 2$ (mod $4$) and $K_{m; i} K_{m; j}^{*} \in \mathbb{R}$, and
\begin{align}
    {\cal Q}\,{\cal L}^{\dag}\,{\cal Q}^{-1} 
    &= + {\cal L}
\end{align}
for $q \equiv 0$ (mod $4$) and $K_{m; i} K_{m; j}^{*} \in \mathbb{R}$.

The symmetry classification for the quadratic dissipators $p=2$ is summarized as the periodic table~\ref{tab: SYK Lindbladian p=2}.
Similarly to the linear dissipators $p=1$, the condition $K_{m; i} K_{m; j}^{*} \in \mathbb{R}$, including the real coupling $K_{m; i} \in  \mathbb{R}$ and the pure-imaginary coupling $K_{m; i} \in  \ii \mathbb{R}$, is relevant to the symmetry classification.
It is also notable that no symmetry appears (i.e., class A) for $N \equiv 2$ (mod $4$) and $K_{m; i} K_{m; j}^{*} \notin \mathbb{R}$ although modular conjugation symmetry $\mathcal{J}$ is respected in arbitrary Lindbladians.
This is because $\mathcal{J}$ anticommutes with fermion parity $\left( -1 \right)^{F^{\pm}}$ for $\left( -1\right)^{N/2} \left( -1 \right)^{\cal F} = -1$
and is no longer symmetry in the subspace of fixed $\left( -1 \right)^{F^{\pm}}$ [see Eq.~(\ref{eq: modular conjugation & sub-fermion parity})].
However, the combination of modular conjugation $\mathcal{J}$ and the antiunitary operation $\mathcal{P}$, $\mathcal{Q}$, $\mathcal{R}$, $\mathcal{S}$ can give rise to symmetry even in the subspace of fixed $\left( -1 \right)^{F^{\pm}}$.
In fact, combining Eqs.~(\ref{eq: modular conjugation symmetry}) and (\ref{eq: p=2 q=0 P}), we have
\begin{align}
    (\mathcal{J}\mathcal{P})\,\mathcal{L}^{\dag}\,(\mathcal{J}\mathcal{P})^{-1} = \mathcal{L},
        \label{eq: p2 pH}
\end{align}
where $\mathcal{J}\mathcal{P}$ is a unitary operator.
This symmetry is called pseudo-Hermiticity~\cite{Mostafazadeh-02-1, *Mostafazadeh-02-2, *Mostafazadeh-02-3}, 
and the Lindbladian $\mathcal{L}$ belongs to class A + $\eta$
in the 38-fold symmetry classification of non-Hermitian operators (or equivalently, $\ii \mathcal{L}$ respects chiral symmetry and belongs to class AIII;
see Appendix~\ref{asec: 38-fold symmetry} for details)~\cite{KSUS-19}.
While $\mathcal{J}$ and $\mathcal{P}$ individually anticommute with fermion parity $\left( -1 \right)^{F^{\pm}}$ [see Eqs.~(\ref{eq: modular conjugation & sub-fermion parity}) and (\ref{eq: fermion parity braket PQRS})], its combination $\mathcal{J}\mathcal{P}$ satisfies
\begin{align}
    \left( -1 \right)^{F^{\pm}}\,(\mathcal{J}\mathcal{P})\,\left( -1 \right)^{F^{\pm}} = \left( -1 \right)^{\cal F} (\mathcal{J}\mathcal{P})
\end{align}
and hence commutes with $\left( -1 \right)^{F^{\pm}}$ for $\left( -1 \right)^{\cal F} = +1$.
Consequently, for $\left( -1 \right)^{\cal F} = +1$ and $K_{m; i} K_{m; j}^{*} \in \mathbb{R}$, pseudo-Hermiticity in Eq.~(\ref{eq: p2 pH}) is respected even for $p=2$ and $N \equiv 2$ (mod $4$).
In Sec.~\ref{sec: dissipative quantum chaos}, we numerically confirm the presence of this symmetry and its consequences on the complex-spectral statistics.

\subsection{Generic dissipator (\texorpdfstring{$p \geq 3$}{p>=3})}
    \label{subsec: generic dissipator}

In the preceding sections, we have considered the symmetry classification of the SYK Lindbladians with the linear and quadratic dissipators $p = 1, 2$.
Here, we study the symmetry classification for generic $p \geq 3$.
For even $p$, the symmetry classification with $p = 2$ is still valid.
Thus, the periodic table~\ref{tab: SYK Lindbladian p=2} is applicable to the generic dissipators as long as $p$ is even.

On the other hand, for generic dissipators with odd $p \geq 3$, the antiunitary symmetry $\mathcal{P}$ is violated while the antiunitary symmetry $\mathcal{Q}$ is respected.
In this sense, the linear dissipator $p=1$ is special, for which both antiunitary operations $\mathcal{P}$ and $\mathcal{Q}$ give rise to symmetry.
For $p = 3$, for example, the dissipation term reads ${\cal D} = {\cal D}^{+-} + {\cal D}^{++} + {\cal D}^{--}$ with
\begin{align}
    {\cal D}^{+-} &\coloneqq \sum_{m, i, j} K_{m; i} K_{m; j}^{*} \left( -\ii {\psi}_{i_1}^{+} {\psi}_{i_2}^{+} {\psi}_{i_3}^{+} {\psi}_{j_1}^{-} {\psi}_{j_2}^{-} {\psi}_{j_3}^{-} \right), \\
    {\cal D}^{++} &\coloneqq -\frac{1}{2} \sum_{m, i, j} K_{m; i} K_{m; j}^{*} \left( {\psi}_{j_3}^{+} {\psi}_{j_2}^{+} {\psi}_{j_1}^{+} {\psi}_{i_1}^{+} {\psi}_{i_2}^{+} {\psi}_{i_3}^{+} \right), \\
    {\cal D}^{--} &\coloneqq -\frac{1}{2} \sum_{m, i, j} K_{m; i} K_{m; j}^{*} \left( {\psi}_{i_3}^{-} {\psi}_{i_2}^{-} {\psi}_{i_1}^{-} {\psi}_{j_1}^{-} {\psi}_{j_2}^{-} {\psi}_{j_3}^{-} \right).
\end{align}
Similarly to $p=1$, we have
\begin{align}
    &{\cal P} {\cal D}^{+-} {\cal P}^{-1} \begin{cases}
    = - {\cal D}^{+-} & \left( K_{m; i} K_{m; j}^{*} \in \mathbb{R} \right); \\
    \neq - {\cal D}^{+-} & \left( K_{m; i} K_{m; j}^{*} \notin \mathbb{R} \right), \\
    \end{cases} \\
    &{\cal P}\,({\cal D}^{+-})^{\dag}\,{\cal P}^{-1} = - {\cal D}^{+-}, \\
    &{\cal Q} {\cal D}^{+-} {\cal Q}^{-1} \begin{cases} 
    = + {\cal D}^{+-} & \left( K_{m; i} K_{m; j}^{*} \in \mathbb{R} \right); \\
    \neq + {\cal D}^{+-} & \left( K_{m; i} K_{m; j}^{*} \notin \mathbb{R} \right), \\
    \end{cases} \\
    &{\cal Q}\,({\cal D}^{+-})^{\dag}\,{\cal Q}^{-1} = {\cal D}^{+-}.
\end{align}
For the other terms ${\cal D}^{++}$ and ${\cal D}^{--}$, we have
\begin{align}
    ({\cal D}^{\pm\pm})^{\dag} &= {\cal D}^{\pm\pm},\\
    {\cal P} {\cal D}^{\pm\pm} {\cal P}^{-1} &= {\cal D}^{\pm\pm} \quad \left( K_{m; i} K_{m; j}^{*} \in \mathbb{R} \right), \\
    {\cal Q} {\cal D}^{\pm\pm} {\cal Q}^{-1} &= {\cal D}^{\pm\pm} \quad \left( K_{m; i} K_{m; j}^{*} \in \mathbb{R} \right).
\end{align}
Thus, the antiunitary symmetry ${\cal Q}$ is respected even for $p = 3$ (and arbitrary odd $p$).
By contrast, the antiunitary symmetry ${\cal P}$ is respected only for $p=1$.
In fact, we have
\begin{align}
    &{\cal P}\,({\cal D}^{++} -\mathrm{tr}\,{\cal D}^{++}/\mathrm{tr}\,I)\,{\cal P}^{-1} \nonumber \\
    &\qquad= - ({\cal D}^{++} -\mathrm{tr}\,{\cal D}^{++}/\mathrm{tr}\,I) \nonumber \\ 
    &\qquad\qquad-\sum_{m, i, j} K_{m; i} K_{m; j}^{*} \delta_{i_1, j_1} \left( {\psi}_{j_3}^{+} {\psi}_{j_2}^{+} {\psi}_{i_2}^{+} {\psi}_{i_3}^{+} \right),
\end{align}
where the sum is taken for $i, j$ that satisfies $i_1 < i_2 < i_3$, $j_1 < j_2 < j_3$ but does not satisfy $i_1 = j_1$, $i_2 = j_2$, $i_3 = j_3$.
For $p = 3$, the remainder term $- \sum_{m, i, j} K_{m; i} K_{m; j}^{*} \delta_{i_1, j_1} \left( {\psi}_{j_3}^{+} {\psi}_{j_2}^{+} {\psi}_{i_2}^{+} {\psi}_{i_3}^{+} \right)$ is in general nonzero.
For and only for $p = 1$, on the other hand, this term vanishes.
Consequently, we have the following symmetry classification for odd $p \geq 3$:
for $q \equiv 0$ (mod $4$) with $K_{m; i} K_{m; j}^{*} \notin \mathbb{R}$ and $q \equiv 2$ (mod $4$), the symmetry class is class AI (or equivalently class D$^{\dag}$) for arbitrary $N$; 
for $q \equiv 0$ (mod $4$) with $K_{m; i} K_{m; j}^{*} \in \mathbb{R}$, the symmetry class is class BDI$^{\dag}$ for arbitrary $N$.

\section{Dissipative quantum chaos of the Sachdev-Ye-Kitaev Lindbladians}
    \label{sec: dissipative quantum chaos}

In this section,
we numerically study the complex-spectral statistics of the SYK Lindbladians and demonstrate that they obey the complex-spectral statistics of non-Hermitian random matrices, 
signaling dissipative quantum chaos. 
The SYK Lindbladians belong to different symmetry classes 
depending on $p$, $q$, $N$, and $K_{m; i}$, 
as summarized in Tables~\ref{tab: SYK Lindbladian p=1} and \ref{tab: SYK Lindbladian p=2}.
We show such symmetry-enriched dissipative quantum chaos for various choices of the parameters.
Throughout this section, we set $q=4$. 

\subsection{Linear dissipator (\texorpdfstring{$p=1$}{p=1}) with complex dissipative coupling}
    \label{sec: dissipative quantum chaos - p=1, K in C}

%

\begin{figure}[t]
\centering




\includegraphics[width=86mm]{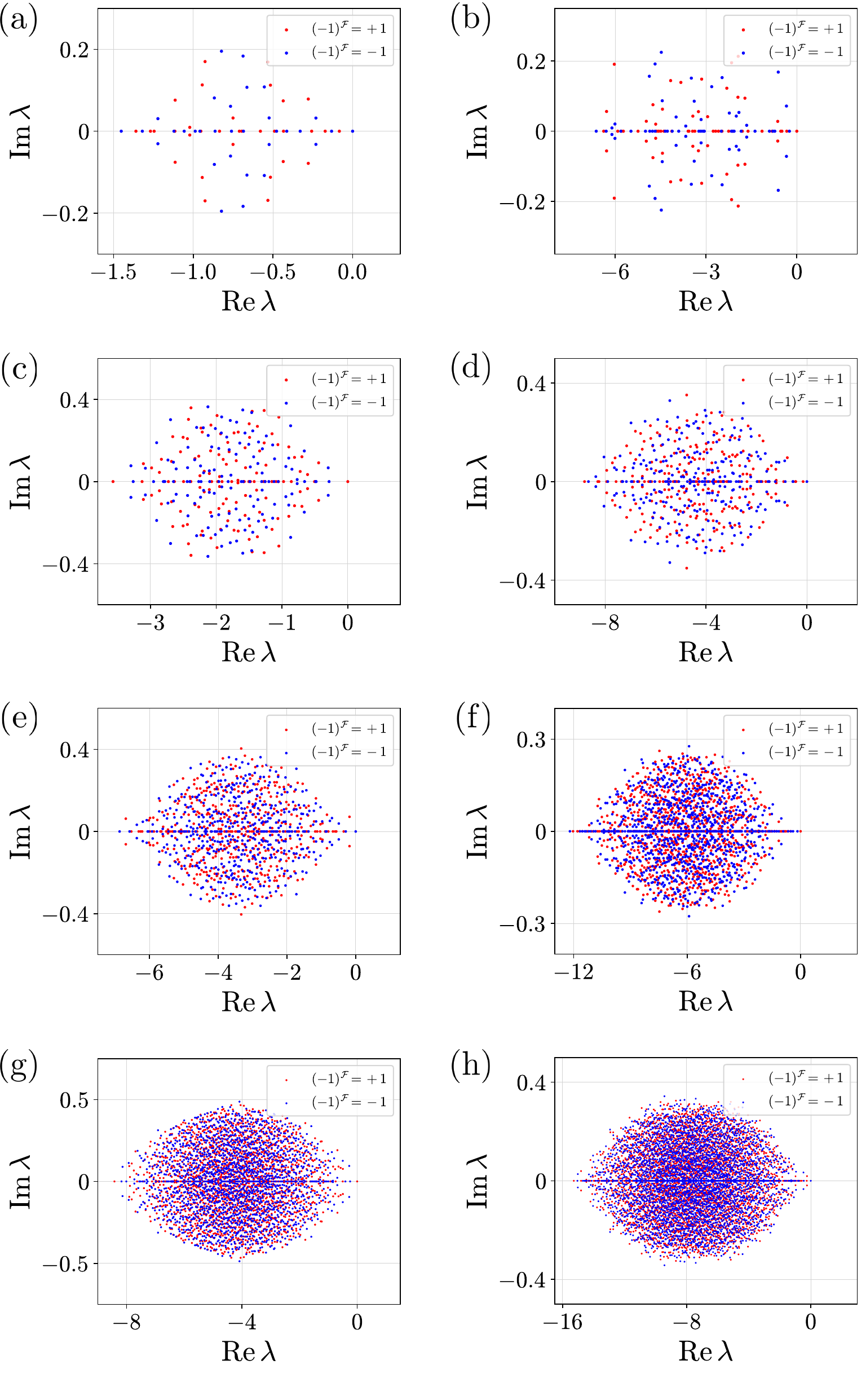} 

\caption{Complex spectrum of a single realization of the Sachdev-Ye-Kitaev (SYK) Lindbladian with linear dissipators $p=1$ and complex coupling $K_{m;i} \in \mathbb{C}$.  
The number $N$ of fermion flavors and the strength $K$ of the dissipators are respectively chosen to be (a)~$N=6$, $K=0.52$, (b)~$N=7$, $K=1$, (c)~$N=8$, $K=0.68$, (d)~$N=9$, $K=1$, (e)~$N=10$, $K=0.78$, (f)~$N=11$, $K=1$, (g)~$N=12$, $K=0.87$, and (h)~$N=13$, $K=1$.
The other parameters are taken as $J=1$ and $M=N$. 
The different colors represent eigenstates with even ($+1$; red dots) or odd ($-1$; blue dots) total fermion parity $\left( -1 \right)^{\cal F}$.
}
	\label{fig: spectrum_p_1_complex_K_2}
\end{figure}

\begin{figure}
    \centering

\includegraphics[width=86mm]{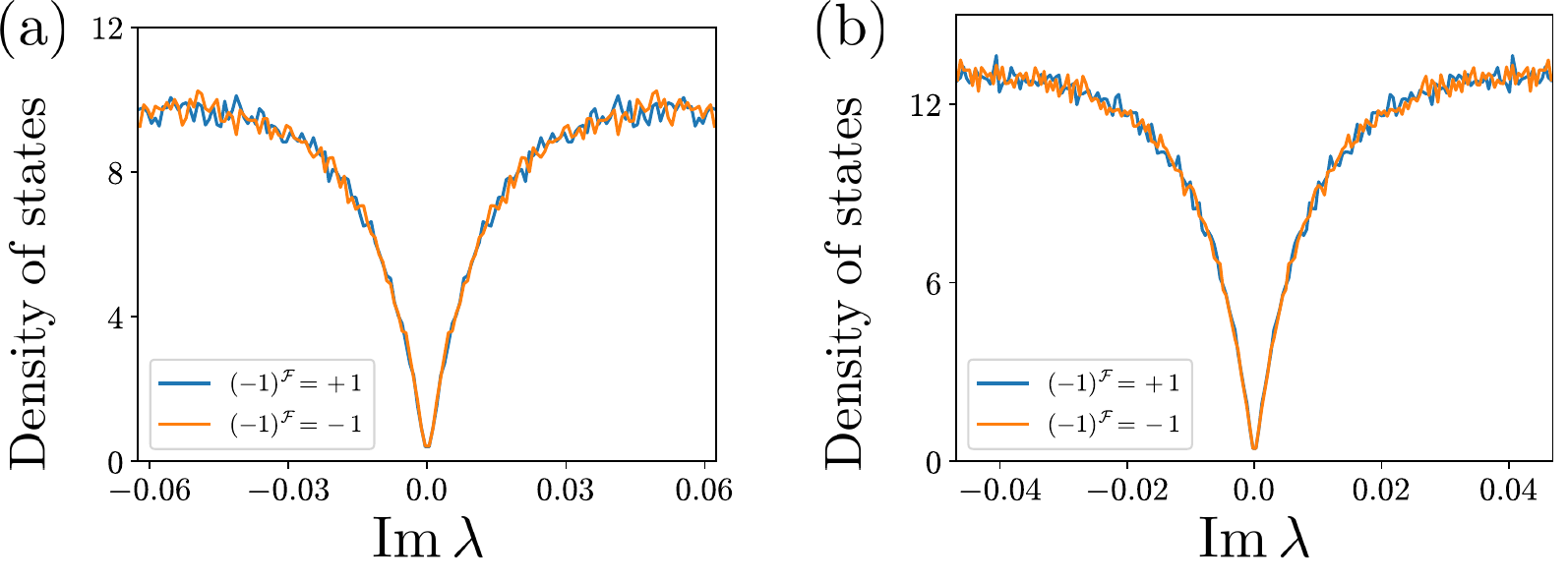} 

    \caption{Density of states across the real axis as a function of the imaginary part of complex eigenvalues $\lambda$ for the Sachdev-Ye-Kitaev (SYK) Lindbladians with linear dissipators $p=1$ and complex coupling $K_{m; i} \in \mathbb{C}$. 
    The number $N$ of fermion flavors and the strength $K$ of the dissipators are respectively chosen to be (a)~$N=10$, $K=0.78$ and (b)~$N=12$, $K=0.87$.
    The other parameters are taken as $J=1$ and $M=N$.
    The double Hilbert space is divided into the two subspaces according to total fermion parity $\left( -1 \right)^{\cal F}$, as shown by the blue curves ($+1$) and orange curves ($-1$).
    We take eigenvalues satisfying $\left| \mathrm{Im}\,\lambda \right| > \epsilon/\sqrt{\mathrm{dim}\,\mathcal{L}}$ ($\epsilon = 10^{-4}$) and exclude real eigenvalues.
    }
    \label{fig:symmLineDecay-p_1-complex_K-fits-linear}
\end{figure}

We begin with the SYK Lindbladians 
with
the linear random dissipators $p=1$ and the complex dissipative coupling $K_{m; i} \in \mathbb{C}$.
Figure~\ref{fig: spectrum_p_1_complex_K_2} shows the complex spectra for $6 \leq N \leq 13$.
In our calculations, 
we choose
the strength $K$ of the dissipators such that the Hamiltonian parts are well balanced with the dissipators to observe the typical behavior of the SYK Lindbladians.
In particular, we choose $K$ by 
demanding
$\mathrm{tr}\,( \mathcal{D}^\dag \mathcal{D} ) = \mathrm{tr}\,( \mathcal{L}_0^\dag \mathcal{L}_0 )$, where $\mathcal{L}_0$ is the SYK Lindbladian without dissipation.
Owing to this choice, the complex spectra of the SYK Lindbladians are similar to the lemon-shaped spectrum of the completely random Lindbladians~\cite{Denisov-19}.
For odd $p$ including $p=1$, 
total fermion parity $\left( -1 \right)^{\cal F}$ in the double Hilbert space is conserved but fermion parity $\left( -1 \right)^{F^{\pm}}$ in each ket or bra space is not conserved.
Consequently, the double Hilbert space is divided into the two subspaces 
with even and odd fermion parity,
$\left( -1 \right)^{\cal F} = \pm 1$.
In the numerical results in Fig.~\ref{fig: spectrum_p_1_complex_K_2} and also the subsequent figures, we plot the complex spectra for both $\left( -1 \right)^{\cal F} = +1$ and $\left( -1 \right)^{\cal F} = -1$ for the sake of completeness, although the physically relevant subspace 
has even total fermion parity $\left( -1 \right)^{\cal F} = +1$.

Because of the complex coupling, only modular conjugation symmetry is respected (i.e., class AI), as shown in Table~\ref{tab: SYK Lindbladian p=1}.
Consistently, the complex spectra of the SYK Lindbladians are symmetric about the real axis, and a subextensive number of eigenvalues accumulate on the real axis.
We also obtain the density of states across the real axis as a function of the imaginary part of complex eigenvalues (Fig.~\ref{fig:symmLineDecay-p_1-complex_K-fits-linear}).
The density of states linearly vanishes toward the real axis, which shows the level repulsion around the real axis.
The linear decay of the density of states around the real axis is consistent with non-Hermitian random matrices with time-reversal symmetry (see Appendix~\ref{asec: NH RMT TRS} for details)~\cite{Ginibre-65, Xiao-22}.

\begin{figure}[t]
\centering

\includegraphics[width=86mm]{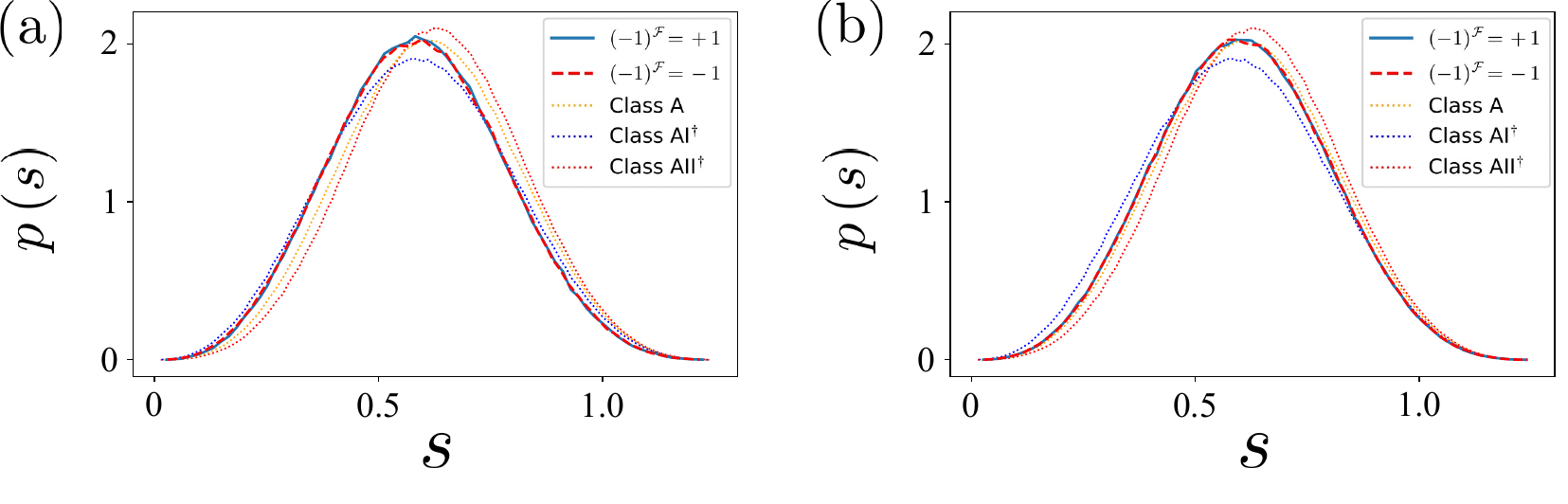} 

\caption{Distributions for normalized nearest spectral spacings $s$ of the Sachdev-Ye-Kitaev (SYK) Lindbladians with linear dissipators $p=1$ and complex coupling $K_{m;i} \in \mathbb{C}$, compared with the random-matrix results. 
The number $N$ of fermion flavors and the strength $K$ of the dissipators are respectively chosen to be (a)~$N=10$, $K=0.78$ and (b)~$N=12$, $K=0.87$.
The other parameters are taken as $J=1$ and $M=N$. 
Each datum is averaged over $2000$ samples. 
}
	\label{fig: spectral_spacing_p_1_complex_K}
\end{figure}

To quantify the chaotic behavior of the SYK Lindbladians, we also study the statistics of complex level spacing $s$
numerically
(Fig.~\ref{fig: spectral_spacing_p_1_complex_K}).
This is defined as the distance with the closest eigenvalues in the complex plane, i.e., 
\begin{equation}
    s \coloneqq \min_{i} \left| \lambda - \lambda_{i} \right|
\end{equation}
for all complex eigenvalues $\lambda_i$ with $\lambda_{i} \neq \lambda$.
The statistics of the complex level spacing $s$ capture the local spectral correlations of non-Hermitian random matrices and open quantum systems~\cite{Haake-textbook, Ginibre-65, Grobe-88, Hamazaki-19, Hamazaki-20, Akemann-19}.
In our calculations, we only focus on complex eigenvalues in the bulk of the spectrum and exclude complex eigenvalues near the edges.
In addition, the SYK Lindbladians are invariant under modular conjugation, which makes the level statistics around the real axis special;
consequently, we also remove real eigenvalues from our calculations by requiring $\left| \mathrm{Im}\,\lambda \right| > \epsilon/\sqrt{\mathrm{dim}\,\mathcal{L}}$, where $\epsilon = 10^{-4}$ is a cutoff constant and $\mathrm{dim}\,\mathcal{L}$ is the dimensions of the double Hilbert space.
Furthermore, while the density of states is uniform in non-Hermitian random matrices, it is not necessarily so in actual physical models including the SYK Lindbladians, and hence we need to unfold the complex spectrum to calculate the complex-level-spacing statistics~\cite{Haake-textbook}.

When the complex eigenvalues are uncorrelated because of integrability, the complex-level-spacing statistics obey the two-dimensional Poisson statistics $p \left( s \right) = \left( \pi s/2 \right) e^{-\,(\pi/4)\,s^2}$.
The level statistics 
in Fig.~\ref{fig: spectral_spacing_p_1_complex_K}
clearly deviate from the Poisson statistics, which indicates the nontrivial spectral correlations and nonintegrability of the SYK Lindbladians.
Instead, they are close to the level-spacing distribution of non-Hermitian random matrices without symmetry (i.e., class A), which is consistent with Table~\ref{tab: SYK Lindbladian p=1}.
However, we see a significant deviation with the random-matrix statistics.
In particular, while the level-spacing distribution for $N=12$ well agrees with the random-matrix distribution in class A, 
this is not the case for $N=10$;
the $N=10$ distribution 
agrees rather well with the random-matrix distribution in class AI$^{\dag}$
in the large and small $s$ regimes,  
while it
may be closer to
the random-matrix distribution in class A 
around the peak. 
This intermediate behavior is understood as partial chaotic behavior of the SYK Lindbladian with the linear dissipators $p=1$.
In fact, while the Hamiltonian parts are completely chaotic, the dissipators consist only of quadratic fermions in the double Hilbert space and hence are not fully chaotic.
In Ref.~\cite{GarciaGarcia-18}, the level-spacing distribution of the SYK Hamiltonian with $q=4$ was shown not to obey the random-matrix distribution 
in the presence of an additional 
two-body SYK Hamiltonian with $q=2$.
Similarly, the SYK Lindbladian consists of the completely chaotic Hamiltonian parts with $q=4$ and the integrable linear dissipators $p=1$ and hence should not exhibit the maximally chaotic behavior.


\begin{figure}[t]
\centering

\includegraphics[width=86mm]{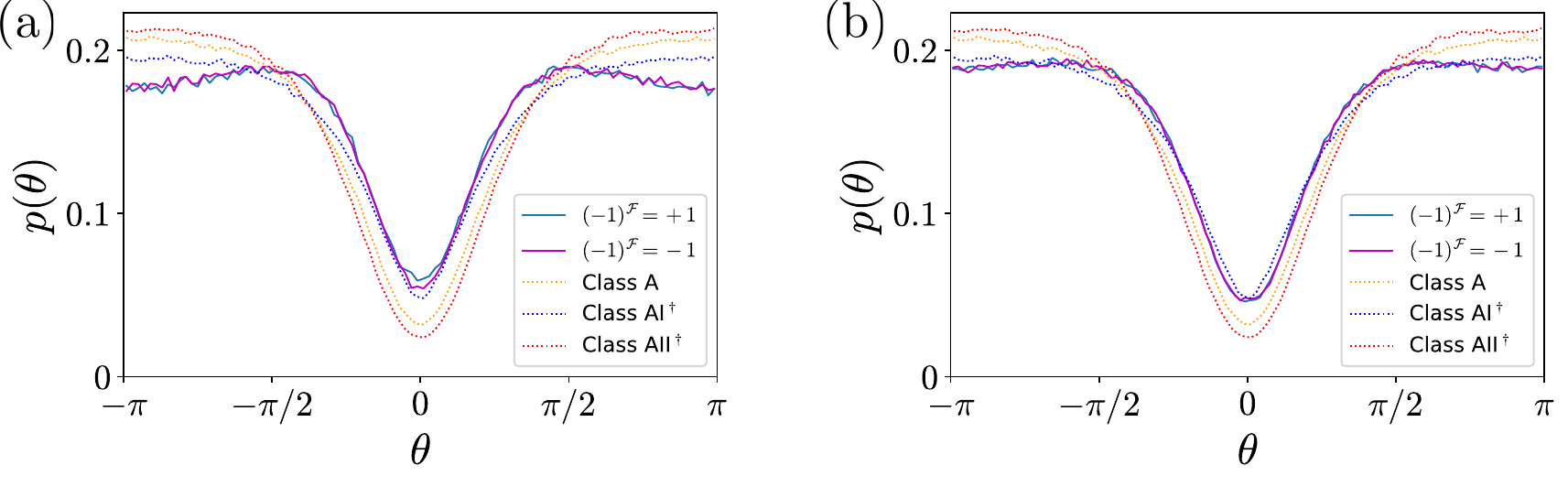} 

\caption{Marginal distributions for argument $\theta$ of the complex-spacing ratios of the Sachdev-Ye-Kitaev (SYK) Lindbladians with linear dissipators $p=1$ and complex coupling $K_{m;i} \in \mathbb{C}$, compared with the random-matrix results. 
The number $N$ of fermion flavors and the strength $K$ of the dissipators are respectively chosen to be (a)~$N=10$, $K=0.78$ and (b)~$N=12$, $K=0.87$.
The other parameters are taken as $J=1$ and $M=N$.
Each datum is averaged over $2000$ samples.
}
	\label{fig: complex_R_p_1_complex_K_2}
\end{figure}

To further study the chaotic behavior of the SYK Lindbladians, we also investigate the statistics of the complex-spacing ratio
(Fig.~\ref{fig: complex_R_p_1_complex_K_2}) 
defined as
\begin{align}
    z \coloneqq \frac{\lambda - \lambda^{\text{NN}}}{\lambda - \lambda^{\text{NNN}}},
\end{align}
where $\lambda$ is a complex eigenvalue picked uniformly at random from the bulk of the spectrum, and $\lambda^{\text{NN}}$ and $\lambda^{\text{NNN}}$ are respectively the nearest and next-nearest eigenvalues to $\lambda$ in the bulk~\cite{Sa-20}.
In contrast to the level spacing $s$, we do not need to unfold the complex spectrum to study the statistics of the level-spacing ratio $z$.
When the Lindbladian is integrable and the complex spectrum obeys the Poisson distribution, the complex-spacing ratio $z$ distributes uniformly in the disk in the complex plane.
By contrast, when the Lindbladian is fully chaotic, signatures of the random-matrix statistics appear in the distribution of $z$.
Here, while $z$ distributes in the two-dimensional complex plane, we focus on the marginal distribution
\begin{align}
    p \left( \theta \right) \coloneqq \int p \left( r, \theta \right) dr
\end{align}
with $r \coloneqq \left| z \right|$ and $\theta \coloneqq \mathrm{arg}\,z$. 
The distribution $p \left( \theta \right)$
clearly exhibits a dip around $\theta = 0$, which is different from the Poisson distribution $p \left( \theta \right) = 1/2\pi$ and indicates the nonintegrability of the SYK Lindbladians.
However, the obtained 
distributions $p \left( \theta \right)$
for $N=10$ and $N=12$ are closer to the random-matrix distribution for class AI$^{\dag}$ instead of that for class A, which is inconsistent with Table~\ref{tab: SYK Lindbladian p=1} and also the numerical results on the level-spacing distributions in Fig.~\ref{fig: spectral_spacing_p_1_complex_K}.
This discrepancy may originate from the partial chaotic behavior due to the linear dissipators $p=1$, as discussed above.
We also observe and discuss similar discrepancies between the level-spacing-ratio distributions and level-spacing distributions, as well as the level-spacing-ratio distributions and another indicator of symmetry, for the SYK Lindbladians with the quadratic ($p=2$) and cubic ($p=3$) dissipators.

\subsection{Linear dissipator (\texorpdfstring{$p=1$}{p=1}) with real dissipative coupling}
    \label{sec: dissipative quantum chaos - p=1, K in R}

\begin{figure}[t]
\centering




\includegraphics[width=86mm]{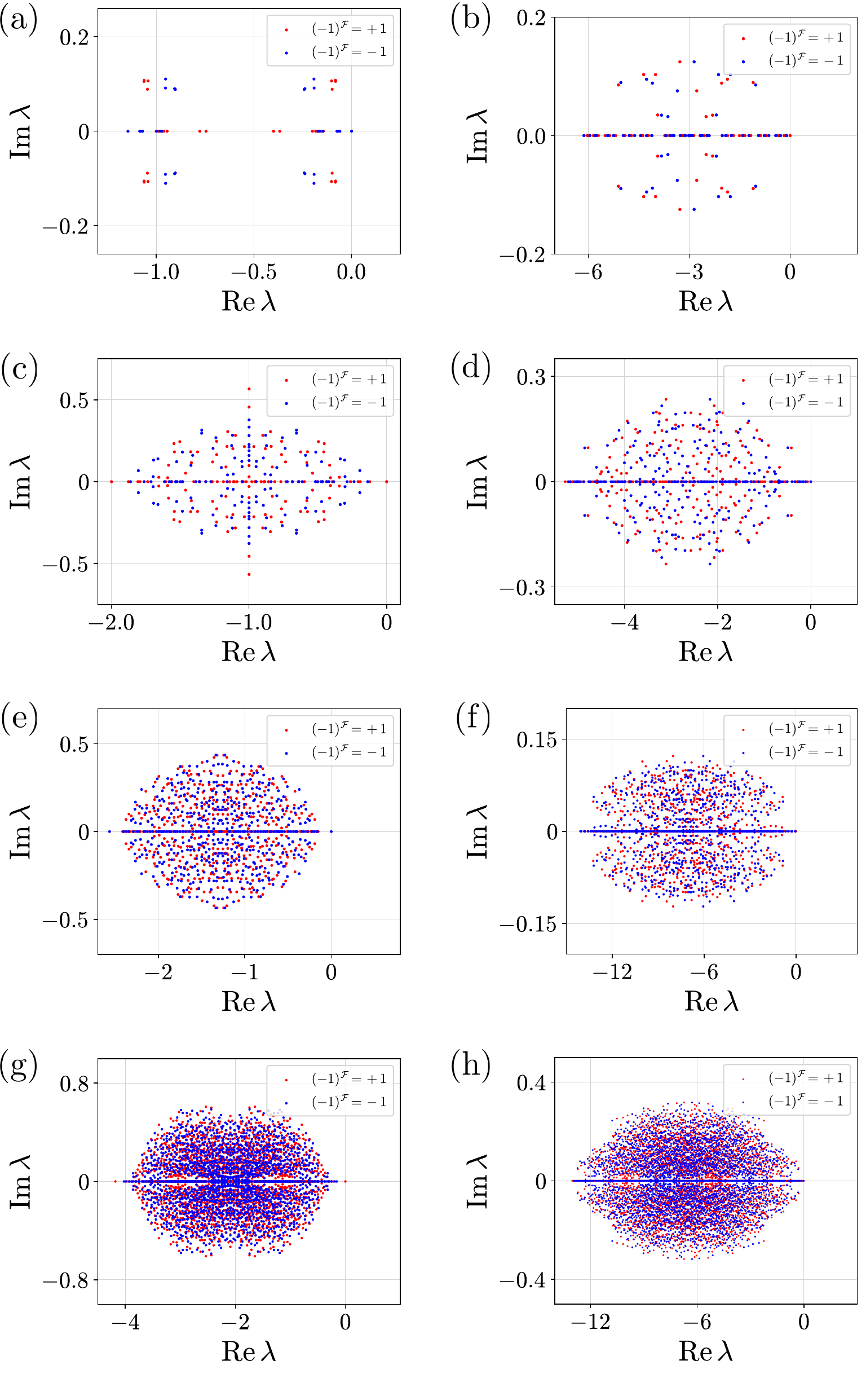}

\caption{Complex spectrum of a single realization of the Sachdev-Ye-Kitaev (SYK) Lindbladian with linear dissipators $p=1$ and real coupling $K_{m;i} \in \mathbb{R}$. 
The number $N$ of fermion flavors and the strength $K$ of the dissipators are respectively chosen to be (a)~$N=6$, $K=0.73$, (b)~$N=7$, $K=1$, (c)~$N=8$, $K=0.96$, (d)~$N=9$, $K=1$, (e)~$N=10$, $K=1.10$, (f)~$N=11$, $K=1$, (g)~$N=12$, $K=1.23$, and (h)~$N=13$, $K=1$.
The other parameters are taken as $J=1$ and $M=N$.
The different colors represent eigenstates with even ($+1$; red dots) or odd ($-1$; blue dots) total fermion parity $\left( -1 \right)^{\cal F}$. 
}
	\label{fig: spectrum_p_1_real_K}
\end{figure}

\begin{figure}
    \centering

    \includegraphics[width=86mm]{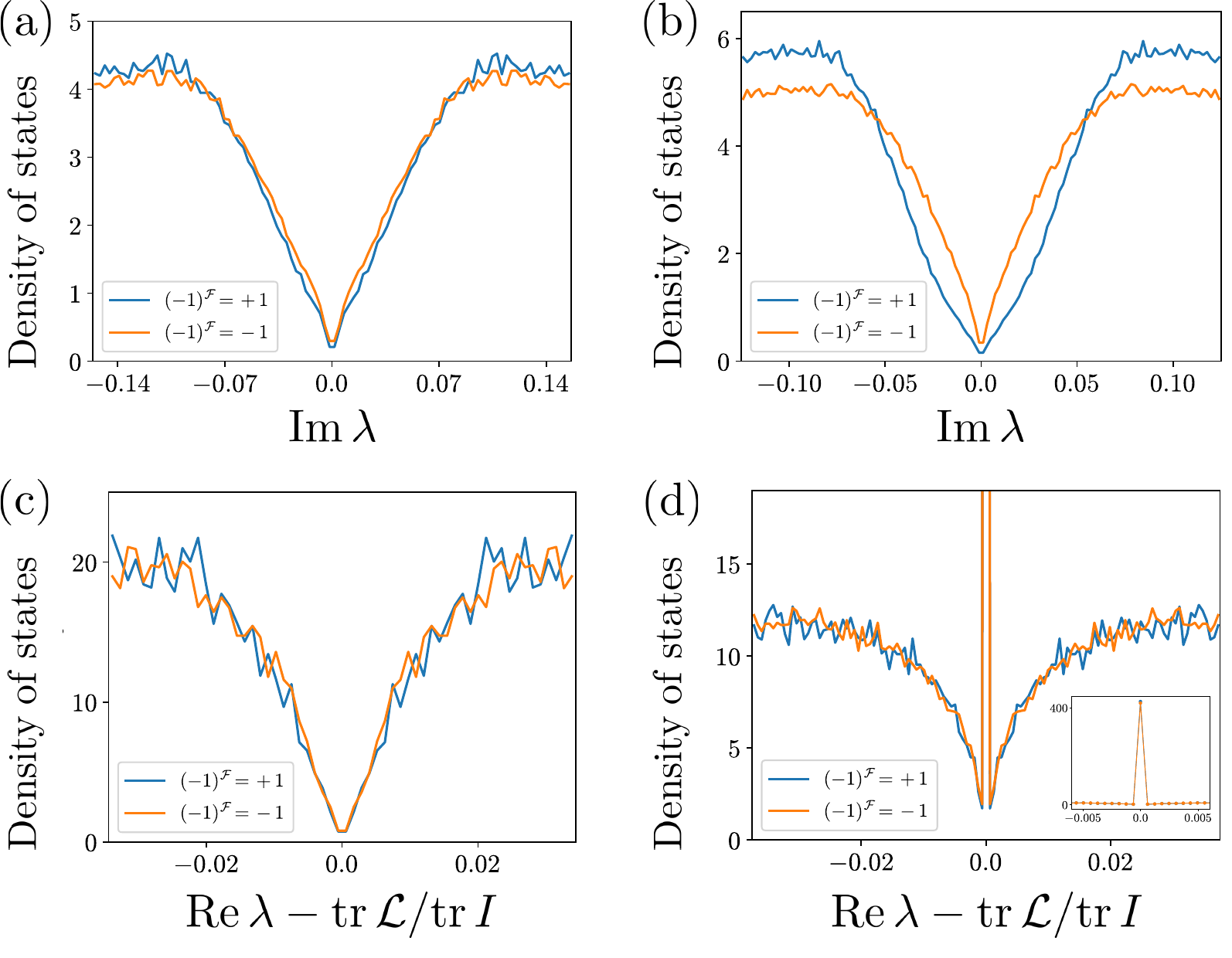} 
        
    \caption{Density of states projected to (a, b)~the imaginary axis and (c, d)~the real axis as a function of  
    the projected eigenvalues
    for the Sachdev-Ye-Kitaev (SYK) Lindbladians with linear dissipators $p=1$ and real coupling $K_{ij} \in \mathbb{R}$.
    The number $N$ of fermion flavors and the strength $K$ of the dissipators are respectively chosen to be (a, c)~$N=10$, $K=1.10$ and (b, d)~$N=12$, $K=1.23$.
    The other parameters are taken as $J=1$ and $M=N$.
    The double Hilbert space is divided into the two subspaces according to total fermion parity $\left( -1 \right)^{\cal F}$, as shown by the blue curves ($+1$) and orange curves ($-1$). 
    In~(a, b), we take eigenvalues satisfying $\left| \mathrm{Im}\,\lambda \right| > \epsilon/\sqrt{\mathrm{dim}\,\mathcal{L}}$ ($\epsilon = 10^{-4}$) and exclude real eigenvalues.
    In~(c, d), on the other hand, we include the eigenvalues on the symmetric line $\mathrm{Re}\,\lambda= \mathrm{tr}\,\mathcal{L}/\mathrm{tr}\,I$.
    In~(c), no eigenvalues appear on the symmetric line.
    The inset in~(d) shows the delta-function peak of the density on the symmetric line. 
    } 
    \label{fig:symmLineDecay-p_1-real_K-fits-quadratic}
\end{figure}

As shown in Sec.~\ref{sec: Lindblad SYK symmetry}, the symmetry classification of the SYK Lindbladians depends on whether the dissipative coupling $K_{m; i}$ satisfies $K_{m; i} K_{m; j}^{*} \in \mathbb{R}$ (see Table~\ref{tab: SYK Lindbladian p=1}).
Then, we study the complex-spectral statistics of the SYK Lindbladians with the linear dissipators $p=1$ and the real coupling $K_{m; i} \in \mathbb{R}$. 
Figure~\ref{fig: spectrum_p_1_real_K} shows the complex spectrum for $6 \leq N \leq 13$.
Similarly to the previous case with the complex dissipative coupling $K_{m; i} \in \mathbb{C}$, the complex spectrum is symmetric about the real axis and a subextensive number of real eigenvalues appear, both of which originate from modular conjugation symmetry.
A unique feature arising from the real dissipative coupling $K_{m; i} \in \mathbb{R}$ is  
the presence of additional time-reversal symmetry
(or equivalently particle-hole symmetry$^{\dag}$) that makes the complex spectrum symmetric also about the line $\mathrm{Re}\,\lambda = \mathrm{tr}\,\mathcal{L}/\mathrm{tr}\,I$ (see Table~\ref{tab: SYK Lindbladian p=1}).
Consistent with our symmetry classification, a subextensive number of eigenvalues accumulate on this symmetric line $\mathrm{Re}\,\lambda = \mathrm{tr}\,\mathcal{L}/\mathrm{tr}\,I$ for $N \equiv 0$ (mod $4$). 
Notably, whereas this spectral symmetry about $\mathrm{Re}\,\lambda = \mathrm{tr}\,\mathcal{L}/\mathrm{tr}\,I$ is respected also in each subspace with fixed total fermion parity $\left( -1 \right)^{\cal F}$ for even $N$, it is no longer respected in each subspace for odd $N$.
Instead, for odd $N$, the spectrum with $\left( -1 \right)^{\cal F} = +1$ and that with $\left( -1 \right)^{\cal F} = -1$ come in pairs and are symmetric about $\mathrm{Re}\,\lambda = \mathrm{tr}\,\mathcal{L}/\mathrm{tr}\,I$.
This difference originates from the different algebraic relation between the antiunitary operation $\mathcal{P}$ and total fermion parity $\left( -1 \right)^{\cal F}$.
In fact, for even $N$, since $\mathcal{P}$ commutes with $\left( -1 \right)^{\cal F}$ [i.e., Eq.~(\ref{eq: total fermion parity & A - even N})], $\mathcal{P}$ remains to be respected in the subspaces of $\left( -1 \right)^{\cal F}$.
For odd $N$, on the other hand, since $\mathcal{P}$ anticommutes with $\left( -1 \right)^{\cal F}$ [i.e., Eq.~(\ref{eq: total fermion parity & A - odd N})], $\mathcal{P}$ flips total fermion parity $\left( -1 \right)^{\cal F}$ and maps the two subspaces onto each other.

We also obtain the density of states along both real and imaginary axes (Fig.~\ref{fig:symmLineDecay-p_1-real_K-fits-quadratic}).
Owing to modular conjugation symmetry, a subextensive number of real eigenvalues appear, and the density of states linearly decays toward the real axis, similar to the previous case with the complex dissipative coupling $K_{m; i} \in \mathbb{C}$.
By contrast, the level statistics exhibit distinct behavior around the other symmetric line $\mathrm{Re}\,\lambda = \mathrm{tr}\,\mathcal{L}/\mathrm{tr}\,I$.
While a subextensive number of complex eigenvalues appear on this symmetric line for $N=12$, no eigenvalues appear on it for $N=10$.
This difference originates from the different signs of antiunitary symmetry $\mathcal{P}$ (see Table~\ref{tab: SYK Lindbladian p=1} and Appendix~\ref{asec: NH RMT TRS}).
In general, antiunitary symmetry $\mathcal{P}$ with $\mathcal{P}^2 = +1$ suppresses the spectral correlations of non-Hermitian random matrices while $\mathcal{P}$ with the opposite sign $\mathcal{P}^2 = -1$ enhances the spectral correlations~\cite{Ginibre-65, Hamazaki-20, Xiao-22}.
This is also similar to Hermitian random matrices, in which time-reversal symmetry with the sign $+1$ ($-1$) suppresses (enhances) the spectral correlations~\cite{Haake-textbook}.
In our SYK Lindbladian with $N \equiv 0$ (mod $4$), antiunitary symmetry $\mathcal{P}$ with the sign $\mathcal{P}^2 = +1$ is present and makes the spectral correlations at the symmetric line $\mathrm{Re}\,\lambda = \mathrm{tr}\,\mathcal{L}/\mathrm{tr}\,I$ special.
As a result, the level repulsion on the symmetric line $\mathrm{Re}\,\lambda = \mathrm{tr}\,\mathcal{L}/\mathrm{tr}\,I$ is weaker than that on generic points in the complex plane, which results in a subextensive number of eigenvalues on $\mathrm{Re}\,\lambda = \mathrm{tr}\,\mathcal{L}/\mathrm{tr}\,I$.
For $N \equiv 2$ (mod $4$), by contrast, while antiunitary symmetry $\mathcal{P}$ is still present, its sign is $\mathcal{P}^2 = -1$.
Hence, antiunitary symmetry $\mathcal{P}$ makes the level repulsion on the symmetric line $\mathrm{Re}\,\lambda = \mathrm{tr}\,\mathcal{L}/\mathrm{tr}\,I$ stronger than that on generic points in the complex plane, leading to the absence of eigenvalues on $\mathrm{Re}\,\lambda = \mathrm{tr}\,\mathcal{L}/\mathrm{tr}\,I$.
Moreover, the density of states almost linearly vanishes toward the symmetric line $\mathrm{Re}\,\lambda = \mathrm{tr}\,\mathcal{L}/\mathrm{tr}\,I$ for both $N=10$ and $N=12$, which is consistent with the linear decay of the density of states for non-Hermitian random matrices in classes BDI$^{\dag}$ and CI$^{\dag}$ (see Table~I of Ref.~\cite{Xiao-22}).
These results of the density of states also provide evidence of symmetry-enriched dissipative quantum chaos in the SYK Lindbladians.

\begin{figure}[t]
\centering

\includegraphics[width=86mm]{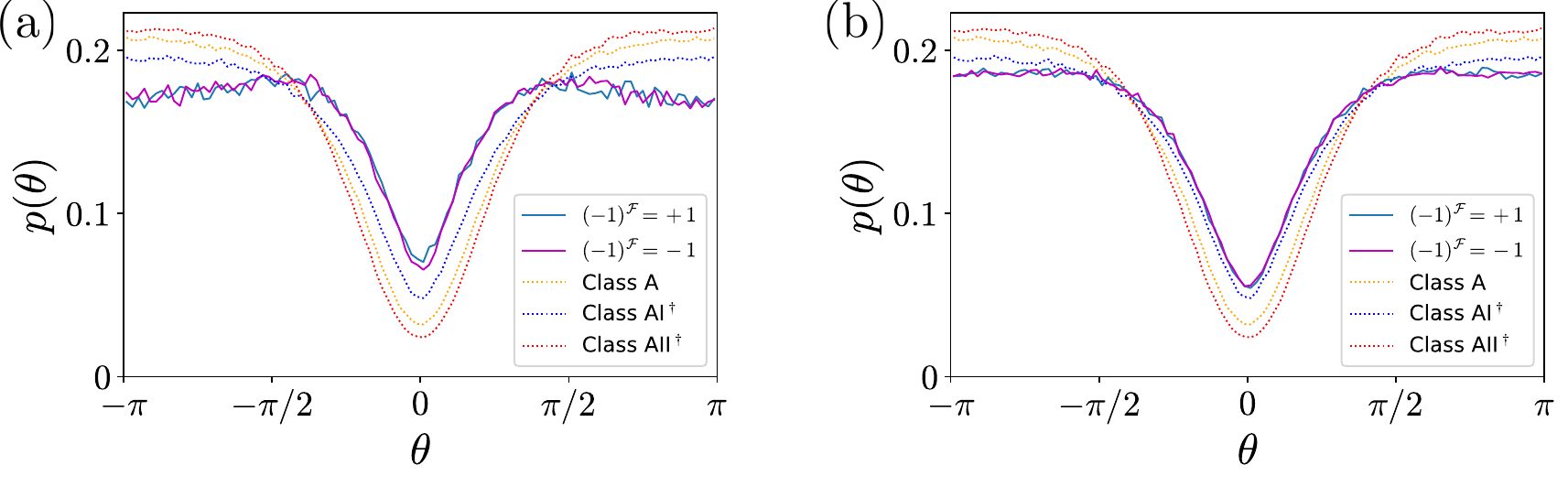} 

\caption{Marginal distributions for argument $\theta$ of the complex-spacing ratios of the Sachdev-Ye-Kitaev (SYK) Lindbladians with linear dissipators $p=1$ and real coupling $K_{m;i} \in \mathbb{R}$, compared with the random-matrix results. 
The number $N$ of fermion flavors and the strength $K$ of the dissipators are respectively chosen to be (a)~$N=10$, $K=1.10$ and (b)~$N=12$, $K=1.23$. 
The other parameters are taken as $J=1$ and $M=N$. 
Each datum is averaged over $3000$ samples.}
	\label{fig: complex_R_p_1_real_K}
\end{figure}

To study the spectral correlations and chaotic behavior, we also obtain the complex-spacing-ratio distributions $p \left( \theta \right)$, as shown in Fig.~\ref{fig: complex_R_p_1_real_K}.
In comparison with the previous case with the complex dissipative coupling $K_{m; i} \in \mathbb{C}$, another unique feature due to the real dissipative coupling $K_{m; i} \in \mathbb{R}$ is the presence of time-reversal symmetry$^{\dag}$ (see Table~\ref{tab: SYK Lindbladian p=1}), which changes the local spectral correlations in the bulk.
Consistent with our symmetry classification, the spacing-ratio distribution for $N=12$ well agrees with the random-matrix distribution for class AI$^{\dag}$.
On the other hand, the spacing-ratio distribution for $N=10$ significantly deviates from the random-matrix distribution for class AI$^{\dag}$ and does not agree with any random-matrix distributions.
This discrepancy seems to arise from the partial chaotic behavior due to the linear dissipators $p=1$, as also discussed in Sec.~\ref{sec: dissipative quantum chaos - p=1, K in C}.
Since we have the flat distribution $p \left( \theta \right) = 1/2\pi$ for the Poisson statistics in integrable open quantum systems, the dip of $p \left( \theta \right)$ around $\theta = 0$ is a direct signature of the level repulsion and hence the nonintegrability.
Thus, the shallower dip of $p \left( \theta \right)$ around $\theta = 0$ for the SYK Lindbladian with $N=10$ (see Fig.~\ref{fig: complex_R_p_1_real_K}) indicates the less chaotic behavior compared with the random-matrix statistics.

\subsection{Quadratic dissipator (\texorpdfstring{$p=2$}{p=2}) with complex dissipative coupling}

%

\begin{figure}[t]
\centering




\includegraphics[width=86mm]{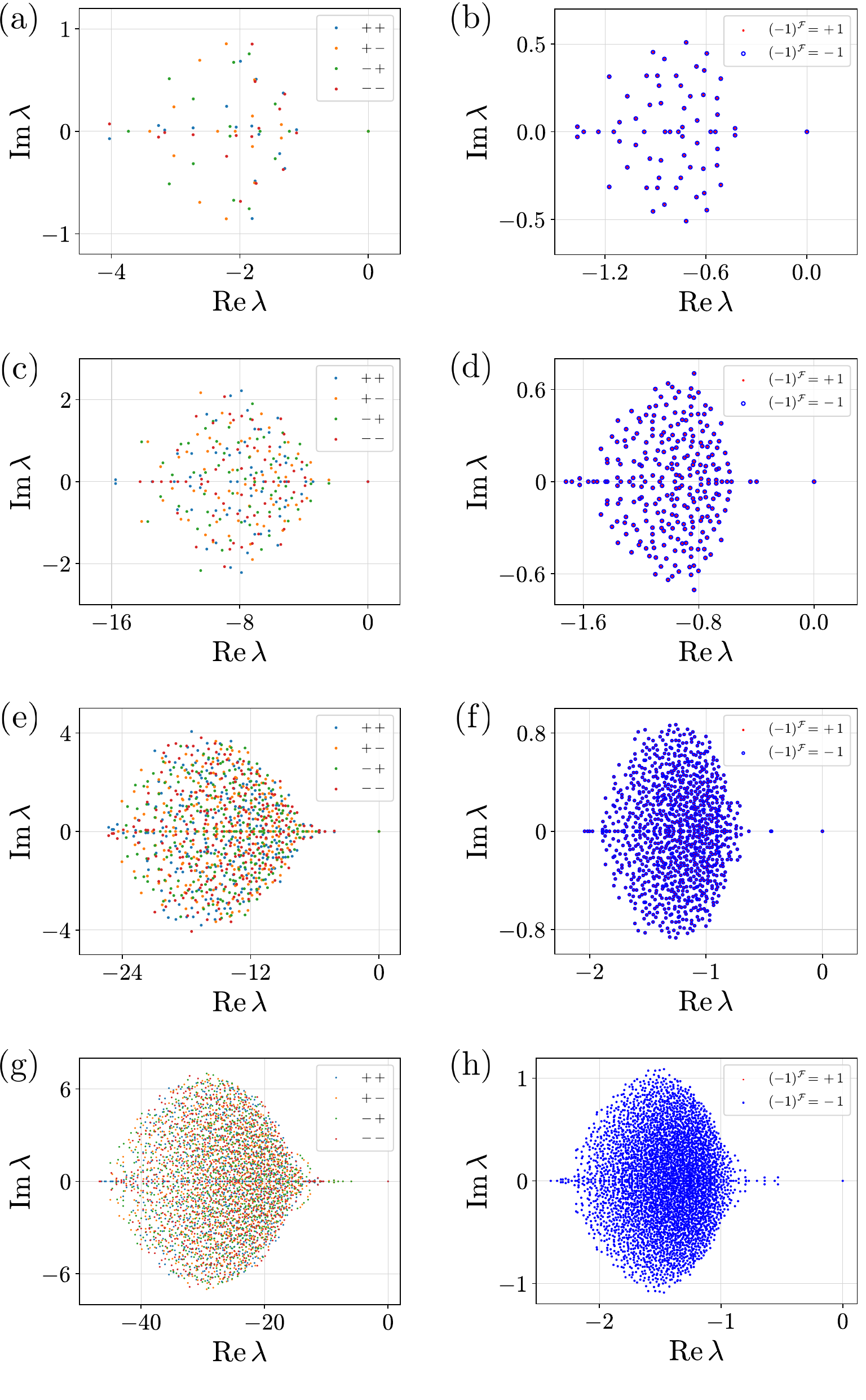} 

\caption{Complex spectrum of a single realization of the Sachdev-Ye-Kitaev (SYK) Lindbladian with quadratic dissipators $p=2$ and complex coupling $K_{m;i} \in \mathbb{C}$. 
The number $N$ of fermion flavors and the strength $K$ of the dissipators are respectively chosen to be (a)~$N=6$, $K=1.99$, (b)~$N=7$, $K=1$, (c)~$N=8$, $K=2.91$, (d)~$N=9$, $K=1$, (e)~$N=10$, $K=3.64$, (f)~$N=11$, $K=1$, (g)~$N=12$, $K=4.47$, and (h)~$N=13$, $K=1$.
The other parameters are taken as $J=1$ and $M=N$. 
For even $N$, the different colors represent eigenstates from even ($+1$) or odd ($-1$) fermion parity $\left( -1 \right)^{F^{+}}$ and $\left( -1 \right)^{F^{-}}$, shown by the blue dots ($++$), orange dots ($+-$), green dots ($-+$), and red dots ($--$). 
For odd $N$, eigenstates with even (red dots) and odd (blue dots) total fermion parity $\left( -1 \right)^{\cal F}$ are degenerate.
}
	\label{fig: spectrum_p_2_complex_K_2}
\end{figure}

\begin{figure}
    \centering

    \includegraphics[width=86mm]{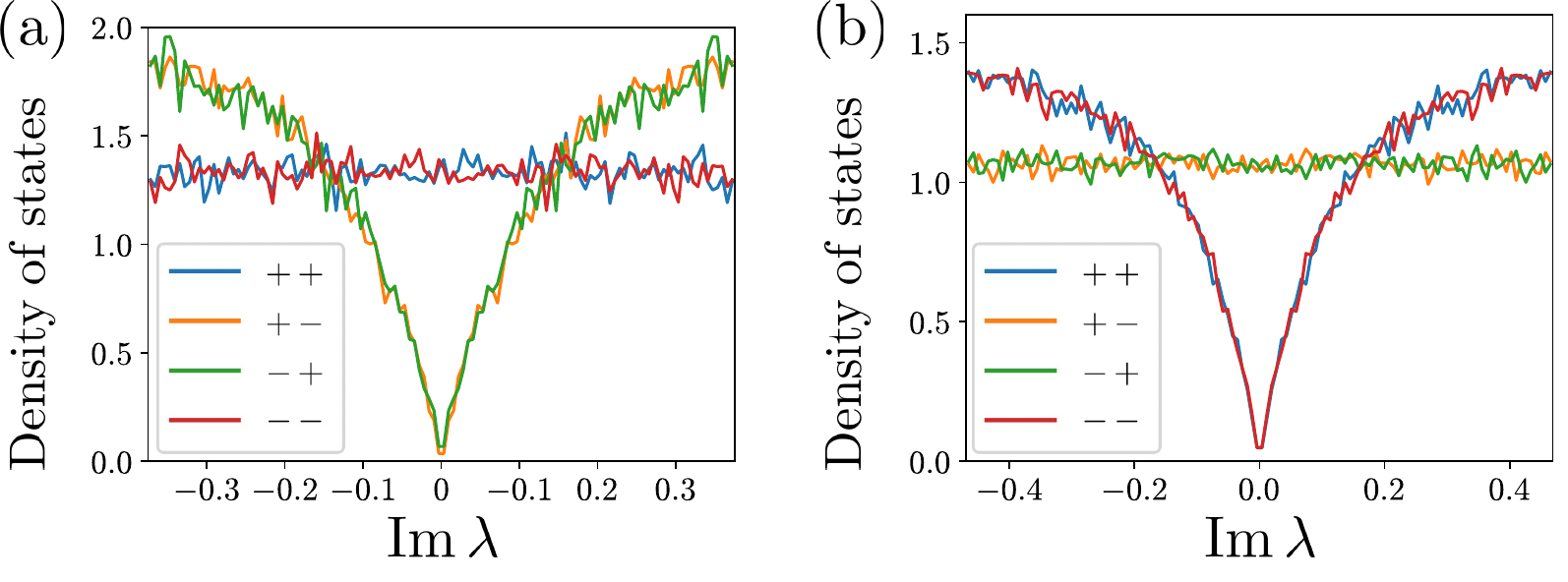} 
        
    \caption{Density of states across the real axis as a function of the imaginary part of complex eigenvalues $\lambda$ for the Sachdev-Ye-Kitaev (SYK) Lindbladians with quadratic dissipators $p=2$ and complex coupling $K_{m; i} \in \mathbb{C}$.
    The number $N$ of fermion flavors and the strength $K$ of the dissipators are respectively chosen to be (a)~$N=10$, $K=3.64$ and (b)~$N=12$, $K=4.74$.
    The other parameters are taken as $J=1$ and $M=N$.
    The double Hilbert space is divided into the four subspaces according to fermion parity $\left( -1 \right)^{F^{\pm}}$, as shown by the blue curves ($++$), orange curves ($+-$), green curves ($-+$), and red curves ($--$).
    We take eigenvalues satisfying $\left| \mathrm{Im}\,\lambda \right| > \epsilon/\sqrt{\mathrm{dim}\,\mathcal{L}}$ ($\epsilon = 10^{-4}$) and exclude real eigenvalues.
    }
    \label{fig:symmLineDecay-p_2-complex_K-fits-log}
\end{figure}

Next, we study the SYK Lindbladians with the quadratic dissipators $p=2$ and the complex dissipative coupling $K_{m; i} \in \mathbb{C}$.
Figure~\ref{fig: spectrum_p_2_complex_K_2} shows the complex spectrum for $6 \leq N \leq 13$.
For even $p$ and even $N$, fermion parity is conserved in both ket and bra spaces, and we investigate the spectral properties in each subspace of fixed fermion parity $\left( -1 \right)^{F^{\pm}}$.
As also discussed in Sec.~\ref{subsec: SYK modular conjugation}, the presence or absence of modular conjugation symmetry depends on $N$.
In the physically relevant subspace with $\left( -1 \right)^{\cal F} = +1$, complex eigenvalues in each subspace of $\left( -1 \right)^{F^{\pm}}$ do not individually respect modular conjugation symmetry for $N \equiv 2$ (mod $4$).
Rather, complex eigenvalues in the subspace of $\left( -1 \right)^{F^{+}} = +1$ and those in the subspace of $\left( -1 \right)^{F^{+}} = -1$ come in complex-conjugate pairs.
These numerical results are consistent with our symmetry classification (see, especially, the entry ``A" in Table~\ref{tab: SYK Lindbladian p=2} with $N \equiv 2$).
By contrast, for $N \equiv 0$ (mod $4$), complex eigenvalues in each subspace of $\left( -1 \right)^{F^{\pm}}$ individually respect modular conjugation symmetry, and a subextensive number of real eigenvalues appear on the real axis, which is also consistent with our symmetry classification (see, especially, the entry ``AI = D$^{\dag}$" in Table~\ref{tab: SYK Lindbladian p=2} with $N \equiv 0$).

We also obtain the density of states
projected to the imaginary axis,
as shown in Fig.~\ref{fig:symmLineDecay-p_2-complex_K-fits-log}.
For $\left( -1 \right)^{\cal F} = +1$ and $N \equiv 2$ (mod $4$), the density of states continuously changes as a function of $\mathrm{Im}\,\lambda$ because of the absence of modular conjugation symmetry in the subspace of fermion parity $\left( -1 \right)^{F^{\pm}}$.
This behavior is consistent with the level statistics of non-Hermitian random matrices without symmetry (i.e., class A).
For $\left( -1 \right)^{\cal F} = +1$ and $N \equiv 0$ (mod $4$), by contrast, modular conjugation symmetry is respected in the individual subspaces with fixed $\left( -1 \right)^{F^{\pm}}$ and changes the spectral correlations around the real axis $\mathrm{Im}\,\lambda = 0$.
Consequently, the density of states linearly vanishes toward the real axis.
The linear decay of the density of states toward the real axis is consistent with the random-matrix statistics with time-reversal symmetry (i.e., class AI)~\cite{Xiao-22}, which confirms the dissipative quantum chaos.
We also note that the behavior of the density of states is opposite in the subspaces with $\left( -1 \right)^{\cal F} = -1$:
the density of states linearly vanishes for $N \equiv 2$ (mod $4$) but exhibits no singular behavior for $N \equiv 0$ (mod $4$).

\begin{figure}[t]
\centering

\includegraphics[width=86mm]{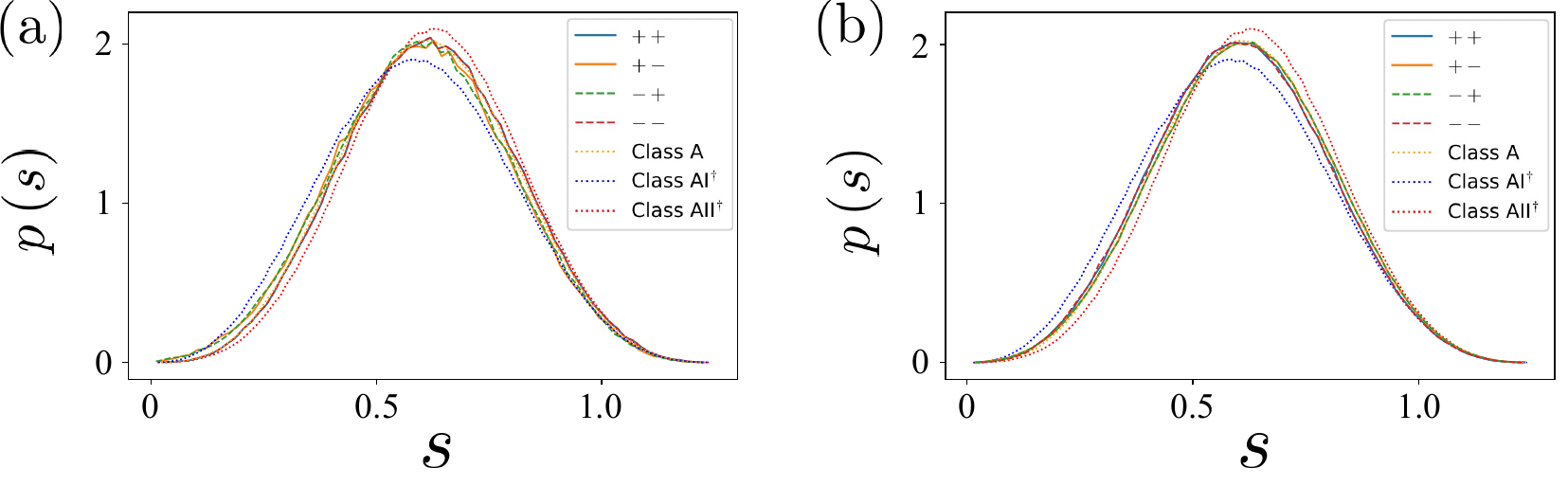} 

\caption{Distributions for normalized nearest spectral spacings $s$ of the Sachdev-Ye-Kitaev (SYK) Lindbladians with quadratic dissipators $p=2$ and complex coupling $K_{m;i} \in \mathbb{C}$, compared with the random-matrix results. 
The number $N$ of fermion flavors and the strength $K$ of the dissipators are respectively chosen to be (a)~$N=10$, $K=3.64$ and (b)~$N=12$, $K=4.47$.
The other parameters are taken as $J=1$ and $M=N$. 
Each datum is averaged over $2000$ samples. 
}
	\label{fig: spectral_spacing_p_2_complex_K}
\end{figure}


\begin{figure}[t]
\centering

\includegraphics[width=86mm]{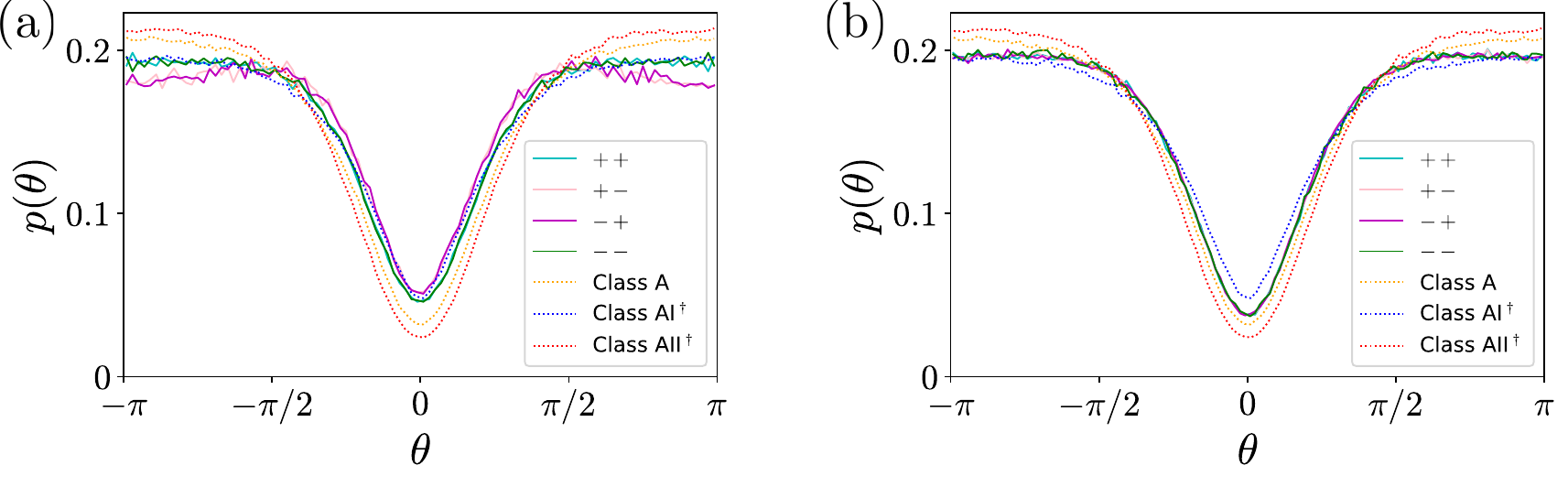} 

\caption{Marginal distributions for argument $\theta$ of the complex-spacing ratio of the Sachdev-Ye-Kitaev (SYK) Lindbladians with quadratic dissipators $p=2$ and complex coupling $K_{m;i} \in \mathbb{C}$, compared with the random-matrix results. 
The number $N$ of fermion flavors and the strength $K$ of the dissipators are respectively chosen to be (a)~$N=10$, $K=3.64$ and (b)~$N=12$, $K=4.47$.
The other parameters are taken as $J=1$ and $M=N$.
Each datum is averaged over 3000 samples.
}
	\label{fig: complex_R_p_2_complex_K_2}
\end{figure}

\begin{table}[t]
	\centering
	\caption{Angle test of time-reversal symmetry$^{\dag}$ (TRS$^{\dag}$) for the Sachdev-Ye-Kitaev (SYK) Lindbladians with quadratic dissipators $p=2$ and $q=4$. 
    The number $N$ of fermion flavors is chosen as $4$, $6$, $8$, $10$, and $12$. 
    The random dissipative coupling $K_{m; i}$ is chosen to be complex ($\mathbb{C}$) or real ($\mathbb{R}$). 
    The angle test is carried out for each subsector with fermion parity $\left( -1 \right)^{F^{+}}$ and $\left( -1 \right)^{F^{-}}$, which is denoted by $\left( \pm, \pm \right)$. 
    The entries ``T" show the presence of TRS$^{\dag}$ while the entries ``F" show the absence of TRS$^{\dag}$.}
     \begin{tabular}{cc|cccc} \hline \hline
     ~~$N$ (mod $4$)~~ & ~$K_{m; i}$~ & ~$\left( +, + \right)$~ & ~$\left( +, - \right)$~ & ~$\left( -, + \right)$~ & ~$\left( -, - \right)$~ \\ \hline
     $4$, $8$, $12$ $\equiv 0$ & $\mathbb{C}$ & F & F & F & F \\ 
     $4$, $8$, $12$ $\equiv 0$ & $\mathbb{R}$ & T & T & T & T \\ \hline 
     $6$, $10$ $\equiv 2$ & $\mathbb{C}$ & F & F & F & F \\ 
     $6$, $10$ $\equiv 2$ & $\mathbb{R}$ & F & F & F & F \\ \hline \hline
    \end{tabular}
	\label{tab: angle test TRS-dag}
\end{table}

Furthermore, we study the spectral correlations and dissipative quantum chaos.
We first obtain the complex spacing distributions, as shown in Fig.~\ref{fig: spectral_spacing_p_2_complex_K}.
Similarly to the previous case with the linear dissipators $p=1$, we focus on the level statistics in the bulk.
The level-spacing distributions for all the four subsectors and both $N=10$ and $N=12$ well coincide with the random-matrix distribution without symmetry (i.e., class A), which is consistent with our symmetry classification in Table~\ref{tab: SYK Lindbladian p=2}.
Notably, the quadratic dissipators $p=2$ reduce to the fermionic quartic terms in the double Hilbert space and hence are sufficiently nonintegrable.
Consequently, the SYK Lindbladians exhibit the completely chaotic behavior, which contrasts with the partial chaotic behavior for the linear dissipators $p=1$ (compare Figs.~\ref{fig: spectral_spacing_p_1_complex_K} and \ref{fig: spectral_spacing_p_2_complex_K}).

We also obtain the level-spacing-ratio distributions, as shown in Fig.~\ref{fig: complex_R_p_2_complex_K_2}.
For $N=12$, the level-spacing-ratio distribution is closer to the random-matrix distribution for class A with a slight deviation, which is compatible with the level-spacing distribution in 
Fig.~\ref{fig: spectral_spacing_p_2_complex_K}
and the symmetry classification in Table~\ref{tab: SYK Lindbladian p=2}.
For $N=10$, however, the level-spacing-ratio distribution is closer to the random-matrix distribution for class AI$^{\dag}$, which is incompatible with both the level-spacing distribution and the symmetry classification.
This result may imply the possible presence of time-reversal symmetry$^{\dag}$ for $N \equiv 2$ (mod $4$).

To clarify the presence or absence of time-reversal symmetry$^{\dag}$, we directly test it in terms of the symmetry constraints on eigenstates instead of the spectral statistics.
Let $\lambda \in \mathbb{C}$ be an eigenvalue of the Lindbladian $\mathcal{L}$, and $\ket{\phi}$ and $\ket{\chi}$ be right and left eigenstates, respectively (i.e., $\mathcal{L} \ket{\phi} = \lambda \ket{\phi}$, $\mathcal{L}^{\dag} \ket{\chi} = \lambda^{*} \ket{\chi}$).
Then, suppose that the Lindbladian $\mathcal{L}$ respects time-reversal symmetry$^{\dag}$ $\mathcal{P} \mathcal{L}^{\dag} \mathcal{P}^{-1} = \mathcal{L}$ with an antiunitary operator $\mathcal{P}$ satisfying $\mathcal{P}^2 = +1$.
Because of time-reversal symmetry$^{\dag}$, the right eigenstate $\ket{\phi}$ and the corresponding left eigenstate $\ket{\chi}$ are related to each other.
In fact, we have
\begin{align}
    \mathcal{L} \left( \mathcal{P} \ket{\chi} \right)
    = \mathcal{P} \mathcal{L}^{\dag} \ket{\chi}
    = \mathcal{P} \left( \lambda^{*} \ket{\chi} \right)
    = \lambda \left( \mathcal{P} \ket{\chi} \right),
\end{align}
which shows that $\mathcal{P} \ket{\chi}$ is also a right eigenstate of $\mathcal{L}$ with the same eigenvalue $\lambda$.
Thus, without any additional symmetry, especially additional spectral degeneracy, we generally have
\begin{align}
    \ket{\phi} \propto \mathcal{P} \ket{\chi},
\end{align}
which is a direct indicator of time-reversal symmetry$^{\dag}$ with the sign $+1$~\cite{KSUS-19}.
Notably, this symmetry test is mathematically well formulated as the angle test~\cite{Balayan-Garcia-angle-test};
it is proved that the Lindbladian $\mathcal{L}$ (or more generally, non-Hermitian operator) without spectral degeneracy respects time-reversal symmetry$^{\dag}$ if and only if we have
\begin{align}
    \braket{\phi_i | \phi_j} \braket{\phi_j | \phi_k} \braket{\phi_k | \phi_{i}} = \left[ \braket{\chi_i | \chi_j} \braket{\chi_j | \chi_k} \braket{\chi_k | \chi_{i}} \right]^{*}
        \label{eq: angle test}
\end{align}
whenever $i \leq j \leq k$ and not all of $i$, $j$, $k$ are equal.
In particular, the violation of this condition even for a few eigenstates means the absence of time-reversal symmetry$^{\dag}$.
While this angle test assumes the absence of spectral degeneracy, the complex spectrum of the SYK Lindbladian in each subspace of fermion parity $\left( -1 \right)^{F^{\pm}}$ is not indeed degenerate.

We carry out the angle test of time-reversal symmetry$^{\dag}$ for the SYK Lindbladians with the quadratic dissipators $p=2$, as summarized in Table~\ref{tab: angle test TRS-dag}.
For all $N$ and all the subsectors of fermion parity $\left( -1 \right)^{F^{\pm}}$, we see that time-reversal symmetry$^{\dag}$ is absent as long as the dissipative coupling $K_{m; i}$ is complex.
In fact, a typical choice of a triplet of eigenstates breaks the condition in Eq.~(\ref{eq: angle test}).
This is consistent with the symmetry classification in Table~\ref{tab: SYK Lindbladian p=2} and the level-spacing distributions in Fig.~\ref{fig: spectral_spacing_p_2_complex_K}, but incompatible with the level-spacing-ratio distribution for $N=10$ in Fig.~\ref{fig: complex_R_p_2_complex_K_2}\,(a).
Notably, while the complete agreement with the spectral statistics of random matrices requires the sufficient nonintegrability, the angle test in Table~\ref{tab: angle test TRS-dag} works for generic non-Hermitian operators even without the nonintegrability.
Consequently, owing to the angle test, time-reversal symmetry$^{\dag}$ should be absent in the SYK Lindbladian with $p=2$ and $K_{m; i} \in \mathbb{C}$, which confirms our symmetry classification in Table~\ref{tab: SYK Lindbladian p=2}.
The discrepancy of the level-spacing-ratio distribution for $N=10$ in Fig.~\ref{fig: complex_R_p_2_complex_K_2} should originate from the special structure that makes the model less chaotic and cannot be captured solely from the internal symmetry classification.
In particular, as shown in Sec.~\ref{subsec: SYK Lindbladian p2 classification}, only a part of the dissipation term $\mathcal{D}$ can break time-reversal symmetry$^{\dag}$ in the SYK Lindbladian.
In such a case, the level statistics should be subject to a severe crossover effect between classes A and AI$^{\dag}$, which may lead to the unusual behavior in Fig.~\ref{fig: complex_R_p_2_complex_K_2}.
It is worthwhile to further study the relationship between the different measures of the complex-spectral statistics, such as the spacing distribution and spacing-ratio distribution, and its relevance to the chaotic behavior of open quantum systems.

\subsection{Quadratic dissipator (\texorpdfstring{$p=2$}{p=2}) with real dissipative coupling}

\begin{figure}
    \centering

    \includegraphics[width=86mm]{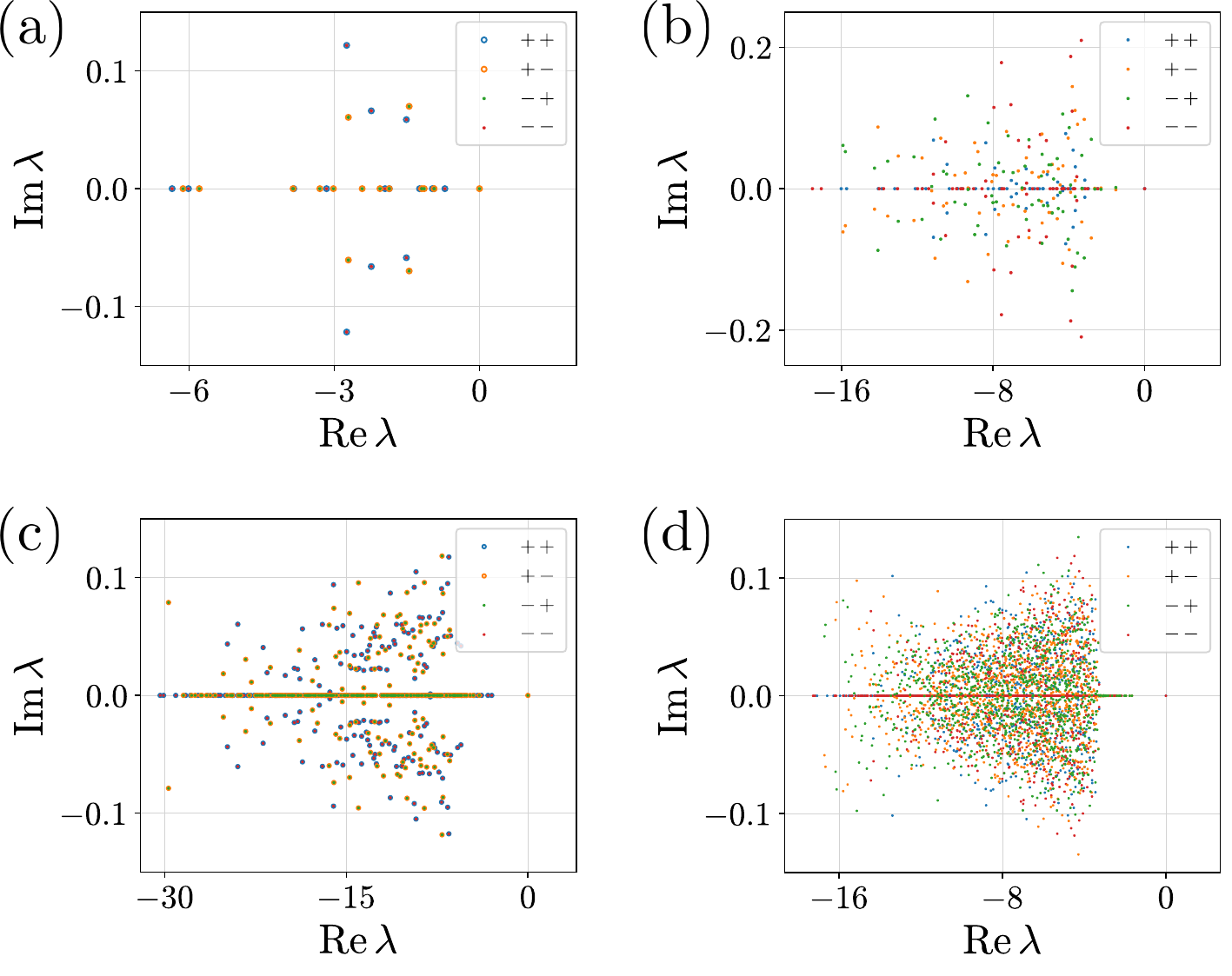} 
    
    \caption{Complex spectrum of a single realization of the Sachdev-Ye-Kitaev (SYK) Lindbladian with quadratic dissipators $p=2$ and real coupling $K_{m; i} \in \mathbb{R}$.
    The number $N$ of fermion flavors and the strength $K$ of the dissipators are respectively chosen to be (a)~$N=6$, $K=2.82$, (b)~$N=8$, $K=4.12$, (c)~$N=10$, $K=5.15$, and (d)~$N=12$, $K=6.33$. 
    The other parameters are taken as $J=1$ and
    $M=N$.
    The double Hilbert space is divided into the four subspaces according to fermion parity $\left( -1 \right)^{F^{+}}$ and $\left( -1 \right)^{F^{-}}$, shown by the blue dots ($++$), orange dots ($+-$), green dots ($-+$), and red dots ($--$).
    In each subspace with fixed fermion parity $\left( -1 \right)^{F^{\pm}}$, modular conjugation symmetry is respected for $\left( -1 \right)^{N/2} \left( -1 \right)^{\cal F} = +1$, while pseudo-Hermiticity is respected for $\left( -1 \right)^{\cal F} = +1$.
    For $N = 6, 10 \equiv 2$ (mod $4$), eigenvalues with the same total fermion parity $\left( -1 \right)^{\cal F}$ are two-fold degenerate.
    }
    \label{fig:spec-p_2-real_K}
\end{figure}

\begin{figure}
    \centering

    \includegraphics[width=86mm]{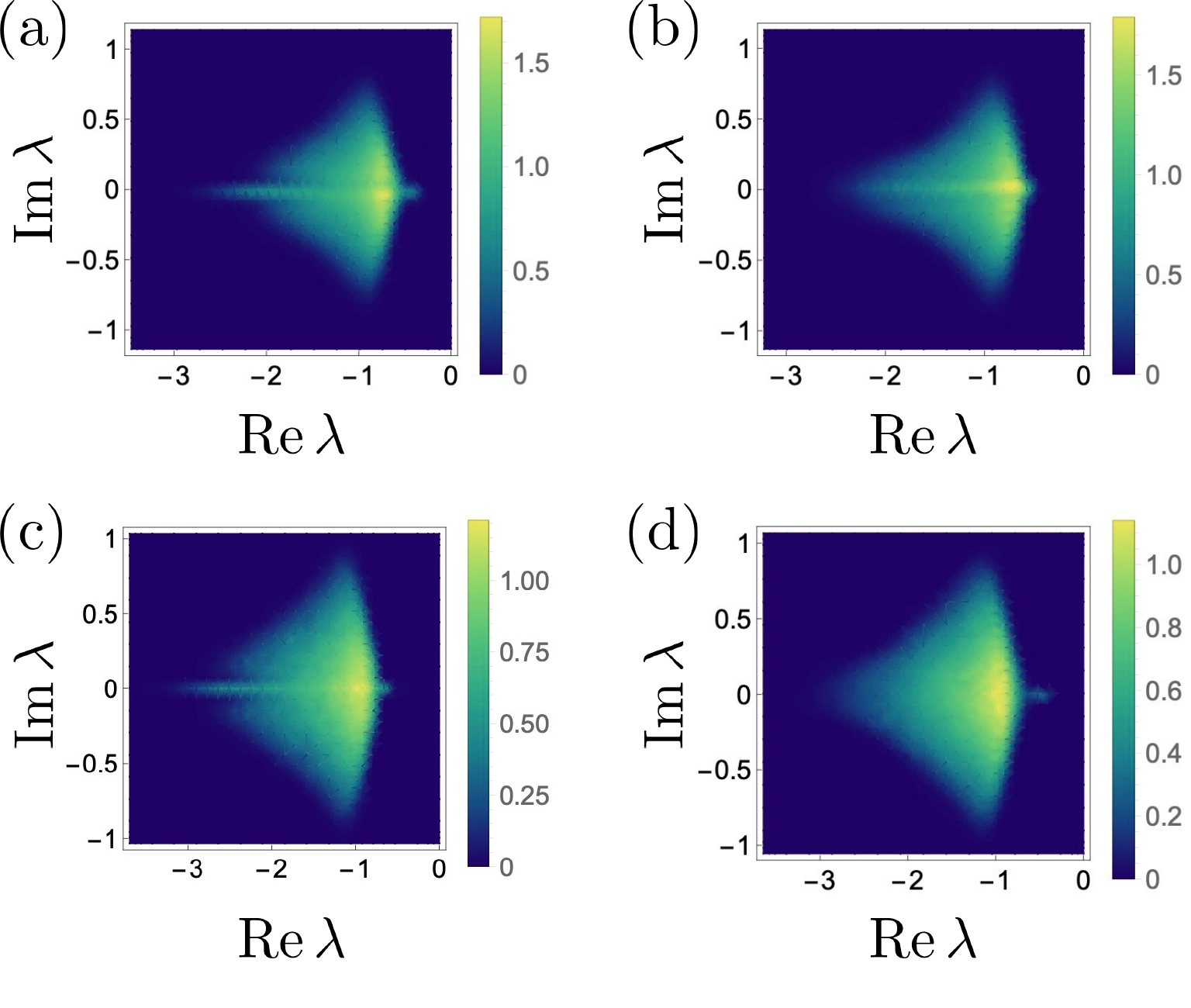} 
    
    \caption{Complex spectral density of the Sachdev-Ye-Kitaev (SYK) Lindbladians with quadratic dissipators $p=2$ and real coupling $K_{m; i} \in \mathbb{R}$ for 
    (a, b)~$N=10$ and (c, d)~$N=12$.
    Fermion parity $\left( -1 \right)^{F^{+}}$ and $\left( -1 \right)^{F^{-}}$ is chosen as (a, c)~$++$ and (b, d)~$+-$.
    The other parameters are chosen to be $J=1$, $K=1$, and $M=N$.
    Each datum is averaged over (a,b)~$1024$ samples for $N=10$ and (c, d)~$166$ samples for $N=12$.}
        \label{fig:spec-p_2-real_K - average}
\end{figure}

\begin{figure}
    \centering

    \includegraphics[width=86mm]{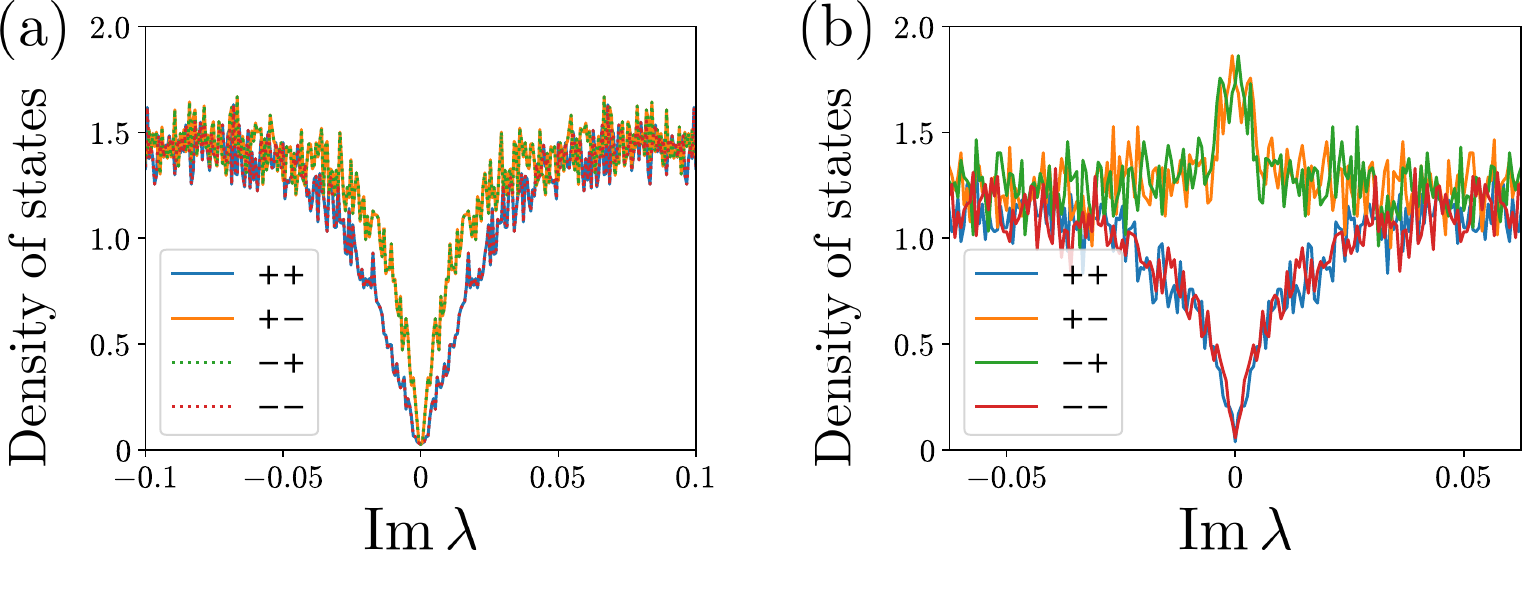} 
    
    \caption{Density of states across the real axis as a function of the imaginary part of complex eigenvalues $\lambda$ for the Sachdev-Ye-Kitaev (SYK) Lindbladians with quadratic dissipators $p=2$ and real coupling $K_{m; i} \in \mathbb{R}$.
    The number $N$ of fermion flavors is chosen to be (a)~$N = 10$ and (b)~$N=12$.
    The other parameters are taken as $J=1$, $K=1$, and $M=N$.
    The double Hilbert space is divided into the four subspaces according to fermion parity $\left( -1 \right)^{F^{\pm}}$, as shown by the blue curves ($++$), orange curves ($--$), green curves ($+-$), and red curves ($-+$).
    For $N \equiv 2$ (mod $4$) including $N=10$, the complex spectrum with the same total fermion parity $\left( -1 \right)^{\cal F}$ is two-fold degenerate.
    We only consider eigenvalues away from the origin ($|\lambda|>10^{-10}$) and away from the real axis ($|\text{Im}\,\lambda|>10^{-10}$).
    }
    \label{fig:symmLineDecay-p_2-real_K}
\end{figure}

As discussed in Sec.~\ref{sec: Lindblad SYK symmetry}, the condition $K_{m; i} K_{m;j}^{*} \in \mathbb{R}$ is relevant to the symmetry classification of the SYK Lindbladians also for even $p$.
Now, we study the SYK Lindbladians with the quadratic dissipators $p=2$ and the real dissipative coupling $K_{m; i} \in \mathbb{R}$.
Figure~\ref{fig:spec-p_2-real_K} shows the complex spectra for $N=6, 8, 10, 12$.
Figure~\ref{fig:spec-p_2-real_K - average} also shows the complex spectral density averaged over different samples of the SYK Lindbladians.
Similarly to the previous case with the complex dissipative coupling $K_{m; i} \in \mathbb{C}$, modular conjugation symmetry is broken in the subspace with fixed fermion parity $\left( -1 \right)^{F^{\pm}}$ for $N \equiv 2$ (mod $4$) and $\left( -1 \right)^{\cal F} = +1$ [see Eq.~(\ref{eq: modular conjugation & sub-fermion parity})].
However, as discussed in Sec.~\ref{subsec: SYK Lindbladian p2 classification} and summarized in Table~\ref{tab: SYK Lindbladian p=2}, the combination of modular conjugation and the antiunitary operation $\mathcal{P}$ gives rise to pseudo-Hermiticity in Eq.~(\ref{eq: p2 pH}) even in the subspace with fixed fermion parity $\left( -1 \right)^{F^{\pm}}$ for 
$\left( -1 \right)^{\cal F} = +1$.
In fact, we numerically confirm the presence of pseudo-Hermiticity, as shown in Fig.~\ref{fig:spec-p_2-real_K}.
On the other hand,
for $N \equiv 2$ (mod $4$) and $\left( -1 \right)^{\cal F} = -1$, as well as $N \equiv 0$ (mod $4$) and $\left( -1 \right)^{\cal F} = +1$,
modular conjugation symmetry is respected even in the individual subspaces of $\left( -1 \right)^{F^{\pm}}$.
Notably, while both modular conjugation and pseudo-Hermiticity make the complex spectrum symmetric about the real axis, they lead to the quantitatively different complex-spectral statistics of non-Hermitian random matrices~\cite{Xiao-22}.
We also obtain the density of states across the real axis, as shown in Fig.~\ref{fig:symmLineDecay-p_2-real_K}.
Similarly to the previous case, the density of states linearly decays toward the real axis, signaling the dissipative quantum chaos also for the real coupling $K_{m; i} \in \mathbb{R}$.

Furthermore, we carry out the angle test for the real dissipative coupling $K_{m; i} \in \mathbb{R}$, as summarized in Table~\ref{tab: angle test TRS-dag}.
While no time-reversal symmetry$^{\dag}$ is 
detected
for $N \equiv 2$ (mod $4$), the presence of time-reversal symmetry$^{\dag}$ is confirmed for $N \equiv 0$ (mod $4$), which completely agrees with the symmetry classification in Table~\ref{tab: SYK Lindbladian p=2}.
Notably, the absence of time-reversal symmetry$^{\dag}$ is due to the nontrivial anticommutation relation between the antiunitary operation $\mathcal{P}$ and fermion parity $\left( -1 \right)^{F^{\pm}}$, although they commute with each other for $N \equiv 0$ (mod $4$) (see Sec.~\ref{subsec: PQRS} for details).

\subsection{Cubic dissipator (\texorpdfstring{$p=3$}{p=3})}

\begin{figure}[t]
\centering


\includegraphics[width=86mm]{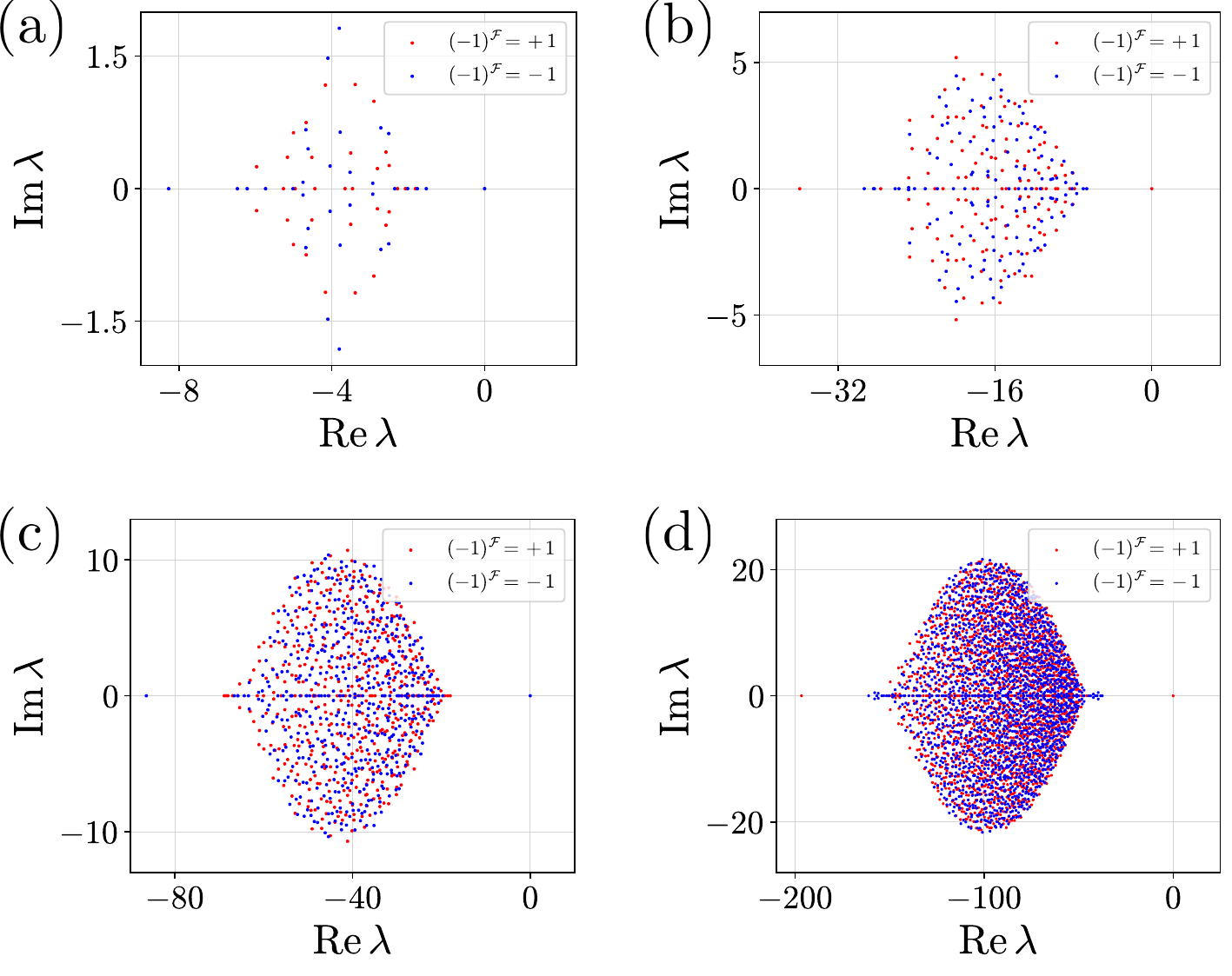} 

\caption{Complex spectrum of a single realization of the Sachdev-Ye-Kitaev (SYK) Lindbladian with cubic dissipators $p=3$ and complex coupling $K_{m;i} \in \mathbb{C}$. 
The number $N$ of fermion flavors and the strength $K$ of the dissipators are respectively chosen to be (a)~$N=6$, $K=5.21$, (b)~$N=8$, $K=8.39$, (c)~$N=10$, $K=11.8$, and (d)~$N=12$, $K=15.4$.
The other parameters are taken as $J=1$ 
and $M=2N$. 
The double Hilbert space is divided into the two subspaces according to total fermion parity $\left( -1 \right)^{\cal F}$, as shown by the red dots (even) and blue dots (odd).
}
	\label{fig: spectrum_p_3_complex_K}
\end{figure}

\begin{figure}
    \centering

    \includegraphics[width=86mm]{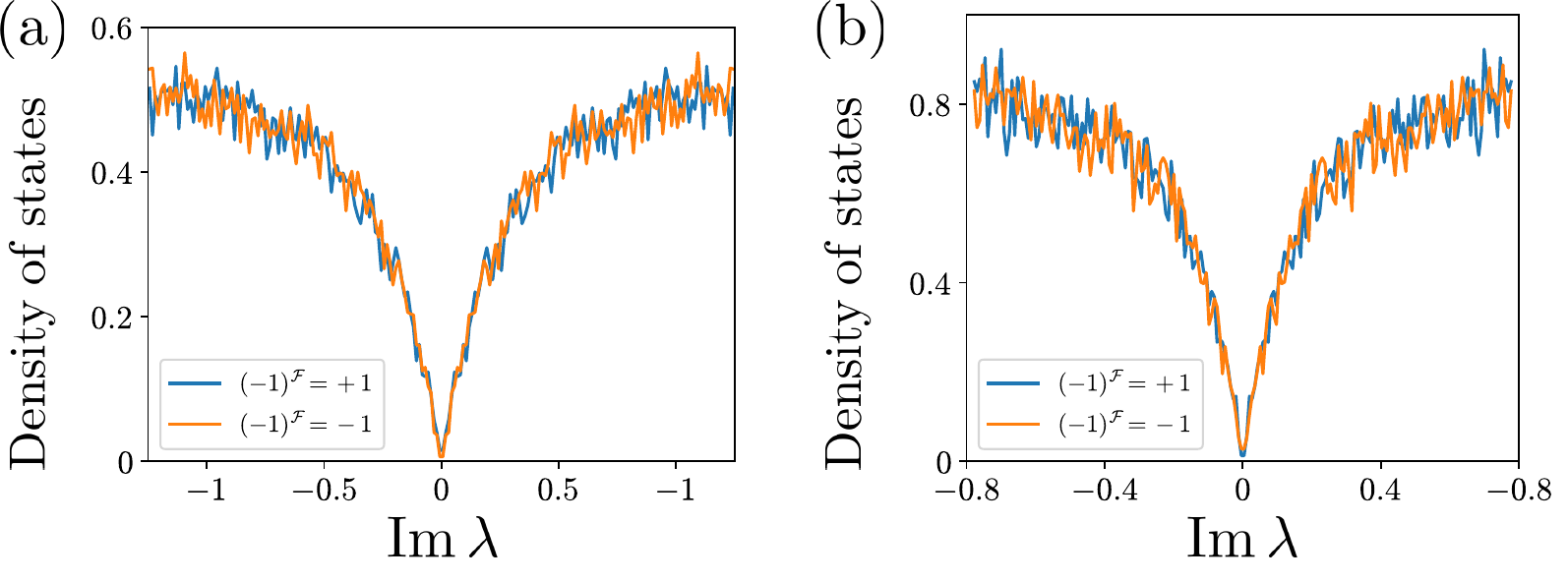} 
    
    \caption{Density of states across the real axis as a function of the imaginary part of complex eigenvalues $\lambda$ for the Sachdev-Ye-Kitaev (SYK) Lindbladians with cubic dissipators $p=3$ and complex coupling $K_{m; i} \in \mathbb{C}$.
    The number $N$ of fermion flavors is chosen to be (a)~$N=10$ and (b)~$N=12$.
    The other parameters are taken as $J=1$, $K=10$, and $M=N$.
    The double Hilbert space is divided into the two subspaces according to total fermion parity $\left( -1 \right)^{\cal F}$, as shown by the blue curves ($+$) and orange curves ($-$).
    We take eigenvalues satisfying $\left| \mathrm{Im}\,\lambda \right| > \epsilon/\sqrt{\mathrm{dim}\,\mathcal{L}}$ ($\epsilon = 10^{-4}$) and exclude real eigenvalues.
    }
    \label{fig:symmLineDecay-p_3-complex_K-fits-log}
\end{figure}

Furthermore, we study the SYK Lindbladians with the cubic dissipators $p=3$ and the complex dissipative coupling $K_{m; i} \in \mathbb{C}$.
We first obtain the complex spectrum, as shown in Fig.~\ref{fig: spectrum_p_3_complex_K}.
Similarly to the linear dissipators $p=1$, only modular conjugation symmetry is respected.
However, the complex spectrum seems more chaotic.
In fact, the complex spectrum in Fig.~\ref{fig: spectrum_p_3_complex_K} looks more like the lemon-shaped spectrum of the completely random Lindbladians~\cite{Denisov-19}.
In Fig.~\ref{fig:symmLineDecay-p_3-complex_K-fits-log}, we also obtain the density of states across the real axis $\mathrm{Im}\,\lambda = 0$.
The density of states linearly vanishes toward the real axis, which is consistent with the random-matrix results with time-reversal symmetry (i.e., class AI)~\cite{Xiao-22}.

\begin{figure}[t]
\centering

\includegraphics[width=86mm]{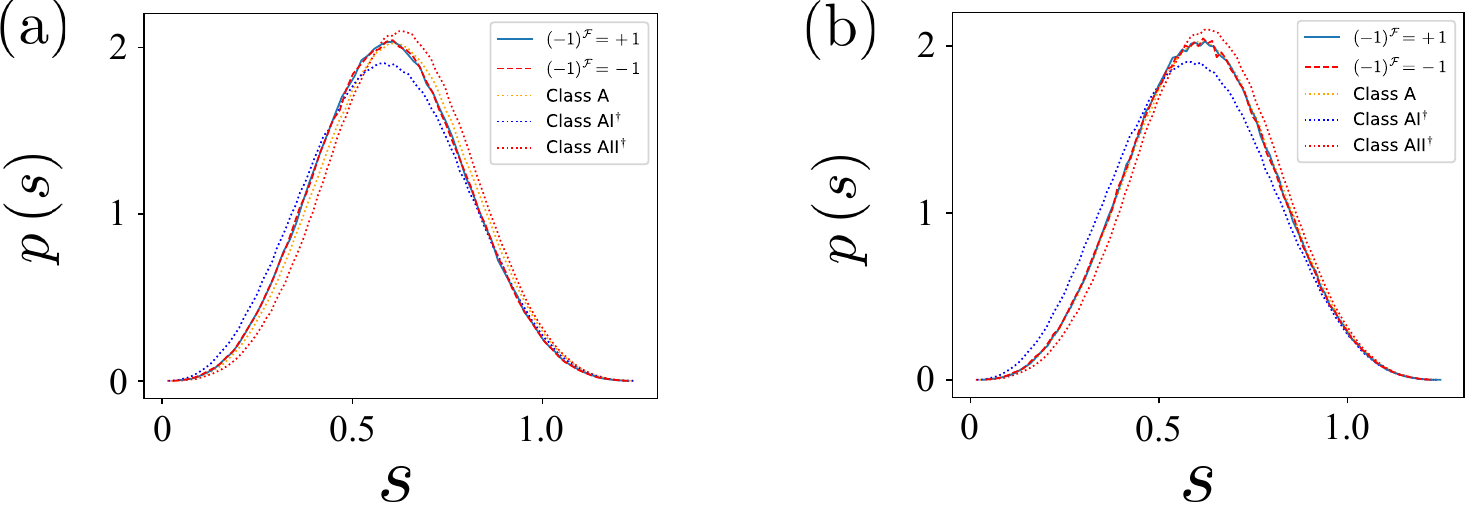} 

\caption{Distributions for normalized nearest spectral spacings $s$ of the Sachdev-Ye-Kitaev (SYK) Lindbladians with cubic dissipators $p=3$ and complex coupling $K_{m;i} \in \mathbb{C}$, compared with the random-matrix results. 
The number $N$ of fermion flavors and the strength $K$ of the dissipators are respectively chosen to be (a)~$N=10$, $K=11.8$ and (b)~$N=12$, $K=15.4$.
The other parameters are taken as $J=1$, and $M=N$. 
Each datum is averaged over (a)~$3000$ and (b)~$800$ samples. 
}
	\label{fig: spectral_spacing_p_3_complex_K}
\end{figure}

\begin{figure}
    \centering

    \includegraphics[width=86mm]{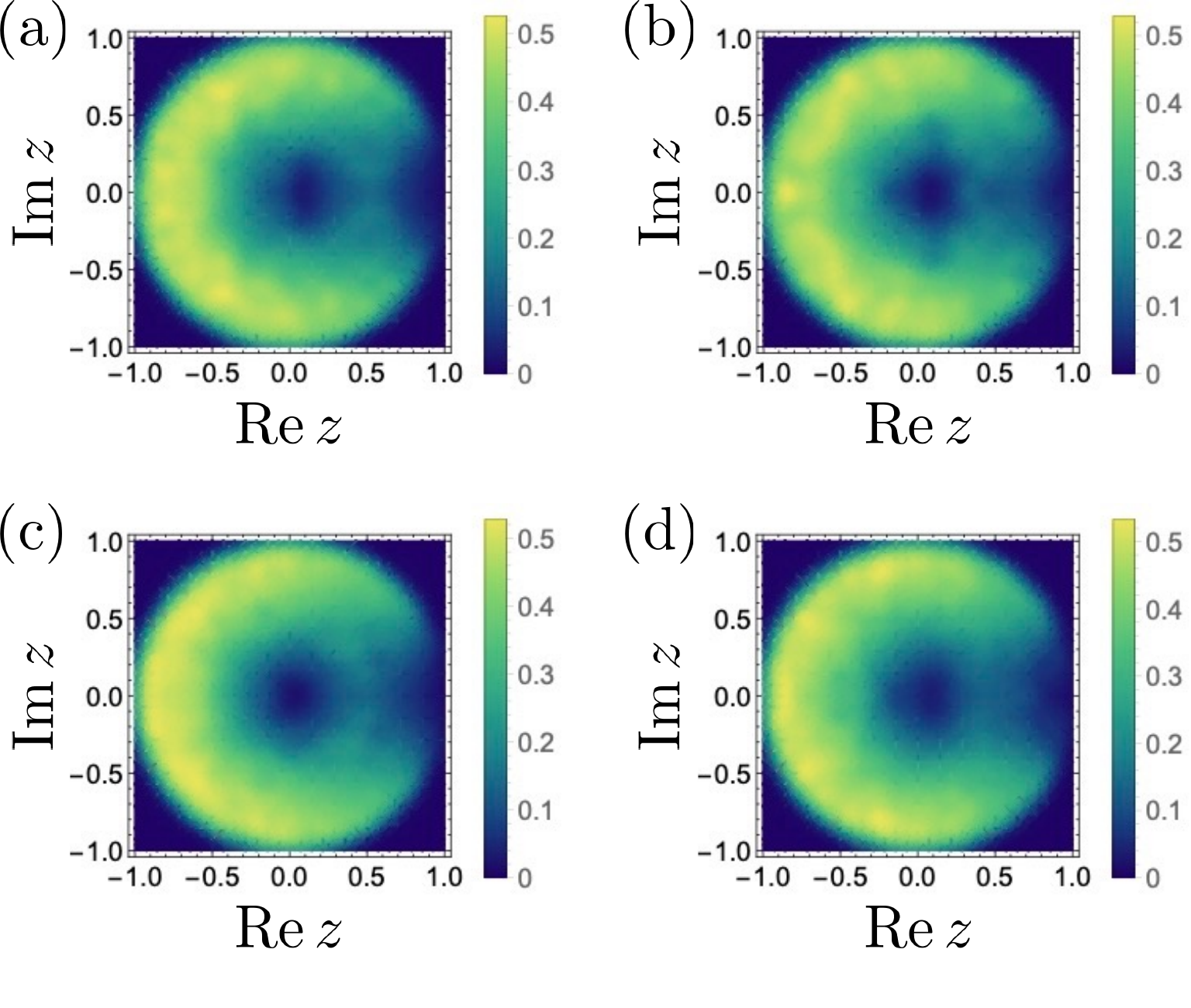}
    
    \caption{Distribution of the complex-spacing ratios of the bulk spectrum for the Sachdev-Ye-Kitaev (SYK) Lindbladians with cubic dissipators $p=3$ and complex coupling $K_{m; i} \in \mathbb{C}$ for (a, b)~$N=10$ and (c, d)~$N=12$.
    Total fermion parity $\left( -1 \right)^{\cal F}$ is chosen as (a, c)~$+$ and (b, d)~$-$. 
    The other parameters are taken as $J=1$, $K=10$, and $M=N$.
    The complex eigenvalues with $|\lambda|>0.01$ and $|\text{Im}\,\lambda|>10^{-5}$ are taken.
    Each datum is averaged over (a, b)~$96$ samples for $N=10$ and (c, d)~$24$ samples for $N=12$.
    }
    \label{cmplx-ratio-p_3-cmplx_K}
\end{figure}

\begin{figure}[t]
\centering

\includegraphics[width=86mm]{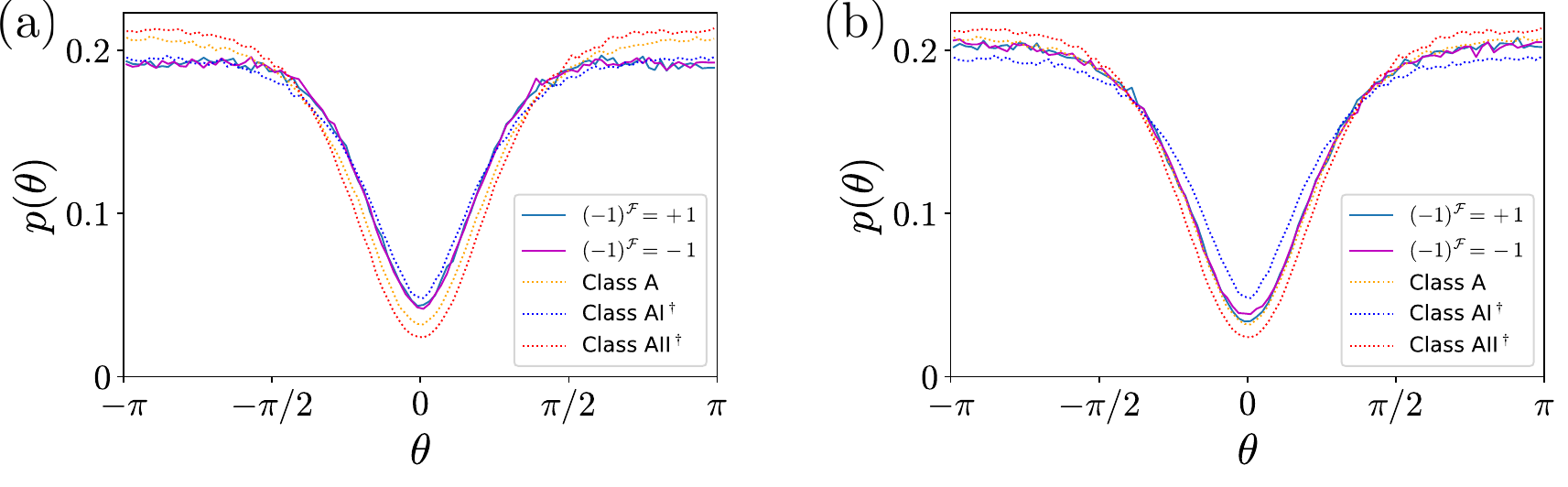}

\caption{Angle distribution of the complex-spacing ratios of the Sachdev-Ye-Kitaev (SYK) Lindbladians with cubic dissipators $p=3$ and complex coupling $K_{m;i} \in \mathbb{C}$, compared with the random-matrix results. 
The number $N$ of fermion flavors and the strength $K$ of the dissipators are respectively chosen to be (a)~$N=10$, $K=11.8$ and (b)~$N=12$, $K=15.4$.
The other parameters are taken as $J=1$ and $M=N$. 
Each datum is averaged over (a)~$3000$ and (b)~$800$ samples. 
}
	\label{fig: complex_R_p_3_complex_K}
\end{figure}

In Fig.~\ref{fig: spectral_spacing_p_3_complex_K}, we also show the complex-level-spacing distributions $p \left( s \right)$ of the SYK Lindbladians with the cubic dissipators $p=3$. 
For $N=12$, $p \left( s \right)$ agrees well with the random-matrix distribution in class A, which is consistent with the symmetry classification in Table~\ref{tab: SYK Lindbladian p=1}.
For $N=10$, on the other hand, while the peak and small-$s$ behavior of $p \left( s \right)$ agree well with the random-matrix distribution in class A, the tail (i.e., large-$s$) behavior of  $p \left( s \right)$ rather agrees with the random-matrix distribution in class AI$^{\dag}$. 
This intermediate behavior should originate from the special structure of the SYK Lindbladians, including a crossover effect between classes A and AI$^{\dag}$, as discussed above.
We also obtain the complex-spacing-ratio distributions in Figs.~\ref{cmplx-ratio-p_3-cmplx_K} and \ref{fig: complex_R_p_3_complex_K}.
Similarly to the level-spacing distributions, the level-spacing-ratio distribution for $N=12$ agrees well with the random-matrix distribution in class A, which is consistent with Table~\ref{tab: SYK Lindbladian p=1}.
For $N=10$, on the other hand, the level-spacing-ratio distribution coincides with the random-matrix distribution in class AI$^{\dag}$ rather than class A, which should again originate from the special structure of the SYK Lindbladians that cannot be captured solely by the internal symmetry classification.

\section{Discussions}
    \label{sec: conclusion}

Symmetry serves as a foundation for universal descriptions of diverse phenomena and plays a key role in physics.
Despite the significance of symmetry in closed quantum systems, symmetry of open quantum systems was not fully understood.
In this work, we have developed a theory of symmetry in open quantum systems.
Building upon the operator-state mapping, we have reduced symmetry of Liouvillian superoperators to symmetry of non-Hermitian operators in the double Hilbert space and applied the 38-fold symmetry classification of non-Hermitian operators~\cite{Bernard-LeClair-02, KSUS-19, Zhou-Lee-19}.
We have found the rich symmetry classification due to the interplay between symmetry in the corresponding closed quantum systems and symmetry inherent in the construction of Liouvillian superoperators.
As an illustrative example of open quantum bosonic systems, we have studied symmetry classes of dissipative quantum spin models.
Furthermore, we have developed the symmetry classification of the SYK Lindbladians as a prototype of open quantum fermionic many-body systems.
We have established the periodic tables~\ref{tab: SYK Lindbladian p=1} and \ref{tab: SYK Lindbladian p=2} and elucidated the difference from the counterparts in closed quantum systems.
We have also numerically studied the complex-spectral statistics of the SYK Lindbladians and demonstrated the dissipative quantum chaos enriched by symmetry.

Owing to the generality and significance of symmetry, our theory applies to a wide variety of open quantum phenomena and leads to their unified understanding.
For example, our theory is relevant to the localization transitions and topological phases of open quantum systems, which we leave for future work.
In particular, the symmetry classification of the SYK Hamiltonians is closely related to the $\mathbb{Z}_8$ topological phases of interacting fermions~\cite{Fidkowski-Kitaev-10, *Fidkowski-Kitaev-11, Turner-11}, as also discussed in Sec.~\ref{subsec: PSA}.
Similarly, our symmetry classification of the SYK Lindbladians is relevant to the topological phases of open interacting fermions.
The symmetry classification of the SYK Hamiltonians is also related to the Jackiw-Teitelboim gravity~\cite{Saad-Shenker-Stanford-19, Stanford-Witten-19}.
It is significant to consider the analogs 
in open quantum systems.
Furthermore, it is worthwhile to investigate the dynamical signatures of different complex-spectral statistics for different symmetry classes.
It also merits further study to investigate the effect of other fundamental constraints in open quantum systems, such as complete positivity, on complex-spectral statistics and dissipative quantum chaos.

Moreover, our theoretical framework is straightforwardly applicable to generic open quantum systems, encompassing non-Markovian Liouvillians, although we have focused on Markovian Liouvillians (i.e., Lindbladians) in this work for the sake of clarity.
While it is generally formidable to analyze non-Markovian Liouvillians, it is significant to study their general properties based on symmetry.
In this respect, it is also notable that generic open quantum systems can be embedded into larger closed quantum systems that consist of the original systems and their surrounding environment.
Thus, symmetry of open quantum systems should be related to symmetry of closed quantum systems in the dilated Hilbert space.
This perspective may lead to a unified understanding of physics in closed and open quantum systems, such as quantum chaos~\cite{Maldacena-Qi-18, Cao-22} and topological phases~\cite{McGinley-20}.

\section*{Acknowledgements}
We thank Lucas S\'a for useful discussions.
K.K. thanks David A. Huse for a comment that stimulated this work.
K.K. is supported by the Japan Society for the Promotion of Science (JSPS) through the Overseas Research Fellowship.
T.N. is supported
by MEXT KAKENHI Grant-in-Aid for Transformative Research Areas A ``Extreme Universe"
Grant No.~22H05248.
S.R. is supported by the National Science Foundation under Award No.~DMR-2001181, and by a Simons Investigator Grant from the Simons Foundation (Award No.~566116).
This work is supported by the Gordon and Betty Moore Foundation through Grant No.~GBMF8685 toward the Princeton theory program.

\bigskip

{\it Note added}.---When we were finalizing 
the draft, we learned 
about the related works~\cite{Sa-Ribeiro-Prosen-22,Garcia-Garcia-22unpub}.

\appendix

\section{38-fold symmetry classification of non-Hermitian operators}
    \label{asec: 38-fold symmetry}
    
\begin{table}[t]
	\centering
	\caption{Altland-Zirnbauer (AZ) and Altland-Zirnbauer$^{\dag}$ ($\text{AZ}^{\dag}$) symmetry classes for non-Hermitian operators. Time-reversal symmetry (TRS) and particle-hole symmetry (PHS) are defined by $\mathcal{T} H \mathcal{T}^{-1} = H$ and $\mathcal{C} H^{\dag} \mathcal{C}^{-1} = - H$ with the antiunitary operators $\mathcal{T}$ and $\mathcal{C}$ satisfying $\mathcal{T}^2 = \pm 1$ and $\mathcal{C}^2 = \pm 1$, respectively. Chiral symmetry (CS) is combined symmetry of TRS and PHS defined by $\Gamma H^{\dag} \Gamma^{-1} = - H$ with the unitary operator $\Gamma$ satisfying $\Gamma^{2} = 1$. The ten-fold AZ symmetry class is divided into two complex classes that only involve CS and eight real classes where TRS and PHS are relevant. Moreover, $\text{TRS}^{\dag}$ and $\text{PHS}^{\dag}$ are respectively defined by $\mathcal{T} H^{\dag} \mathcal{T}^{-1} = H$ and $\mathcal{C} H \mathcal{C}^{-1} = - H$ with the antiunitary operators $\mathcal{T}$ and $\mathcal{C}$ satisfying $\mathcal{T}^2 = \pm 1$ and $\mathcal{C}^2 = \pm 1$, which constitute the $\text{AZ}^{\dag}$ symmetry classes. Class AI (AII) in the real AZ symmetry class and class $\text{D}^{\dag}$ ($\text{C}^{\dag}$) in the real $\text{AZ}^{\dag}$ symmetry class are equivalent to each other.}
		\label{tab: AZ}
     \begin{tabular}{ccccccc} \hline \hline
    \multicolumn{2}{c}{~Symmetry class~} & ~TRS~ & ~PHS~ & ~$\text{TRS}^{\dag}$~ & ~$\text{PHS}^{\dag}$~ & ~CS~ \\ \hline
    \multirow{2}{*}{~Complex AZ~} 
    & A & $0$ & $0$ & $0$ & $0$ & $0$ \\
    & AIII & $0$ & $0$ & $0$ & $0$ & $1$ \\ \hline
    \multirow{9}{*}{Real AZ} 
    & AI & $+1$ & $0$ & $0$ & $0$ & $0$ \\
    & BDI & $+1$ & $+1$ & $0$ & $0$ & $1$ \\
    & D & $0$ & $+1$ & $0$ & $0$ & $0$ \\
    & DIII & $-1$ & $+1$ & $0$ & $0$ & $1$ \\
    & AII & $-1$ & $0$ & $0$ & $0$ & $0$ \\
    & CII & $-1$ & $-1$ & $0$ & $0$ & $1$ \\
    & C & $0$ & $-1$ & $0$ & $0$ & $0$ \\
    & CI & $+1$ & $-1$ & $0$ & $0$ & $1$ \\ \hline
    \multirow{9}{*}{Real $\text{AZ}^{\dag}$} 
    & $\text{AI}^{\dag}$ & $0$ & $0$ & $+1$ & $0$ & $0$ \\
    & $\text{BDI}^{\dag}$ & $0$ & $0$ & $+1$ & $+1$ & $1$ \\
    & $\text{D}^{\dag}$ & $0$ & $0$ & $0$ & $+1$ & $0$ \\
    & $\text{DIII}^{\dag}$ & $0$ & $0$ & $-1$ & $+1$ & $1$ \\
    & $\text{AII}^{\dag}$ & $0$ & $0$ & $-1$ & $0$ & $0$ \\
    & $\text{CII}^{\dag}$ & $0$ & $0$ & $-1$ & $-1$ & $1$ \\
    & $\text{C}^{\dag}$ & $0$ & $0$ & $0$ & $-1$ & $0$ \\
    & $\text{CI}^{\dag}$ & $0$ & $0$ & $+1$ & $-1$ & $1$ \\ \hline \hline
    \end{tabular}
\end{table}

\newcommand{\SLS}{\mathcal{S}}

\begin{table}[t]
	\centering
	\caption{Possible types [$t=0, 1$ (mod $2$)] of sublattice symmetry as additional symmetry in the complex Altland-Zirnbauer (AZ) symmetry class [$s=0, 1$ (mod $2$)]. The subscript of $\SLS_{\pm}$ specifies the commutation ($+$) or anticommutation ($-$) relation to chiral symmetry: $\Gamma \SLS_{\pm} = \pm \SLS_{\pm} \Gamma$.}
	\label{tab: symmetry - complex AZ + SLS}
	\small
     \begin{tabular}{cccc} \hline \hline
    ~$s$~ & ~AZ class~ & ~$t=0$~ & ~$t=1$~ \\ \hline
    $0$ & A & & $\SLS$ \\
    $1$ & AIII & $\SLS_{+}$ & $\SLS_{-}$ \\ \hline \hline
  \end{tabular}
\end{table}

\begin{table}[t]
	\centering
	\caption{Possible types [$t=0, 1, 2, 3$ (mod $4$)] of sublattice symmetry as additional symmetry in the real Altland-Zirnbauer (AZ) symmetry class [$s=0, 1, \cdots, 7$ (mod $8$)]. The subscript of $\SLS_{\pm}$ specifies the commutation ($+$) or anticommutation ($-$) relation between $\SLS_{\pm}$ and time-reversal symmetry (TRS) and/or particle-hole symmetry (PHS). For the symmetry classes that involve both TRS and PHS (BDI, DIII, CII, and CI), the first subscript specifies the relation to TRS and the second one to PHS. Classes AI with $\SLS_{-}$, BDI with $\SLS_{-+}$ or $\SLS_{--}$, and CII with $\SLS_{-+}$ or $\SLS_{--}$ are equivalent to classes AII with $\SLS_{-}$, DIII with $\SLS_{-+}$ or $\SLS_{--}$, and CI with $\SLS_{-+}$ or $\SLS_{--}$, respectively.}
	\label{tab: symmetry - real AZ + SLS}
	\small
     \begin{tabular}{cccccc} \hline \hline
    ~$s$~ & ~AZ class~ & ~$t=0$~ & ~$t=1$~ & ~$t=2$~ & ~$t=3$~ \\ \hline
    $0$ & AI &  & $\SLS_{-}$ &  & $\SLS_{+}$ \\
    $1$ & BDI & $\SLS_{++}$ & $\SLS_{-+}$ & $\SLS_{--}$ & $\SLS_{+-}$ \\
    $2$ & D &  & $\SLS_{+}$ &  & $\SLS_{-}$ \\
    $3$ & DIII & $\SLS_{--}$ & $\SLS_{-+}$ & $\SLS_{++}$ & $\SLS_{+-}$ \\
    $4$ & AII &  & $\SLS_{-}$ &  & $\SLS_{+}$ \\
    $5$ & CII & $\SLS_{++}$ & $\SLS_{-+}$ & $\SLS_{--}$ & $\SLS_{+-}$ \\
    $6$ & C &  & $\SLS_{+}$ &  & $\SLS_{-}$ \\
    $7$ & CI & $\SLS_{--}$ & $\SLS_{-+}$ & $\SLS_{++}$ & $\SLS_{+-}$ \\ \hline \hline
  \end{tabular}
\end{table}

In general, Hermitian operators $H$ are classified according to the two types of antiunitary symmetry, one of which is time-reversal symmetry (TRS)
\begin{equation}
    \mathcal{T} H \mathcal{T}^{-1} = H,
        \label{aeq: TRS}
\end{equation}
and the other of which is particle-hole symmetry (PHS)
\begin{equation}
    \mathcal{C} H \mathcal{C}^{-1} = - H.
        \label{aeq: PHS-dag}
\end{equation}
Here, $\mathcal{T}$ and $\mathcal{C}$ are antiunitary operators satisfying $\mathcal{T}^2 =\pm 1$ and  $\mathcal{C}^2 = \pm 1$.
As a combination of TRS and PHS, we can also introduce chiral symmetry (CS), or equivalently sublattice symmetry (SLS), by
\begin{equation}
    \mathcal{S} H \mathcal{S}^{-1} = - H,
        \label{aeq: SLS}
\end{equation}
where $\mathcal{S}$ is a unitary operator satisfying $\mathcal{S}^2 = 1$.
These two antiunitary symmetry and one unitary symmetry form the tenfold Altland-Zirnbauer (AZ) symmetry classification~\cite{AZ-97}, which determines the universality classes of Anderson localization~\cite{Evers-review} and topological phases~\cite{CTSR-review} in Hermitian systems.
We note that $H$ is assumed not to have any unitary symmetry that commutes with it (i.e., $\mathcal{U} H \mathcal{U}^{-1} = H$); 
if $H$ of interest respects such unitary symmetry, we perform the block diagonalization and then study the internal symmetry in each subspace.

In contrast to the 10-fold AZ symmetry classification for Hermitian operators, non-Hermitian operators are generally classified according to the 38-fold internal symmetry~\cite{KSUS-19}.
First, the two types of antiunitary symmetry in Eqs.~(\ref{aeq: TRS}) and (\ref{aeq: PHS-dag}) remain symmetry even for non-Hermitian operators $H$, each of which is respectively denoted by TRS and PHS$^{\dag}$~\cite{KSUS-19}.
Similarly, the unitary symmetry in Eq.~(\ref{aeq: SLS}) is still symmetry and denoted by SLS.
In addition to these symmetry, we can consider symmetry that relates non-Hermitian operators $H$ to their Hermitian conjugates $H^{\dag}$.
We can introduce two such antiunitary symmetry by
\begin{equation}
    \mathcal{T} H^{\dag} \mathcal{T}^{-1} = H,
        \label{aeq: TRS-dag}
\end{equation}
and
\begin{equation}
    \mathcal{C} H^{\dag} \mathcal{C}^{-1} = - H,
        \label{aeq: PHS}
\end{equation}
where $\mathcal{T}$ and $\mathcal{C}$ are antiunitary operators satisfying $\mathcal{T}^2 = \pm 1$ and $\mathcal{C}^2 = \pm 1$.
These symmetry is respectively denoted by time-reversal symmetry$^{\dag}$ (TRS$^{\dag}$) and particle-hole symmetry (PHS) because they are obtained by additional Hermitian conjugation to TRS and PHS$^{\dag}$ in Eqs.~(\ref{aeq: TRS}) and (\ref{aeq: PHS-dag}).
In a similar manner, we can also consider Eq.~(\ref{aeq: SLS}) with additional Hermitian conjugation by
\begin{equation}
    \Gamma H^{\dag} \Gamma^{-1} = - H,
        \label{aeq: CS}
\end{equation}
where $\Gamma$ is a unitary operator satisfying $\Gamma^2 = 1$.
Unitary symmetry in Eq.~(\ref{aeq: SLS}) is called SLS for non-Hermitian operators while unitary symmetry in Eq.~(\ref{aeq: CS}) is called CS~\cite{KSUS-19}.
Here, while CS and SLS are equivalent to each other for Hermitian operators, they are different for non-Hermitian operators.
The combination of TRS and PHS, as well as the combination of TRS$^{\dag}$ and PHS$^{\dag}$, gives rise to CS;
on the other hand, the combination of TRS and PHS$^{\dag}$, as well as the combination of TRS$^{\dag}$ and PHS, gives rise to SLS.

In a similar manner to the 10-fold AZ symmetry class for Hermitian operators, TRS in Eq.~(\ref{aeq: TRS}), PHS in Eq.~(\ref{aeq: PHS}), and CS in Eq.~(\ref{aeq: CS}) form the 10-fold symmetry class for non-Hermitian operators (Table~\ref{tab: AZ}).
Moreover, TRS$^{\dag}$ in Eq.~(\ref{aeq: TRS-dag}), PHS$^{\dag}$ in Eq.~(\ref{aeq: PHS-dag}), and CS in Eq.~(\ref{aeq: CS}) form another 10-fold symmetry class, which is called the AZ$^{\dag}$ symmetry class for non-Hermitian operators.
In these AZ and AZ$^{\dag}$ symmetry classes, SLS in Eq.~(\ref{aeq: SLS}) is not included.
Taking SLS into consideration as additional symmetry (Tables~\ref{tab: symmetry - complex AZ + SLS} and \ref{tab: symmetry - real AZ + SLS}), we have the 38-fold symmetry class for non-Hermitian operators~\cite{KSUS-19}.
In non-Hermitian random matrix theory, TRS$^{\dag}$ in Eq.~(\ref{aeq: TRS-dag}) changes the spectral statistics in the bulk~\cite{Hamazaki-20} while the other symmetry changes the spectral statistics at the symmetric point or lines~\cite{Xiao-22}.
TRS$^{\dag}$ physically means reciprocity of open quantum systems and changes the universality classes of the Anderson transitions~\cite{KR-21}.
The 10-fold classification of quadratic Lindbladians based on the AZ$^{\dag}$ symmetry class was also developed in Ref.~\cite{Lieu-20}.

Some symmetry classes in Tables~\ref{tab: AZ}, \ref{tab: symmetry - complex AZ + SLS}, and \ref{tab: symmetry - real AZ + SLS} give the equivalent symmetry classes, which are not double counted in the 38-fold symmetry classification.
For example, if a non-Hermitian operator $H$ respects TRS in Eq.~(\ref{aeq: TRS}), another non-Hermitian operator $\ii H$ respects PHS$^{\dag}$ in Eq.~(\ref{aeq: PHS-dag}), both of which exhibit essentially the same universal spectral statistics.
Consequently, classes AI and AII are respectively equivalent to classes D$^{\dag}$ and C$^{\dag}$ and characterized by the same universality classes (Table~\ref{tab: AZ}).
Similarly, if a non-Hermitian operator $H$ respects pseudo-Hermiticity $\eta H^{\dag} \eta^{-1} = H$ ($\eta^2 = 1$, $\eta^{\dag} = \eta$)~\cite{Mostafazadeh-02-1, *Mostafazadeh-02-2, *Mostafazadeh-02-3} and belongs to class A + $\eta$, another non-Hermitian operator $\ii H$ respects CS $\eta \left( \ii H \right) \eta^{-1} = - \left( \ii H \right)$ in Eq.~(\ref{aeq: CS}) and hence belongs to class AIII.
It is also notable that some examples of open quantum bosonic and fermionic systems studied in this work belong to class BDI$^{\dag}$ + $\mathcal{S}_{++}$, in which the shifted Lindbladians respect TRS$^{\dag}$ in Eq.~(\ref{aeq: TRS-dag}), PHS$^{\dag}$ in Eq.~(\ref{aeq: PHS-dag}), and SLS in Eq.~(\ref{aeq: SLS}).
In such a case, they also respect TRS in Eq.~(\ref{aeq: TRS}), PHS in Eq.~(\ref{aeq: PHS}), and SLS in Eq.~(\ref{aeq: SLS}), and hence belongs to class BDI + $\mathcal{S}_{++}$.
In a similar manner, class CI$^{\dag}$ + $\mathcal{S}_{+-}$ is equivalent to class BDI + $\mathcal{S}_{-+}$ (see also Table~XIII in Ref.~\cite{KSUS-19} for the correspondence between symmetry classes).

\section{Non-Hermitian random matrices with time-reversal symmetry}
    \label{asec: NH RMT TRS}

Time-reversal symmetry plays an important role in the universal spectral statistics of Hermitian and non-Hermitian random matrices.
In open quantum systems, time-reversal symmetry also leads to the unique complex-spectral statistics, as shown by the dissipative quantum spin models in Sec.~\ref{subsec: boson} and the SYK Lindbladians in Sec.~\ref{sec: dissipative quantum chaos}.
Here, we discuss the complex-spectral statistics of non-Hermitian random matrices with time-reversal symmetry, using two-by-two matrices in a similar manner to the Wigner surmise.
While time-reversal symmetry with the sign $+1$ (i.e., class AI) leads to the linear decay of the density of states toward the real axis and a number of eigenvalues on the real axis, time-reversal symmetry with the sign $-1$ (i.e., class AII) leads to the quadratic decay of the density of states toward the real axis and the absence of eigenvalues on the real axis.
These behaviors also appear in generic non-Hermitian random matrices with time-reversal symmetry~\cite{Ginibre-65}.
More extensive numerical and analytical studies for generic non-Hermitian random matrices are found in Ref.~\cite{Xiao-22}.

\subsection{Class A}

The density of states of non-Hermitian random matrices without symmetry is known to be uniform in the complex plane (i.e., circular law~\cite{Ginibre-65, Girko-85}):
\begin{equation}
\rho 
= \frac{N}{\pi r^2},
\end{equation}
where $N$ is the number of complex eigenvalues, and $r$ is the radius of the complex spectrum.
Hence, the density of states along the imaginary axis is obtained as (i.e., Wigner semicircle law~\cite{Wigner-51, *Wigner-58, Dyson-62})
\begin{equation}
\rho \left( s \right)
= \int_{-\sqrt{r^2-s^2}}^{\sqrt{r^2-s^2}} dx~\rho
= \frac{2 N \sqrt{r^2 - s^2}}{\pi r^2},
\end{equation}
where $s$ denotes the imaginary part of the complex spectrum.
Thus, the density of states reaches the maximum at the real axis $s=0$ and gradually decreases away from the real axis.
No level repulsion or singular behavior appears around the real axis $s=0$, which contrasts with the subsequent cases with time-reversal symmetry.

The density of states in open quantum systems does not completely coincide with that of non-Hermitian random matrices in the corresponding symmetry class even if the systems are nonintegrable.
Nevertheless, qualitatively similar behavior appears in nonintegrable open quantum systems.
For example, the SYK Lindbladians with the even number $p$ of dissipators belong to class A for $N \equiv 2$ (mod $4$), $q \equiv 0$ (mod $4$), and $K_{m; i} K_{m; j}^{*} \notin \mathbb{R}$ (Table~\ref{tab: SYK Lindbladian p=2}).
As shown in Fig.~\ref{fig:symmLineDecay-p_2-complex_K-fits-log}, the density of states across the real axis continuously changes and does not exhibit the level repulsion, which contrasts with the singular behavior of the density of states around the real axis in the presence of time-reversal symmetry.

\subsection{Class AI}

A generic two-by-two non-Hermitian matrix $h$ in class AI, which is required to respect time-reversal symmetry
\begin{equation}
h^{*} = h,
\end{equation}
is given as
\begin{equation}
h = c + x \sigma_x + \ii y \sigma_y + z \sigma_{z}
\end{equation}
with $c, x, y, z \in \mathbb{R}$.
The two eigenvalues of $h$ are obtained as
\begin{equation}
E_{\pm} = c \pm \sqrt{x^2 - y^2 + z^2}.
\end{equation}
The constant term $c$ just shifts the spectrum along the real axis, and hence we omit it in the following.

We calculate the density of states along the imaginary axis, $\rho \left( s \right) \coloneqq \braket{\delta \left( s - \mathrm{Im}\,E_{+} \right)} + \braket{\delta \left( s - \mathrm{Im}\,E_{-} \right)}$, where the brackets denote the ensemble averages over the different realizations of $h$.
Suppose that $h$ obeys the Gaussian probability distribution function
\begin{equation}
\propto e^{-\beta\,\mathrm{tr}\,[ h^{\dag} h ]/2} = e^{-\beta\,( x^2 + y^2 + z^2 )}
    \label{aeq: Gaussian PDF}
\end{equation}
with a constant $\beta > 0$. 
Here, $\rho \left( s \right)$ is an even function in terms of $s$.
Then, $\rho \left( s \right)$ reads
\begin{align}
&\rho \left( s \right) 
= \frac{1}{N} \int_{-\infty}^{\infty} dx  \int_{-\infty}^{\infty} dy \int_{-\infty}^{\infty} dz \nonumber \\
&\qquad\delta \left( s - \mathrm{Im} \left[ \sqrt{x^2 - y^2 + z^2}\right] \right) e^{-\beta\,( x^2 + y^2 + z^2 )}
\end{align}
for $s \geq 0$,
with the normalization constant
\begin{align}
    N \coloneqq \int_{-\infty}^{\infty} dx  \int_{-\infty}^{\infty} dy \int_{-\infty}^{\infty} dz~e^{-\beta\,( x^2 + y^2 + z^2 )}
= \left( \frac{\pi}{\beta} \right)^{3/2}.
    \label{aeq: RMT - normalization constant}
\end{align}
Introducing the polar coordinate by
\begin{equation}
x \eqqcolon r \cos \theta,\quad
z \eqqcolon r \sin \theta
\end{equation}
with $r>0$ and $\theta \in \left[0, 2\pi\right)$,
we have
\begin{align}
&\rho \left( s \right) = \frac{2\pi}{N} \int_{0}^{\infty} dr~r \int_{-\infty}^{\infty} dy \nonumber \\
&\qquad\qquad\qquad\delta \left( s - \mathrm{Im} \left[ \sqrt{r^2 - y^2}\right] \right) e^{-\beta\,( r^2 + y^2 )}.
\end{align}
Importantly, this integral is singular at the real axis $s=0$.

Let us first calculate $\rho \left( s \right)$ away from the real axis $s >0$. 
For $\delta\,( s - \mathrm{Im}\,[ \sqrt{r^2 - y^2} ] ) \neq 0$ with $s >0$, we need $r < \left| y\right|$ and $y = y_{\pm} \coloneqq \pm \sqrt{r^2+s^2}$. Then, we have
\begin{align}
&\delta \left( s - \mathrm{Im} \left[ \sqrt{r^2 - y^2}\right] \right) \nonumber \\
&\qquad= \frac{s}{\sqrt{r^2+s^2}} \left[ \delta \left( y - y_{+} \right) + \delta \left( y - y_{-} \right) \right],
\end{align}
and hence
\begin{align}
\rho \left( s \right)
&= \frac{4\pi s e^{\beta s^2}}{N} \int_{0}^{\infty} dr~\frac{r e^{-2\beta\,(r^2+s^2)}}{\sqrt{r^2+s^2}} \nonumber \\
&= \frac{4\pi s e^{\beta s^2}}{N\sqrt{2\beta}} \int_{\sqrt{2\beta}s}^{\infty} dt\,e^{-t^2}\quad \left( t \coloneqq \sqrt{2\beta \left( r^2 + s^2 \right)} \right) \nonumber \\
&= \sqrt{2} \beta s e^{\beta s^2}\,\mathrm{erfc}\,( \sqrt{2\beta} s ),
\end{align}
where 
\begin{equation}
\mathrm{erfc} \left( x \right) \coloneqq \frac{2}{\sqrt{\pi}} \int_{x}^{\infty} dt~e^{-t^2}
= 1 - \mathrm{erf} \left( x \right)
\end{equation}
is the complementary error function. 
Using the expansion $\mathrm{erfc} \left( x \right) \simeq 1 - 2x/\sqrt{\pi} + \mathcal{O}\,( x^2 )$, we have
\begin{equation}
\rho \left( s \right) \simeq \sqrt{2} \beta s
\end{equation}
for small $s$, which implies the level repulsion around the real axis. 
In addition, from the formula
\begin{equation}
\int_{0}^{\infty} dx~\left[ x e^{x^2}\,\mathrm{erfc}\,( \sqrt{2} x ) \right]
= \frac{\sqrt{2} - 1}{2},
\end{equation}
we have 
\begin{align}
\int_{0}^{\infty} ds~\rho \left( s \right)
&= \sqrt{2} \int_{0}^{\infty} dx~\left[ x e^{x^2}\,\mathrm{erfc}\,( \sqrt{2} x ) \right] \nonumber \\
&= 1 - \frac{1}{\sqrt{2}} \neq 1.
\end{align}
The remaining probability is compensated by the singular behavior at $s = 0$. 
In fact, for $s = 0$ and $\delta\,( s - \mathrm{Im}\,[ \sqrt{r^2 - y^2} ] ) \neq 0$, we need $r \geq \left| y\right|$. 
Then, the density of states diverges at the real axis $s=0$.
Hence, $\rho \left( s \right)$ contains the delta function $N_0 \delta \left( s \right)$, 
where the normalization constant $N_0 = \sqrt{2}$ is determined so that we will have $\int_{0}^{\infty} ds~\rho \left( s \right) = 1$.

In summary, the density of states along the imaginary axis in class AI is given as
\begin{equation}
\rho \left( s \right) = \sqrt{2} \beta 
\left| s \right|
e^{\beta s^2}\,\mathrm{erfc}\,( \sqrt{2\beta} 
\left| s \right|
) + \sqrt{2}\,\delta \left( s \right)
\end{equation}
for arbitrary $s$.
A unique feature due to time-reversal symmetry is the level repulsion around the real axis $s=0$.
In fact, $\rho \left( s \right)$ linearly decays toward the real axis.
Another characteristic feature is the presence of a number of real eigenvalues, as represented by the delta function in $\rho \left( s \right)$.
This behavior is understood by the change of the spectral correlations due to time-reversal symmetry.
In general, time-reversal symmetry with the sign $+1$ suppresses the level repulsion~\cite{Haake-textbook}. 
Thus, the level repulsion on the real axis is weaker than the level repulsion between generic complex eigenvalues away from the real axis, which results in the presence of a number of real eigenvalues.
While the above results of the density of states are obtained for the two-by-two matrix, qualitatively similar results appear also for generic non-Hermitian random matrices in class AI~\cite{Ginibre-65, Xiao-22}.

As discussed in Secs.~\ref{sec: general - boson} and \ref{sec: general - fermion}, generic bosonic and fermionic Lindbladians are invariant under modular conjugation.
Consequently, in sufficiently nonintegrable open quantum systems, the density of states vanishes toward the real axis, and a subextensive number of eigenvalues appear on the real axis.
We confirm this universal behavior for the SYK Lindbladians with the different numbers $p$ of dissipators and the different numbers $N$ of fermion flavors (see Sec.~\ref{sec: dissipative quantum chaos}).
The same behavior arises also in open quantum bosonic systems, including the dissipative quantum spin models studied in Sec.~\ref{subsec: boson}.
Notably, the dephasing XYZ model in Eqs.~(\ref{eq: spin-dephasing}) and (\ref{eq: XYZ}) respects additional time-reversal symmetry (or equivalently, particle-hole symmetry$^{\dag}$) in Eq.~(\ref{eq: spin-PHS-dag}).
As a consequence of this symmetry, a number of complex eigenvalues appear on the symmetric line $\mathrm{Re}\,\lambda = \mathrm{tr}\,\mathcal{L}/\mathrm{tr}\,I$, and the level repulsion around this symmetric line is observed [Fig.~\ref{fig: XYZ}\,(a)].
By contrast, in the presence of an additional magnetic field that breaks this symmetry, the level repulsion and the concomitant singularity of the density of states around the symmetric line disappear [Fig.~\ref{fig: XYZ}\,(b)], which is consistent with the universal behavior of non-Hermitian random matrices in classes A and AI.

\subsection{Class AII}

A generic two-by-two non-Hermitian matrix in class AII, which is required to respect time-reversal symmetry
\begin{equation}
\sigma_{y} h^{*} \sigma_{y} = h,
\end{equation}
is given as
\begin{equation}
h = c + \ii \left( x \sigma_x + y \sigma_y + z \sigma_{z} \right)
\end{equation}
with $c, x, y, z \in \mathbb{R}$.
In contrast to class AI, the sign of time-reversal symmetry is $\sigma_y \sigma_y^{*} = -1$, which results in the different level statistics.
The two eigenvalues of $h$ are obtained as
\begin{equation}
E_{\pm} = c \pm \ii \sqrt{x^2 + y^2 + z^2}.
\end{equation}
Similarly to the previous case for class AI, we omit the constant term $c$ in the following. 
Notably, $h$ is an anti-Hermitian matrix; 
$\ii h$ is a generic Hermitian matrix in class A. 
Consequently, the density of states along the imaginary axis is the same as the density of states for two-by-two Hermitian random matrices in class A. 

Suppose that $h$ obeys the probability distribution function in Eq.~(\ref{aeq: Gaussian PDF}). 
Then, the density of states along the imaginary axis, $\rho \left( s \right) \coloneqq \braket{\delta \left( s - \mathrm{Im}\,E_{+} \right)} + \braket{\delta \left( s - \mathrm{Im}\,E_{-} \right)}$, reads
\begin{align}
&\rho \left( s \right) 
= \frac{1}{N} \int_{-\infty}^{\infty} dx  \int_{-\infty}^{\infty} dy \int_{-\infty}^{\infty} dz \nonumber \\
&\qquad\qquad\delta \left( s - \sqrt{x^2 + y^2 + z^2} \right) e^{-\beta\,( x^2 + y^2 + z^2 )}
\end{align}
with the normalization constant in Eq.~(\ref{aeq: RMT - normalization constant}).
Introducing the polar coordinate with $r \coloneqq \sqrt{x^2 + y^2 + z^2}$, we have
\begin{align}
\rho \left( s \right)
&= \frac{4\pi}{N} \int_{0}^{\infty} dr~r^2~\delta \left( s - r \right) e^{-\beta r^2} \nonumber \\
&= 4 \sqrt{\frac{\beta^3}{\pi}} s^2 e^{-\beta s^2}
\end{align}
for arbitrary $s$.
Consistently, we have
\begin{equation}
\int_{-\infty}^{\infty} ds~\rho \left( s \right) 
= \frac{8}{\sqrt{\pi}} \int_{0}^{\infty} dx~x^2 e^{-x^2}
= 2.
\end{equation}

An important feature in class AII is the quadratic decay of the density of states toward the real axis $s=0$, which contrasts with the linear decay in class AI.
Furthermore, no real eigenvalues are generally present, and the density of states exhibits no delta-function peak at the real axis $s=0$.
In fact, time-reversal symmetry with the sign $-1$ enhances the level repulsion~\cite{Haake-textbook}.
Thus, the level repulsion on the real axis is stronger than the level repulsion between generic complex eigenvalues away from the real axis, which results in the absence of real eigenvalues.
The qualitatively similar behavior arises also for generic non-Hermitian random matrices in class AII~\cite{Ginibre-65, Xiao-22}.

Sufficiently nonintegrable open quantum systems in class AII also exhibit the similar universal behavior for the density of states.
As a prime example, the SYK Lindbladians with the linear dissipators $p=1$  respect time-reversal symmetry $\mathcal{P}$ with the sign $-1$ for $N \equiv 2$ (mod $4$), $q \equiv 0$ (mod $4$), and $K_{m; i} K_{m; j}^{*} \in \mathbb{R}$ (Table~\ref{tab: SYK Lindbladian p=1}).
As shown in Figs.~\ref{fig: spectrum_p_1_real_K} and \ref{fig:symmLineDecay-p_1-real_K-fits-quadratic}, no eigenvalues appear on the symmetric line $\mathrm{Re}\,\lambda = \mathrm{tr}\,\mathcal{L}/\mathrm{tr}\,I$, which is consistent with the random-matrix behavior in class AII.
While the density of states decays toward the symmetric line linearly instead of quadratically, this linear decay is due to additional time-reversal symmetry$^{\dag}$ (see ``class CI$^{\dag}$" in Table~I of Ref.~\cite{Xiao-22}).

\section{Symmetry classification of the Sachdev-Ye-Kitaev Hamiltonians}
    \label{asec: symmetry classification - SYK Hamiltonian}

\begin{table}[t]
	\centering
	\caption{Periodic table of the Sachdev-Ye-Kitaev (SYK) Hamiltonians for $q \equiv 0, 2$ (mod $4$) and the number $N$ (mod $8$) of Majorana fermions. For the entries of the antiunitary symmetry $P$ and $R$, the signs $\pm 1$ mean $P^2$ and $R^2$.}
     \begin{tabular}{c|cccccccc} \hline \hline
     ~~$N$ (mod $8$)~~ & ~~$0$~~ & ~~$1$~~ & ~~$2$~~ & ~~$3$~~ &  ~~$4$~~ & ~~$5$~~ &  ~~$6$~~ & ~~$7$~~ \\ \hline
     ${P}$ & $+1$ & $+1$ & $+1$ & $-1$ & $-1$ & $-1$ & $-1$ & $+1$\\
     ${R}$ & $+1$ & $+1$ & $-1$ & $-1$ & $-1$ & $-1$ & $+1$ & $+1$ \\ \hline
     $q \equiv 0$ (mod $4$) & AI & AI & A & AII & AII & AII & A & AI \\
     $q \equiv 2$ (mod $4$) & D & D & A & C & C & C & A & D \\ \hline \hline 
    \end{tabular}
	\label{tab: SYK Hamiltonian}
\end{table}

We describe the symmetry classification of the SYK Hamiltonians (Table~\ref{tab: SYK Hamiltonian})~\cite{You-17, Fu-16, GarciaGarcia-16, Cotler-17, Li-17, Kanazawa-17, Behrends-19, Sun-20}.
We consider the following SYK Hamiltonian that includes generic all-to-all $q$-body Majorana fermions $\psi$:
\begin{equation}
    H = \ii^{q/2} \sum_{1\leq i_1 < \cdots < i_q \leq N} J_{i_1, \cdots, i_q} \psi_{i_1} \cdots \psi_{i_q},
\end{equation}
where $q$ is assumed to be even, and $J_{i_1, \cdots, i_q}$'s are the real random coupling drawn from the Gaussian distribution.
This Hamiltonian includes $N$ Majorana fermions, which satisfy 
\begin{align}
    \{ \psi_{i}, \psi_{j} \} = \delta_{ij},\quad
    \psi_{i}^{\dag} = \psi_{i}.
\end{align}
Here, we choose $\psi_i$ with odd $i$ to be real and symmetric and $\psi_i$ with even $i$ to be pure imaginary and antisymmetric so that the corresponding complex fermion operators will be real.
The Hamiltonian part of the SYK Lindbladian [i.e., Eq.~(\ref{eq: SYK Hamiltonian})] coincides with the SYK Hamiltonian.
In the following, we explicitly provide the relevant antiunitary symmetry operations of the SYK Hamiltonian $H$, depending on whether $N$ is even or odd.

\subsection{Even \texorpdfstring{$N$}{N}}

Let us introduce the antiunitary operators
\begin{align}
    P &\coloneqq \left( \prod_{i=1}^{N/2} \sqrt{2} \psi_{2i-1} \right) \mathcal{K}, \\
    R &\coloneqq \left( \prod_{i=1}^{N/2} \ii \sqrt{2} \psi_{2i} \right) \mathcal{K},
\end{align}
where $\mathcal{K}$ denotes complex conjugation.
The antiunitary operator $P$ contains all the Majorana fermions $\psi_{n}$'s with odd $n$ while $R$ contains all $\psi_n$'s with even $n$.
These antiunitary operators satisfy
\begin{align}
    P^2 &= \left( -1 \right)^{\left( N/4 \right) \left( N/2 - 1 \right)} \nonumber \\
    &= \begin{cases}
    +1 & \left[ N \equiv 0,2 \left( \text{mod}~8 \right) \right]; \\
    -1 & \left[ N \equiv 4,6 \left( \text{mod}~8 \right) \right], \\
    \end{cases} \\
    R^2 &= \left( -1 \right)^{\left( N/4 \right) \left( N/2 + 1 \right)} \nonumber \\
    &= \begin{cases}
    +1 & \left[ N \equiv 0,6 \left( \text{mod}~8 \right) \right]; \\
    -1 & \left[ N \equiv 2,4 \left( \text{mod}~8 \right) \right]. 
    \end{cases}
\end{align}
Thus, the signs of $P^2$ and $R^2$ change according to $N$ (mod $8$), summarized as Table~\ref{tab: SYK Hamiltonian}.
Moreover, we have
\begin{align}
    P \psi_{i} P^{-1} &= \left( -1 \right)^{N/2+1} \psi_{i}, \\
    R \psi_{i} R^{-1} &= \left( -1 \right)^{N/2} \psi_{i},
\end{align}
leading to
\begin{align}
    P H P^{-1} 
    &= \left( -1 \right)^{q \left( N+3 \right)/2} H \nonumber \\
    &= \begin{cases}
    + H & \left[ q \equiv 0 \left( \text{mod}~4 \right)\right]; \\
    - H & \left[ q \equiv 2 \left( \text{mod}~4 \right)\right],
    \end{cases} \\
    R H R^{-1} 
    &= \left( -1 \right)^{q \left( N+1 \right)/2} H \nonumber \\
    &= \begin{cases}
    + H & \left[ q \equiv 0 \left( \text{mod}~4 \right) \right]; \\
    - H & \left[ q \equiv 2 \left( \text{mod}~4 \right) \right].
    \end{cases}
\end{align}
Thus, the antiunitary operators $P$ and $R$ individually give time-reversal symmetry for $q \equiv 0$ (mod $4$) and particle-hole symmetry for $q \equiv 2$ (mod $4$).

It is also notable that the product of the two antiunitary operators $P$ and $R$ gives rise to fermion parity symmetry:
\begin{equation}
    PR \propto \left( -1 \right)^{F} \coloneqq \prod_{i=1}^{N/2} 2 \ii \psi_{2i-1} \psi_{2i},
\end{equation}
satisfying 
\begin{align}
    \left( -1 \right)^{F} \psi_{i} \left( -1 \right)^{F} = - \psi_{i}
\end{align}
and 
\begin{equation}
    \left( -1 \right)^{F} H \left( -1 \right)^{F} = H.
\end{equation}
Since the fermion parity gives $\mathbb{Z}_2$ unitary symmetry that commutes with the Hamiltonian $H$, we need to consider the block-diagonalized Hamiltonian to study the symmetry classes and spectral statistics.
If we calculate the spectral statistics for all eigenenergies with both $\left( -1 \right)^{F} = +1$ and $\left( -1 \right)^{F} = -1$, we always obtain the Poisson statistics since the eigenenergies with $\left( -1 \right)^{F} = +1$ and those with $\left( -1 \right)^{F} = -1$ are uncorrelated with each other.
Now, we have the commutation and anticommutation relations
\begin{align}
    \left( -1 \right)^{F} P \left( -1 \right)^{F} &= \left( -1 \right)^{N/2} P, \\
    \left( -1 \right)^{F} R \left( -1 \right)^{F} &= \left( -1 \right)^{N/2} R.
\end{align}
Consequently, the antiunitary operations $P$ and $R$ act on each subspace of fermion parity $\left( -1 \right)^{F}$ and remain symmetry for even $N/2$ [i.e., $N \equiv 0, 4$ (mod $8$)];
by contrast, they switch fermion parity between the two subspaces of $\left( -1 \right)^{F}$ and are no longer symmetry in the subspace with fixed $\left( -1 \right)^{F}$ for odd $N/2$ [i.e., $N \equiv 2, 6$ (mod $8$)].
These considerations lead to the periodic table~\ref{tab: SYK Hamiltonian} for even $N$.

\subsection{Odd \texorpdfstring{$N$}{N}}

For odd $N$, we need to modify the symmetry operations.
We introduce the antiunitary operators 
\begin{align}
    P &\coloneqq \left( \prod_{i=1}^{\left( N+1 \right)/2}  \sqrt{2} \psi_{2i-1} \right) \mathcal{K}, \\
    R &\coloneqq \left( \prod_{i=1}^{\left(N-1\right)/2} \ii \sqrt{2} \psi_{2i} \right) \mathcal{K},
\end{align}
which satisfy 
\begin{align}
    P^2 = R^2 
    &= \left( -1 \right)^{( N^2-1 )/8} \nonumber \\
    &= \begin{cases}
    +1 & \left[ N \equiv 1,7 \left( \text{mod}~8 \right) \right]; \\
    -1 & \left[ N \equiv 3,5 \left( \text{mod}~8 \right) \right]. \\
    \end{cases}
\end{align}
In addition, we have 
\begin{align}
    P \psi_{i} P^{-1} = R \psi_{i} R^{-1} = \left( -1 \right)^{\left( N-1 \right)/2} \psi_{i},
\end{align}
and hence
\begin{align}
    P H P^{-1} = R H R^{-1}
    &= \left( -1 \right)^{qN/2} H \nonumber \\
    &= \begin{cases}
    + H & \left[ q \equiv 0 \left( \text{mod}~4 \right)\right]; \\
    - H & \left[ q \equiv 2 \left( \text{mod}~4 \right)\right].
    \end{cases}
\end{align}
While the antiunitary operations $P$ and $R$ act differently for even $N$, they act in a similar manner for odd $N$.

For odd $N$, the meaning of fermion parity is ambiguous.
Correspondingly, the product of the two antiunitary operators $P$ and $R$ does not give rise to fermion parity symmetry. 
Rather, we have
\begin{align}
    (PR)\,\psi_{i}\,(PR)^{-1} = \psi_i,
\end{align}
which means that $PR$ acts as the identify operator.
As a result, the antiunitary operators $P$ and $R$ act equivalently to each other. 
The symmetry classification for odd $N$ is also summarized as the periodic table~\ref{tab: SYK Hamiltonian}.

\bibliography{SYK-Lindblad}

\end{document}